**Title:** The Genomic Signature of Crop-Wild Introgression in Maize

**Short Title:** Crop-Wild Introgression in Maize

**Authors:** Matthew B. Hufford[1], Pesach Lubinksy[2], Tanja Pyhäjärvi[1], Michael T. Devengenzo[1], Norman C. Ellstrand[3], and Jeffrey Ross-Ibarra[1,4]

**Affiliations:** **1** Department of Plant Sciences, University of California, Davis, California, United States of America, **2** Foreign Agricultural Service, United States Department of Agriculture, Washington, D.C., United States of America, **3** Department of Botany and Plant Sciences, University of California, Riverside, California, United States of America, **4** Genome Center and Center for Population Biology, University of California, Davis, California, United States of America

**Corresponding Author:** Jeffrey Ross-Ibarra;  Email: rossibarra@ucdavis.edu

**Abstract**

The evolutionary significance of hybridization and subsequent introgression has long been appreciated, but evaluation of the genome-wide effects of these phenomena has only recently become possible.  Crop-wild study systems represent ideal opportunities to examine evolution through hybridization.  For example, maize and the conspecific wild teosinte *Zea mays* ssp. *mexicana*, (hereafter, *mexicana*) are known to hybridize in the fields of highland Mexico. Despite widespread evidence of gene flow, maize and *mexicana* maintain distinct morphologies and have done so in sympatry for thousands of years. Neither the genomic extent nor the evolutionary importance of introgression between these taxa is understood. In this study we assessed patterns of genome-wide introgression based on 39,029 single nucleotide polymorphisms genotyped in 189 individuals from nine sympatric maize-*mexicana* populations and reference allopatric populations. While portions of the maize and *mexicana* genomes were particularly resistant to introgression (notably near known cross-incompatibility and domestication loci), we detected widespread evidence for introgression in both directions of gene flow. Through further characterization of these regions and preliminary growth chamber experiments, we found evidence suggestive of the incorporation of adaptive *mexicana* alleles into maize during its expansion to the highlands of central Mexico. In contrast, very little evidence was found for adaptive introgression from maize to *mexicana*. The methods we have applied here can be replicated widely, and such analyses have the potential to greatly informing our understanding of evolution through introgressive hybridization. Crop species, due to their exceptional genomic resources and frequent histories of spread into sympatry with relatives, should be particularly influential in these studies.

**Author Summary**

Hybridization and introgression have been shown to play a critical role in the evolution of species. These processes can generate the diversity necessary for novel adaptations and continued diversification of taxa. Previous research has suggested that not all regions of a genome are equally permeable to introgression. We have conducted one of the first genome-wide assessments of patterns of reciprocal introgression in plant populations. We found evidence that suggests domesticated maize received adaptations to highland conditions from a wild relative, teosinte, during its spread to the high elevations of central Mexico. Gene flow appeared asymmetric, favoring teosinte introgression into maize, and was widespread across populations at putatively adaptive loci. In contrast, regions near known domestication and cross-incompatibility loci appeared particularly resistant to introgression in both directions of gene flow. Crop-wild study systems should play an important role in future studies of introgression due to their well-developed genomic resources and histories of reciprocal gene flow during crop expansion.

**Introduction**

Hybridization and subsequent introgression have long been appreciated as agents of evolution. Adaptations can be transferred through these processes upon secondary contact of uniquely adapted populations or species, in many instances producing the variation necessary for further diversification [1]. Early considerations of adaptive introgression discussed its importance in the context of both domesticated and wild species [2,3], viewing both anthropogenic disturbance and naturally heterogeneous environments as ideal settings for hybridization. More recently, studies of adaptation through introgression have focused primarily on wild species ([4,5] but see [6,7]). Well-studied examples include increased hybrid fitness of Darwin's finches following environmental changes that favor beak morphology intermediate to that found in extant species [8,9] and the introgression of traits related to herbivore resistance [10] and drought escape [11] in species of wild sunflower [12,13]. Molecular and population genetic analyses have also clearly identified instances of adaptive introgression across species at individual loci, including examples such as the *RAY* locus controlling floral morphology and outcrossing rate in groundsels [14]) and the *optix* gene controlling wing color in mimetic butterflies [15,16]. Despite long-standing interest in introgression, however, genome-wide analyses are rare and have been primarily conducted in model systems [17-22].

Studies of natural introgression in cultivated species have been limited in genomic scope and have largely ignored the issue of historical adaptive introgression, focusing instead on contemporary transgene escape and/or the evolution of weediness [23-27]. One notable exception is recent work documenting introgression between different groups of cultivated rice in genomic regions containing loci involved in domestication [19,28-30]. Few studies, however, have investigated the potential for introgression to transfer adaptations between crops and natural populations of their wild relatives post-domestication. Subsequent to domestication, most crops spread from centers of origin into new habitats, potentially encountering locally adapted populations of their wild progenitors and closely related species (*e.g*., [31-33]). These crop expansions provide compelling opportunities to study evolution through introgressive hybridization.

Here, we use a dense SNP genotyping array to investigate the genomic signature of gene flow between cultivated maize and its wild relative *Zea mays* ssp. *mexicana* (hereafter, *mexicana*) and examine evidence for adaptive introgression.

Maize was domesticated approximately 9,000 BP in southwest Mexico from the lowland teosinte taxon *Zea mays* ssp. *parviglumis* (hereafter, *parviglumis*; [34-36]). Following domestication, maize spread to the highlands of central Mexico [34,37], a migration that involved adaptation to thousands of meters of changing elevation and brought maize to substantially cooler (~7°C change in annual temperature) and drier (~300mm change in annual precipitation) climes [38]. During this migration maize came into sympatry with *mexicana*, a highland teosinte that diverged from *parviglumis* ~60,000 BP [39].

Convincing morphological evidence for introgression between maize and *mexicana* has been reported [40,41], and traits putatively involved in adaptation to the cooler highland environment such as dark-red and highly-pilose leaf sheaths [42] are shared between *mexicana* and highland maize landraces [40,43]. These shared morphological features could suggest adaptive introgression [44] but could also reflect parallel or convergent adaptation to highland climate or retention of ancestral traits [45]. Though hybrids are frequently observed, phenological isolation due to flowering time differences [40,46] and cross-incompatibility loci [47-49] are thought to limit the extent of introgression, particularly acting as barriers to maize pollination of *mexicana*. Experimental estimates of maize-*mexicana* pollination success are quite low, ranging from <1-2% depending on the direction of the cross [50,51]. Nevertheless, theory suggests that alleles received through hybridization can persist and spread despite such barriers to gene exchange, particularly when they prove adaptive [52,53].

Molecular analyses over the last few decades have provided increasingly strong evidence for reciprocal introgression between *mexicana* and highland maize landraces. Early work identified multiple allozyme alleles common in highland Mexican maize and *mexicana* but rare in closely related taxa or maize outside of the region [54]. Likewise, sequencing of the putative domestication locus *barren stalk1* (*ba1*) revealed a haplotype unique to *mexicana* and highland Mexican maize [55]. Multiple studies have found further support for bidirectional gene flow and have estimated that ~2-10% of highland maize genomes were derived from *mexicana* [34,56] and 4-8% of *mexicana* genomes were derived from maize [57]. A more recent study including several hundred markers revealed that admixture with *mexicana* may approach 20% in highland Mexican maize [36].

Similar to introgression studies in many other plant species (*e.g.*, [31,58-61]), morphological and molecular studies have only provided rough estimates of the extent of introgression between *mexicana* and maize. Little is known regarding genome-wide patterns in the extent and directionality of gene flow. A genomic picture of introgression could greatly expand our understanding of evolution through hybridization, revealing how particular alleles, genes and genomic regions are disproportionately shaped by and/or resistant to these processes [62,63]. Additionally, assessment of introgression in crop species during post-domestication expansion can provide insight into the genetic architecture of adaptation to newly encountered abiotic and biotic conditions. Here, we provide the most in-depth analysis to date of the genomic extent and directionality of introgression in sympatric collections of maize and its wild relative, *mexicana,* based on high-density single nucleotide polymorphism (SNP) data. We find evidence for pervasive yet asymmetric gene flow in sympatric populations. Across the genome, several regions introgressed from *mexicana* into maize are shared across most populations, while little consistency in introgression is observed in gene flow in the opposite direction. These data, combined with analysis of environmental associations and a growth chamber experiment, suggest that maize colonization of highland environments in Mexico may have been facilitated by adaptive introgression from local *mexicana* populations.

**Results**

*Polymorphism and Differentiation*

To assess the extent of hybridization and introgression we collected nine sympatric population pairs of maize and *mexicana* and one allopatric *mexicana* population from across the highlands of Mexico (Table S1; Figure 1) and genotyped 189 individuals for 39,029 SNPs (see Materials and Methods). Genotype data at the same loci were obtained from Chia *et al.* [64] for a reference allopatric maize population. Average expected heterozygosity ($H_E$), percent polymorphic loci (*%P*), and the proportion of privately segregating sites were higher in maize than *mexicana* (t-test, p≤0.012 for all comparisons, Table S2), likely influenced by the absence of *mexicana* from the discovery panel used to develop the genotyping platform. However, substantial variation in diversity was observed across populations within taxa (*e.g.*, *%P* ranged from 52-88% in maize and from 44-79% in *mexicana* (Table S2)) and meaningful comparisons can be made at this level. Our analysis of diversity identified the Ixtlan maize population as an extreme outlier, containing 31% fewer polymorphic markers than any other maize population. Discussion with farmers during our collection revealed that Ixtlan maize was initially a commercial variety whose seed had been replanted for a number of generations. Excluding this population, diversity in *mexicana* populations varied much more substantially than in maize (*e.g.*, variance in *%P* across *mexicana* populations was 7-fold higher; Table S2)

At the population level, summary statistics of diversity and differentiation were consistent with sympatric gene flow (*i.e.,* local gene flow based on current plant distributions) between maize and *mexicana* (Figure S1). First, *%P* was positively correlated between sympatric population pairs ($R^2$ = 0.65; p = 0.016; Figure S1A), though this trend could also reflect local conditions affecting diversity in both taxa. Second, in a subset of populations, the proportion of shared polymorphisms was higher and pairwise differentiation ($F_{ST}$) was lower between sympatric population pairs than in allopatric comparisons. Finally, an individual-based STRUCTURE analysis assuming two groups (K=2) revealed strong membership of reference allopatric individuals of maize and *mexicana* in their appropriate groups (96% and 99% respectively), yet appreciable admixture in sympatric populations. Four recent hybrids were identified (3 mexicana and 1 maize) with<60% membership in their respective groups. STRUCTURE analysis also indicated that gene flow was asymmetric, with more highland maize germplasm derived from *mexicana* (19% versus 12% of *mexicana* germplasm from maize). Assignment at higher K values continued to indicate admixture in *mexicana* populations but not in maize, suggesting gene flow from *mexicana* into maize may have been more ancient (Figure S2). Collectively, these population-level summaries are suggestive of historical gene flow from *mexicana* into maize and, in a subset of populations, of ongoing sympatric gene flow from maize into *mexicana*.

*Variation in Introgression Levels Across the Genome*

Meaningful information regarding the evolutionary significance of introgression can often be obscured in population-level summaries. However, the high density of our SNP data allowed us to assess variation in the extent of introgression across the genome. We made use of two complementary methods. First, we employed the hidden Markov model of HAPMIX [65] to infer ancestry of chromosomal segments along the genomes of individuals from maize and *mexicana* populations through comparison to reference allopatric populations. Subsampling of the reference allopatric populations (see Materials and Methods) revealed considerable signal of introgression in the maize reference panel, particularly in low recombination regions near centromeres (correction for this signal is illustrated in Figure 2 and Figure S3). While this signal could represent genuine introgression predating allopatry, it could also indicate potential false positives in genomic regions with high linkage disequilibrium or low density of data. We therefore added a complementary analysis using the linkage model of STRUCTURE [66,67] to conduct site-by-site assignment across the genomes of *mexicana* and maize. Because STRUCTURE takes allele frequencies across all populations into account during assignment, the approach is robust to deviations of individual reference populations from ancestral frequencies.

Both methods allowed quantification of introgression along the genome for individual samples. Rather than investigate every putative introgression, however, we focused further analyses on regions of high frequency introgression across populations, requiring an average of one chromosome or 50% assignment to the opposite taxon per individual in a given population (Figure 2; Figure S3; referred to as "introgressed regions" hereafter). Approximately 19.1% and 9.8% of the genome met this criterion in the HAPMIX and STRUCTURE scans respectively for *mexicana* introgression into maize. In the opposite direction, we observed lower proportions at this threshold (11.4% in the case of HAPMIX and 9.2% using STRUCTURE), corroborating asymmetric gene flow favoring *mexicana* introgression into maize. Both scans showed a disproportionate number of introgressed regions shared across populations in *mexicana*-to-maize gene flow. Roughly 50% of regions introgressed from *mexicana* into maize were shared across seven or more populations in the HAPMIX scan, whereas only 4% of introgressed regions had this level of sharing from maize into *mexicana*; similar asymmetry was observed using STRUCTURE (12% versus <1%).

By comparing composite likelihood scores from HAPMIX across individuals within each population, we were able to characterize relative times since admixture (see Materials and Methods). We observed qualitative differences between maize and *mexicana*. The likelihood of the admixture time parameter began to decrease markedly after an average of 83 generations in *mexicana* populations, whereas the decrease in maize was much more gradual and did not occur until after an average of 174 generations (Figure S4; averages exclude Ixtlan) suggesting older introgression from *mexicana* into maize. A notable exception to this trend was observed in the Ixtlan sympatric population pair, where the maize population was likely derived in the recent past from a commercial variety and introgression appeared to be more recent from *mexicana* into maize (Figure S4).

For further population genetic characterization, we focused on the subset of introgressed regions identified in both the HAPMIX and STRUCTURE scans, an approach that should be robust to the individual assumptions of the two methods. These regions spanned an average of 3.6% of the genome in the case of *mexicana*-to-maize introgression and 3.2% for maize-to-*mexicana* introgression (Figure 2C; Figure S3). As expected, differentiation between sympatric maize and *mexicana* was reduced in these introgressed regions in both directions of gene flow (mean 25% reduction of $F_{ST}$ *mexicana*-to-maize, 33% reduction maize-to-*mexicana*, t-test, p<0.001 for all population-level comparisons of introgressed vs. non-introgressed regions in both directions of gene flow). Introgressed regions also showed more shared and fewer fixed and private SNPs than identified in the remainder of the genome (Table S3), as well as longer regions of identity by state (IBS) between maize and *mexicana* (t-test, p<<0.001). Consistent with these results, diversity in introgressed regions was generally different from non-introgressed regions in the recipient taxon and instead comparable to diversity in non-introgressed regions in the taxon of origin (Table S3).

In total, we identified nine regions of introgression from *mexicana* to maize found by both methods and present in ≥ 7 sympatric population pairs (Table S4). Three of these shared regions of introgression span the centromeres of chromosomes 5, 6, and 10 (Figure S3), suggesting that maize from the highlands of Mexico may in fact harbor *mexicana* centromeric or pericentromeric regions. No regions with this level of sharing across populations were found in the opposite direction of gene flow (maize into *mexicana*).

Finally, we characterized regions of the genome that were notably lacking evidence of introgression. We refer to these regions with ≤ 5% probability of introgression confirmed by both scans in ≥ 7 populations as regions resistant to introgression (Figure S5). In both directions of gene flow, we found these regions to have elevated differentiation, decreased diversity, fewer shared variants, more fixed differences, and a higher number of privately segregating SNPs in the opposite taxon (Table S3).

*Evaluating Evidence for Adaptive Introgression*

Two non-mutually exclusive hypotheses of adaptive introgression can be readily discerned for gene flow between *mexicana* and maize: 1) as its natural habitat was transformed, *mexicana* received maize alleles conferring adaptation to the agronomic setting and 2) as it diffused to the highlands of central Mexico from the lowlands of southwest Mexico, maize received alleles conferring highland adaptation from *mexicana,* which was already adapted to these conditions. To evaluate evidence for the first hypothesis we gauged enrichment of 484 domestication candidate genes [68] in regions of introgression. We hypothesized that if maize donated alleles adaptive for the agronomic setting to *mexicana*, we would detect enrichment of domestication loci in regions introgressed from maize into *mexicana*. However, compared to the rest of the genome, introgressed regions in both directions of gene flow harbored significantly fewer domestication candidates (permutation test, p≤0.001), while regions resistant to introgression showed an excess of domestication candidates (permutation test, p=0.121 maize to *mexicana*, p=0.008 *mexicana* to maize; Figure S5). For example, two well-characterized domestication genes affecting branching architecture, *grassy tillers1* (*gt1*;[69]) and *teosinte branched1* (*tb1*;[70]) showed very little evidence of introgression (Figure S5). Introgression also appeared to be rare from maize into *mexicana* across much of the short arm of chromosome 4, a region that includes the domestication loci *teosinte glume architecture1* (*tga1*;[71]), *sugary1* (*su1*;[72]) and *brittle endosperm2* (*bt2*; [72]) and the well characterized pollen-pistil incompatibility locus *teosinte crossing barrier1* (*tcb1*; [73]) that is known to serve as a hybridization barrier between maize and *mexicana* (Figure S5). These results suggest selection against introgression at domestication loci and crossing barriers.

In support of the hypothesis that maize received introgression conferring highland adaptation from *mexicana*, our scan results revealed asymmetric gene flow favoring *mexicana* introgression into maize as well as substantially more sharing of introgressed regions across populations in this direction of gene flow. Significantly reduced diversity and single haplotypes found across all introgressed individuals extending for hundreds of kilobases in these regions are consistent with selection and spread of single introgression events (Figure S6). Additionally, we used the method of Coop *et al.* [74] to detect associations of population allele frequencies with 76 environmental variables (see Materials and Methods). Environmental variables were reduced in dimensionality to four principal components that captured 95% of environmental variation. We found significant enrichment (permutation test, p=0.017) of loci associated with the second principal component (loaded primarily by temperature seasonality) in regions introgressed from *mexicana* into maize, but no significant environmental associations in regions introgressed from maize into *mexicana*. We then compared the nine regions of introgression found in ≥7 populations of maize to QTL for anthocyanin content and leaf macrohairs—traits putatively adaptive to highland conditions—that were identified in a previous study based on a mapping population developed from a cross of *parviglumis* (lowland teosinte) and *mexicana* (highland teosinte) [42]. Six of the introgressed regions overlapped with five of the six QTL detected for these traits.

Two of the regions of shared introgression that overlap with QTL are of particular interest due to their previous characterization. One region, on chromosome 4, overlaps with QTL for both pigment intensity and macrohairs [42], and maps to the same position as a recently identified putative inversion polymorphism showing significant differentiation between *parviglumis* and *mexicana* (Pyhäjärvi *et al.*, unpublished data; Figure 3A). The second region, on chromosome 9, overlaps with a QTL for macrohairs [42] and includes the *macrohairless1* (*mhl1*) locus [75] that promotes macrohair formation on the leaf blade and sheath of maize (Figure 3B). The two lowest elevation maize populations in our study (Puruandiro and Ixtlan) showed a conspicuous lack of introgression in these two regions (Figure 3A & 3B). Analysis of pairwise differentiation ($F_{ST}$) between these populations and two populations showing fixed introgression in the two regions (Opopeo and San Pedro; Figure 3A & 3B) revealed substantial differentiation at both loci: the region on chromosome 4 contained the only fixed SNP differences genome-wide (Puruandiro/Ixtlan versus Opopeo/San Pedro) and a SNP in the region on chromosome 9 was an extreme outlier in the distribution of $F_{ST}$. To explore the potential phenotypic effects of these regions we conducted growth chamber experiments including ten maize plants from each of these four populations. Under temperature and day-length conditions typical of the highlands of Mexico (see Materials and Methods), the leaf sheaths of plants from populations where introgression was detected in the two regions had 21-fold more macrohairs (t-test, p=0.0002; Figure 3C & 3D), and showed greater pigmentation (t-test, p=6E$^{-06}$; Figure 3C & 3D). Introgressed plants were also ~25 cm taller (t-test, p=6E$^{-06}$; Figure 3D), a finding consistent with adaptation to highland conditions and potentially associated with increased fitness. No significant difference in plant height was observed in a separate experiment under lowland conditions (t=test, p=0.51), and a significant interaction was observed between introgression status and environmental treatment (ANOVA, F=4.151, p=0.045), with a disproportionate increase in plant height under lowland conditions in populations lacking introgression (Figure S7).

*Contribution of mexicana to Modern Maize Lines*

While our scans for introgression clearly indicated that *mexicana* has made genomic contributions to maize landraces in the highlands of Mexico, the broader contribution of *mexicana* to modern maize lines remained unclear. Our HAPMIX and STRUCTURE analyses had low power to detect introgression distributed broadly in maize (see Discussion). Therefore, to assess potential ancestral contribution of *mexicana* to modern maize, we evaluated patterns of IBS between *mexicana*, *parviglumis* (Pyhäjärvi *et al.*, unpublished data) and a global diversity panel of 279 modern maize lines [76,77] using the program GERMLINE ([78]; Figures 4, S8 & S9). Substantial IBS was found between *mexicana* and modern lines at a number of genomic locations. To assess whether this IBS merely reflected shared ancestral haplotypes, we compared IBS between modern maize and *parviglumis* to IBS between modern maize and *mexicana* on a site-by-site basis, identifying regions in which various maize groups distinguished by Flint-Garcia et al. [77] showed stronger IBS with *mexicana* relative to *parviglumis* (see Materials and Methods; Figure 4A; Figure S9). As each of the groups identified by Flint-Garcia have distinct evolutionary histories, it is possible that *mexicana* contributed differentially to the founders of each group. For example, the tropical-subtropical, non-stiff-stalk, and mixed groups showed more regions with stronger IBS with *mexicana* (versus *parviglumis*) than found in the stiff-stalk, popcorn, and sweetcorn groups (~31% of sites with greater IBS with *mexicana* in the first group versus ~23% in the latter group; Figure 4B & 4C).

**Discussion**

Despite known pre-zygotic and phenological barriers to hybridization between maize and *mexicana* [46,48-50], we have found evidence consistent with substantial reciprocal introgression. Based on our population genetic analyses, several observations regarding the nature of this gene flow can be made: 1) Gene flow appears to be ongoing and asymmetric, favoring *mexicana* introgression into maize. 2) Gene flow from *mexicana* into maize is generally older than gene flow in the opposite direction. 3) Evidence of nine regions of *mexicana*-into-maize introgression shared across ≥ 7 populations and very low haplotype diversity in these regions suggest single, ancient introgression followed by spread across the Mexican highlands. 4) Introgression from *mexicana* into maize is restricted at domestication loci but enriched at loci putatively involved in highland adaptation. 5) Regions of *mexicana*/maize IBS within a global diversity panel of maize hint at a possible broader contribution of *mexicana* to modern improved maize.

Several of these observations are in line with previous research. For example, the asymmetric gene flow we detect from *mexicana* to maize is consistent with findings of substantially higher pollination success in this direction [50]. Asymmetric gene flow would also be expected based on phenology: in Mexico, maize typically flowers earlier than *mexicana* [46] and pollen shed in both taxa precedes silking (female flowering). Therefore, when maize silks are receptive, *mexicana* could potentially be shedding pollen, whereas when *mexicana* silks are receptive, maize tassels are more likely to be senescent. Under these conditions, F1 progeny would be more likely to have a maize mother and a teosinte father and subsequent inadvertent planting of F1's in maize fields would bias the direction of gene flow.

Our data also provide support for previous assertions that shared morphological features between *mexicana* and maize represent adaptations derived from *mexicana* [44] rather than from maize [41]. We have found significant environmental correlations in regions of *mexicana*-to-maize introgression and overlap with QTL and fine-mapped loci for highland *Zea* traits (*e.g.,* leaf sheath macrohairs and pigmentation) are predominantly found in the direction of *mexicana* to maize gene flow. Two of these regions, on chromosomes 4 and 9, showed particularly strong evidence of introgression. Moreover, these regions were more common in higher elevation maize populations in our sample, and maize populations with and without introgression in these regions showed differential morphology and greater plant height—a proxy for fitness—when grown under highland conditions. In contrast, we found little evidence of adaptive introgression in the opposite direction of gene flow. For example, domestication loci appeared resistant to gene flow from maize into *mexicana,* contradicting previous suggestions that gene flow from maize may have been required for *mexicana* to adapt to an agronomic setting [41]. Instead it appears likely that *mexicana,* like other wild teosintes [79], was a ruderal species adapted to open and disturbed environments even before the transformation of its natural habitat by maize cultivation.

Our detection of haplotype sharing between *mexicana* and a diverse panel of modern maize is consistent with previous findings suggesting the spread of introgressed *mexicana* haplotypes in maize outside of the highlands of Mexico [68].

Both the STRUCTURE and HAPMIX methods we used to identify regions of introgression would likely not detect introgression found ubiquitously in modern maize. Widespread *mexicana* introgression into maize would result in poor resolution between reference populations of these taxa in the HAPMIX analysis, and extensive haplotype sharing across maize and *mexicana* would result in a weak signature of introgression in STRUCTURE. Further analysis of representative panels of *mexicana*, *parviglumis* and maize haplotypes at greater marker density should help clearly distinguish *mexicana* from *parviglumis* haplotypes and determine whether *mexicana* haplotypes are indeed widespread in maize.

While our results are consistent with previous research and the putative history of the spread of maize, our power to detect introgression may be limited for a number of reasons. First, our analysis conservatively focused on regions of introgression identified by two independent methods and shared across individuals within populations, undoubtedly missing a number of genuine instances of more limited introgression. Second, our markers were ascertained in a panel consisting entirely of maize. In addition to inflating the diversity of maize relative to *mexicana*, this ascertainment scheme likely limited our ability to distinguish among *mexicana* haplotypes and thus to detect local introgression from *mexicana* into maize. Third, the resolution of our data was on average one SNP per 80 kb, which could result in a bias toward detection of more recent introgression and introgression in low recombination regions of the genome. Finally, *mexicana* only rarely occurs allopatric from maize [40], and most populations have likely experienced gene flow at some point in time complicating estimation of ancestral *mexicana* haplotypes and allele frequencies.

Many aspects of *mexicana*'s contribution to highland adaptation in maize remain to be resolved. While our growth chamber experiment was suggestive of adaptive introgression, the loci conferring these traits are still ambiguous. Repetition of these experiments with *mexicana*/lowland maize near-isogenic introgression lines will be necessary to bolster the case for adaptive introgression. Additionally, a particularly interesting comparison can be made between highland maize in central Mexico, a region sympatric with *mexicana*, and highland maize in the Andes of South America where no inter-fertile wild *Zea* species can be found. Future research should address whether highland adaptation in South American maize occurred in parallel to maize from Mexico [37] or whether pre-adapted highland maize was transported through Central America as some have suggested [80].

The potential for adaptive introgression during crop expansion is of course not limited to maize. Data from several crops (*e.g.*, rice [19,81], barley [82,83], common bean [84], and wheat [32,85]) suggest defined centers of origin within a broader distribution of wild relatives. The distributions of these crop-wild pairs span continents and a wide range of environments, and many are known to hybridize (for a review, see [24]). The methods we have applied here to maize and *mexicana* can therefore be replicated widely, perhaps revealing unexpected aspects of crop evolution and providing insight regarding the genetic architecture of local adaptation based on conserved regions of introgression.

Crops and related wild taxa can also be seen more broadly as models for the study of evolution through hybridization. If crops are viewed as human-facilitated invasive species, clear connections can be made to theoretical work on introgression during invasion and range expansion. For example, our finding of asymmetric gene flow from *mexicana* into maize is consistent with simulations showing that invaders should receive much higher levels of introgression from local species than occurs in the opposite direction due to differences in population density at the time of invasion [86,87]. Theoretical research has also explored the divergence threshold for successful hybridization and introgression [52,88]. Crop expansions are ideal systems to test such predictions because, as ancient agriculturalists moved crops away from their centers of origin, these domesticates came into sympatry with relatives spanning a range of divergence times. For example, *parviglumis*, the progenitor of maize, has a divergence time from *mexicana* estimated at 60,000 years, from other members of the genus on the order of 100,000-300,000 years, and from the outgroup *Tripsacum dactyloides* approximately 1 million years [39]. While *parviglumis* is currently physically isolated from these taxa and likely was at the time of domestication [38], maize has subsequently come into sympatry with virtually all of its close relatives, providing extensive opportunities for hybridization. These newly-formed hybrid zones can be seen as testing grounds of the fitness of hybrids across a range of divergence and opportunities to study the evolution of barriers to hybridization.

**Materials and Methods**

*Sample Collection and Genotyping*

Samples were collected from nine sympatric population pairs of *mexicana* and maize that spanned the known distribution of *mexicana* in Mexico, as well as a single allopatric population of *mexicana* (Table S1; Figure 1). Seed samples from 12 maternal individuals per *mexicana* population (N=120) were selected for genotyping. A single kernel was also sampled from each of 6-8 maize ears collected from sympatric maize fields (N=69). The tenth kernel down from the tip of each ear was chosen to help control for potential variation in outcrossing rate along the ear. Seeds were treated with fungicide, germinated on filter paper and grown in standard potting mix to the five-leaf stage. Freshly harvested leaf tips were stored at -80° C overnight and lyophilized for 48 hours. Tissue was then homogenized with a Mini-Beadbeater-8 (BioSpec Products, Inc., Bartlesville, OK, USA) and DNA was isolated using a modified CTAB protocol [89]. Purity of DNA isolations was determined with a NanoDrop spectrophotometer (NanoDrop Technologies, Inc., Wilmington, DE, USA). Samples with 260:280 ratios ≥ 1.8 were deemed acceptable for genotyping. Concentrations of DNA isolations were determined with a Wallac VICTOR2 fluorescence plate reader (Perkin-Elmer Life and Analytical Sciences, Torrance, CA, USA) using the Quant-iT™ Picogreen® dsDNA Assay Kit (Invitrogen, Grand Island, NY, USA). Single nucleotide polymorphism genotypes were generated using the Illumina MaizeSNP50 Genotyping BeadChip platform and were clustered separately for the two taxa based on the default algorithm of the GenomeStudio Genotyping Module v1.0 (Illumina Inc., San Diego, CA, USA). Clustering for each SNP in each taxon was visually inspected and manually adjusted. Of the total of 56,110 markers contained on the chip, 39,029 SNPs that were polymorphic within the entire sample of maize and *mexicana* and contained less than 10% missing data in both taxa were used for further analysis.

*Diversity Analyses*

Observed ($H_O$) and expected ($H_E$) heterozygosities were summarized for each taxon in each sympatric population pair using the "genetics" package in R [90]. Polymorphisms were further characterized as shared, fixed, or segregating privately within one of each pair of sympatric populations using the sharedPoly program of the libsequence C++ library [91]. Pairwise differentiation between populations ($F_{ST}$) was calculated based on the method of Weir and Cockerham [92] using custom R scripts and the "hierfstat" package of R [93].

*Detecting Introgression*

To characterize patterns of introgression across the genome in each population we used two complementary methods: 1) Identification of ancestry across chromosomal segments with the hidden Markov model approach of HAPMIX [65]; and 2) A site-by-site analysis of assignment probabilities using the Bayesian linkage model in the program STRUCTURE [66,67]. For both HAPMIX and STRUCTURE analyses, we used a subset of 38,262 SNPs anchored in a genetic map based on the Intermated B73 X Mo17 (IBM) population of maize ([94]; J.P. Gerke et al., unpublished data). The IBM population has been widely used for genetic map development and for determining the genetic architecture of complex traits in maize [95].

Patterns of introgression were assessed using the program HAPMIX by comparing unphased data from putatively admixed individuals from our sympatric populations to phased data from reference ancestral populations. To represent ancestral *mexicana* haplotypes, we chose a population near the town of Amatlán, Morelos state, Mexico that is currently allopatric to maize. An Americas-wide sample of maize landraces collected largely outside the distribution of *mexicana* was chosen as the maize reference population [64]. In order to assess putative introgression and/or false positives in these reference populations, we removed each individual and evaluated introgression through comparison to remaining reference samples using a jackknife approach. Evidence for introgression was assessed in both putatively admixed and reference individuals using HAPMIX as described below.

Initial estimates of ancestry proportions for HAPMIX models were based on a previous admixture analysis of *mexicana* and highland Mexican maize (~20% introgression of *mexicana* into maize and ~10% introgression of maize into *mexicana*; [36]). The number of generations since the time of admixture was varied from 1-5000 and the maximum composite likelihood across individuals in a population was used to compare relative time since admixture on a population-by-population basis (Figure S4). Subsequent analyses of HAPMIX output were based on introgression estimates from the highest likelihood run.

Prior to analysis in STRUCTURE, SNP data were phased using the program fastPHASE (version 1.4.0; [96]). Because STRUCTURE does not account for linkage disequilibrium (LD) due to physical linkage, SNPs were grouped into haplotypes separated by at least 5kb. After grouping, our dataset consisted of 20,035 loci with an average of 3.92 alleles per locus across all sympatric and reference allopatric individuals. We ran the linkage model in STRUCTURE with 5,000 steps of admixture burn-in, a total burn-in of 10,000 steps, and 100,000 subsequent steps retained for analysis. Convergence along the chain and consistency across replicate runs were assessed to ensure an adequate number of steps were included in the analysis. Assignment was carried out for K=2 groups (*i.e.*, maize and *mexicana*) for each chromosome separately. Probability of assignment was summarized locus by locus across individuals from each population for each taxon.

*Local Adaptation at Introgressed Loci*

To identify SNPs associated with environmental variables, we employed the association method of BAYENV [74], using a covariance matrix of allele frequencies estimated using 10,000 random SNPs to control for population structure. Seventy-six climatic and soil variables were summarized as four principal components that captured 95% of the variance among *mexicana* populations. BAYENV was run five times with 1,000,000 iterations for each SNP. A given SNP was considered a candidate if its Bayes factor was consistently in the 95[th] percentile across all five independent runs and its average Bayes factor was in the 99[th] percentile.

Enrichment of significant SNPs in introgressed regions was determined based on bootstrap resampling for each environmental PC.

*Haplotype Sharing*

Analyses of haplotype sharing/identity by state between *mexicana*, *parviglumis*, and modern maize lines were conducted using the program GERMLINE [78] with haplotypes generated by the program fastPHASE [96] from samples of *parviglumis* (Pyhäjärvi *et al.*, unpublished data) and modern maize [76]. Shared haplotypes were identified with a seed of identical genotypes at five SNPs that were extended until mismatch. Analyses were then based on segments with a minimum size of 3 cM.

*Growth Chamber Experiment*

Ten seeds were germinated from each of four maize populations showing little evidence of introgression (Ixtlan and Puruandiro) or fixed introgression (Opopeo and San Pedro) at two loci (one on chromosome 4 and one on chromosome 9; Table S4) putatively linked to highland adaptation ([42]; Pyhäjärvi *et al.*, unpublished data) and showing little evidence of false positives in our reference populations. Plants were grown under highland conditions with 12.5 hours of light at an intensity of 680 µmol/m$^2$∗s, a daytime temperature of 23° C and a nighttime temperature of 11° C. Daytime relative humidity was set at 60% and nighttime relative humidity at 80%. Height measurements were taken at 15, 30, and 50 days. Pigment extent was measured on the second leaf sheath from the top of the plant as the proportion of the total sheath showing pigment. Macrohairs were also measured on this leaf sheath as the total count one third of the way down from the leaf blade within the field of a dissecting microscope at 2X magnification. In order to contrast plant height from our highland treatment to those under conditions more comparable to the lowlands of western Mexico, we conducted a separate growth chamber experiment with a daytime temperature of 32° C and a nighttime temperature of 25° C and measured plant height at 30 days. All other conditions were identical to those of the highland treatment.

**Acknowledgments**

We thank Lauren Sagara and Pui Yan Ho for assistance with genotyping and the growth chamber experiments and Elena Alvarez-Buylla for assistance during sample collection. Sofiane Mezmouk and Shohei Takuno provided comments on a previous version of the manuscript. Graham Coop, Peter Morrell, and John Novembre offered helpful discussion.

**Figure Legends**

**Figure 1.** Map of collection sites. Light red dots indicate known *mexicana* populations and larger, dark red dots indicate populations included in the current study.

**Figure 2.** Detection of introgression across chromosome 4. (A) Stacked bar plots of the HAPMIX introgression scan across sympatric populations. Population labels are indicated between plots for maize (gold) and *mexicana* (maroon). Lighter colors indicate introgression initially detected in each population and darker colors show these values after subtracting introgression proportions from jackknife samples of the allopatric reference populations that may be due to false positives. A dotted black line indicates the position of the centromere. (B) Stacked bar plots of the STRUCTURE introgression scan across sympatric populations. The *mexicana* group

is indicated by maroon and the maize group is indicated by gold. The y-axis for each population in (A) and (B) indicates the average admixture proportion across individuals. (C) Regions in maize populations showing greater than 50% membership in *mexicana*.

**Figure 3.** Growth chamber experiment. (A) Region of *mexicana*-to-maize introgression on chromosome 4 (indicated by blue hash on x-axis) shared across seven populations. Patterns of introgression seen in San Pedro (blue solid line) and Opopeo (blue dashed line) versus Puruandiro (red solid line) and Ixtlan (red dashed line). (B) Region of *mexicana*-to-maize introgression on chromosome 9 (indicated by blue hash on x-axis) shared across seven populations. Populations are as in Figure 3A. (C) Five leaf sheaths from each of four maize populations grown in a growth chamber under highland conditions. (D) Distribution of maize trait values (macrohairs, pigment extent and plant height at 50 days) from growth chamber experiment emulating highland conditions in populations with and without introgression.

**Figure 4.** Contribution of *mexicana* germplasm to a global maize diversity panel. (A) The difference between the average IBS proportion with *mexicana* individuals minus the average IBS proportion with *parviglumis* individuals calculated across the six groups identified by Flint-Garcia *et al.* [77] in the maize association population. Positive values indicate greater IBS with *mexicana*. (B) Average IBS across chromosome 3 in each line in the maize association population compared to both

*mexicana* and *parviglumis*. The one-to-one line is indicated by the dashed line. Colors are as in Figure 4A. (C) The proportion of sites across the genome showing greater IBS with *mexicana* than with *parviglumis* for each of the six maize association population groups.

**Supplementary Figure Legends**

**Supplementary Figure 1.** Population-level polymorphism and differentiation. (A) Correlation of percent polymorphic loci in sympatric populations of *mexicana* and maize. (B) Proportion of shared and privately segregating polymorphisms in *mexicana* and maize and fixed differences between taxa. Letters above bars indicate sympatric maize/*mexicana* comparisons (S), maize from a given population versus allopatric *mexicana* (Ax) and *mexicana* from a given population versus allopatric maize (Az). (C) Pairwise differentiation ($F_{ST}$) in sympatric and allopatric comparisons of *mexicana* and maize. (D) Bar plot of assignment proportions from STRUCTURE analysis at K=2 for *mexicana* (maroon) and maize (gold) individuals. The Ixtlan maize population was excluded from this figure and the STRUCTURE analysis.

**Supplementary Figure 2.** Bar plot of assignment proportions from STRUCTURE analysis at K=2-K=10 for *mexicana* and maize individuals. The Ixtlan maize population was excluded from this figure and the STRUCTURE analysis.

**Supplementary Figure 3.** HAPMIX and STRUCTURE plots of introgression for each chromosome. Colors and axes are as in Figure 2.

**Supplementary Figure 4.** Likelihood plots across generations since admixture for each population for both *mexicana* (A) and maize (B).

**Supplementary Figure 5.** Proportion of populations showing resistance to introgression across each chromosome for maize-to-*mexicana* (A) and *mexicana*-to-maize (B) introgression. Thirteen well known domestication loci (red) and three characterized pollen-pistil cross-incompatibility loci (blue) are indicated with dashed lines and labeled above the plots.

**Supplementary Figure 6.** Evidence supporting single haplotypes in shared regions of *mexicana*-to-maize introgression. (A) Expected heterozygosity in introgressed (blue) and non-introgressed (red) individuals. Expected heterozygosity plotted across chromosomes 4 (B) and 9 (C) in introgressed individuals. Regions of introgression shared across populations are indicated by a red dashed line.

**Supplementary Figure 7.** Plant height at 30 days in maize populations with (blue) and without (red) introgression at loci depicted in Figure 3 under highland and lowland conditions. Confidence interval is +/- 1 standard error.

**Supplementary Figure 8.** Identity by State (IBS) of modern maize lines with *mexicana* and *parviglumis* across each chromosome. All plots are as in Figure 4B.

**Supplementary Figure 9.** The difference between IBS modern maize/*mexicana* and IBS modern maize/*parviglumis* across each chromosome. All plots are as in Figure 4A. Dashed lines indicate regions of *mexicana* introgression into highland Mexican maize conserved across ≥ 7 populations.

**Supplementary Table Legends**

**Supplementary Table 1.** Sampling information for *mexicana* and maize populations.

**Supplementary Table 2.** Summaries of diversity across *mexicana* and maize populations. $H_E$ = expected heterozygosity, $\%P$ = percent polymorphic loci, $H_O$ = observed heterozygosity, $F_{IS}$ = inbreeding coefficient calculated as $(H_E - H_O)/H_E$.

**Supplementary Table 3.** Population genetic summaries from introgressed regions and regions resisting introgression. Parameters are as in Table S2. Significant differences (permutation or t-test, p<0.05) between introgressed and non-introgressed regions are indicated as bold values.

**Supplementary Table 4.** Genomic coordinates of shared introgression regions.

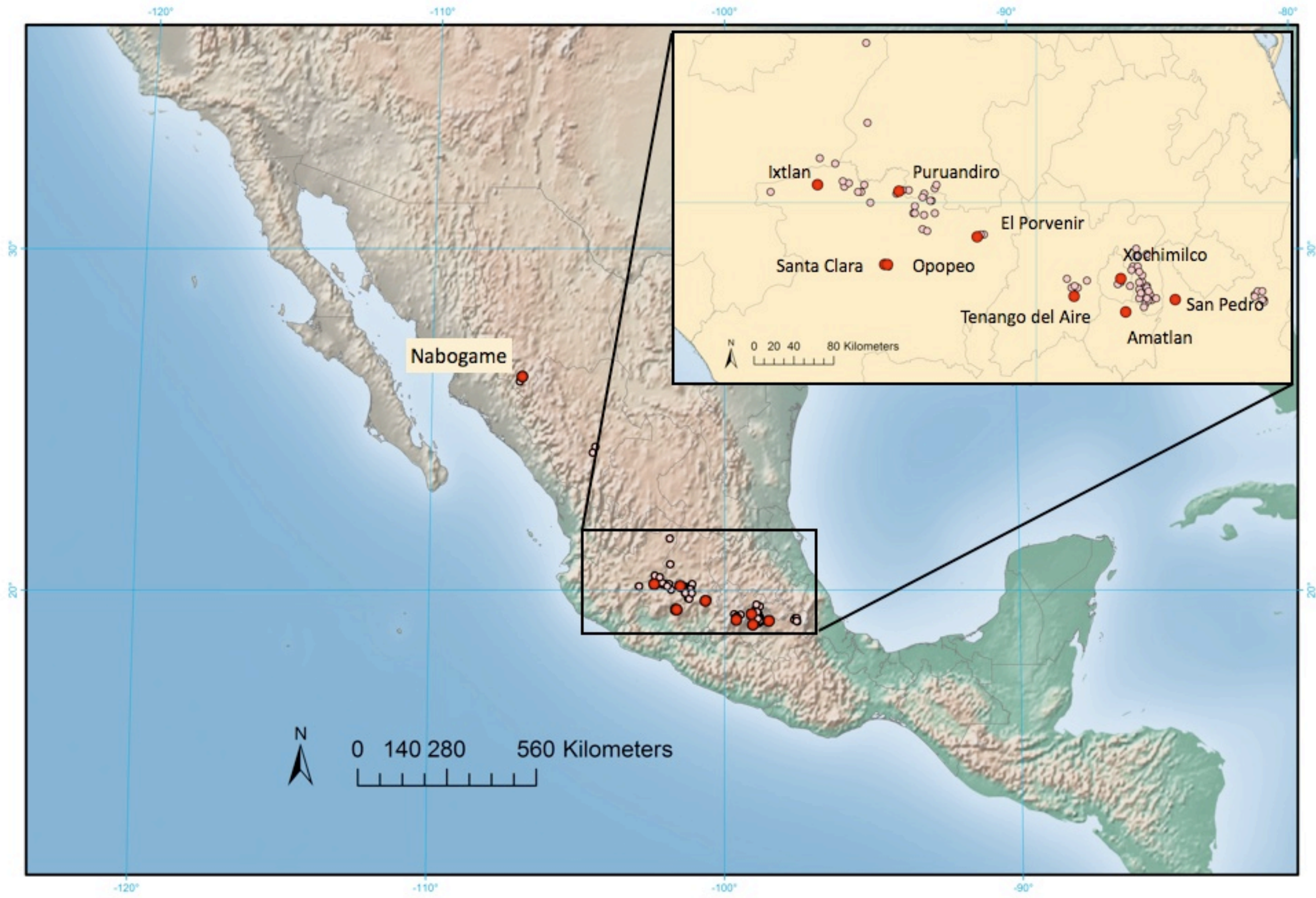

Figure 1

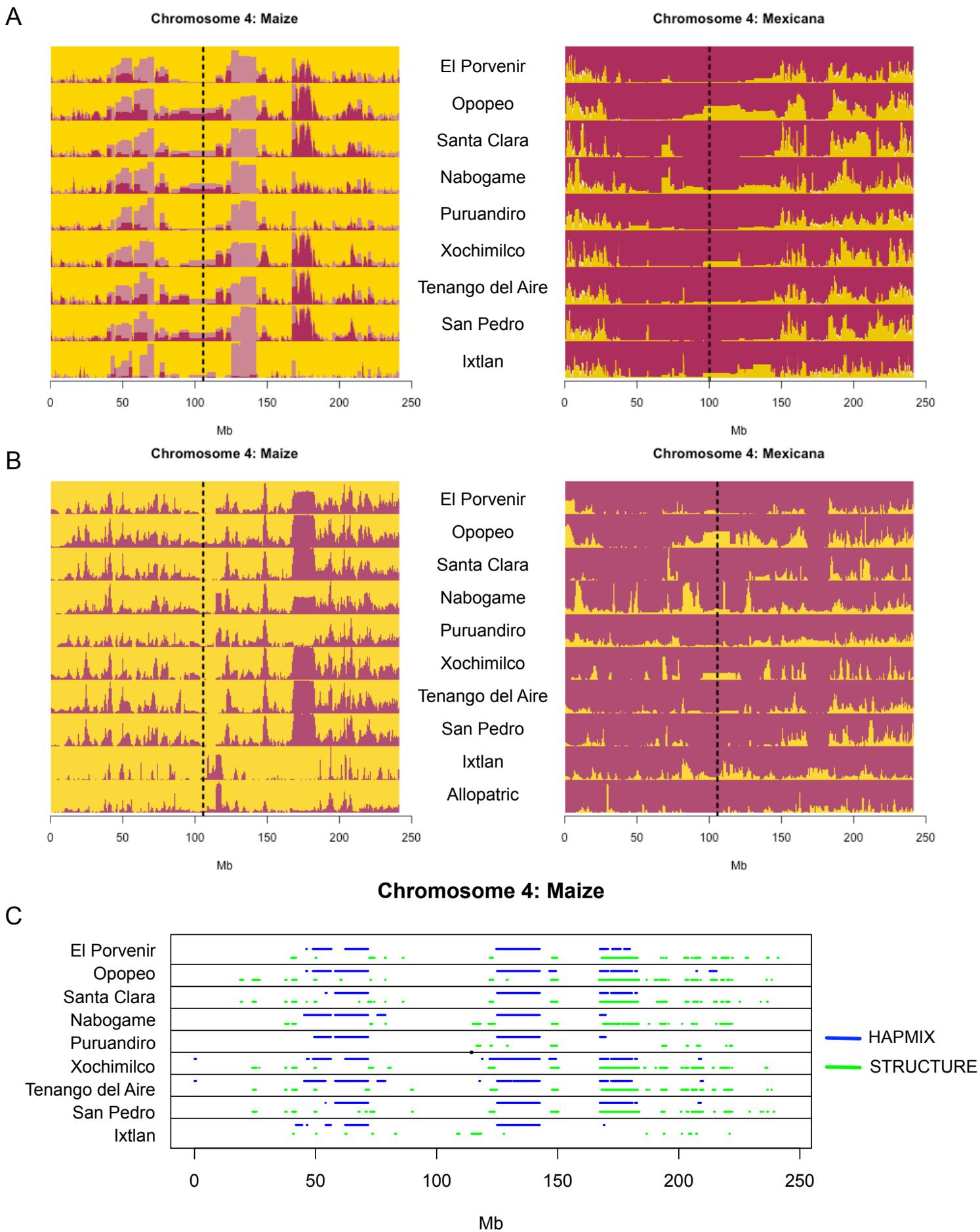

Figure 2

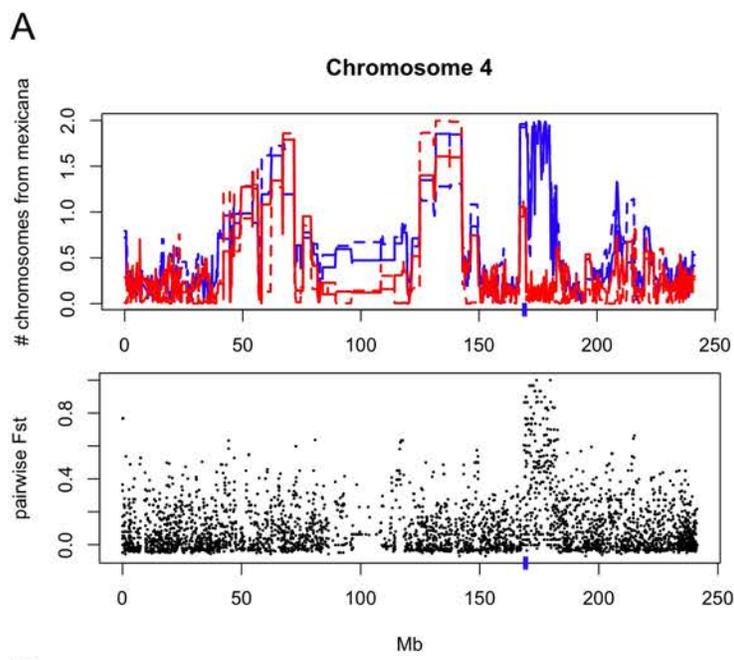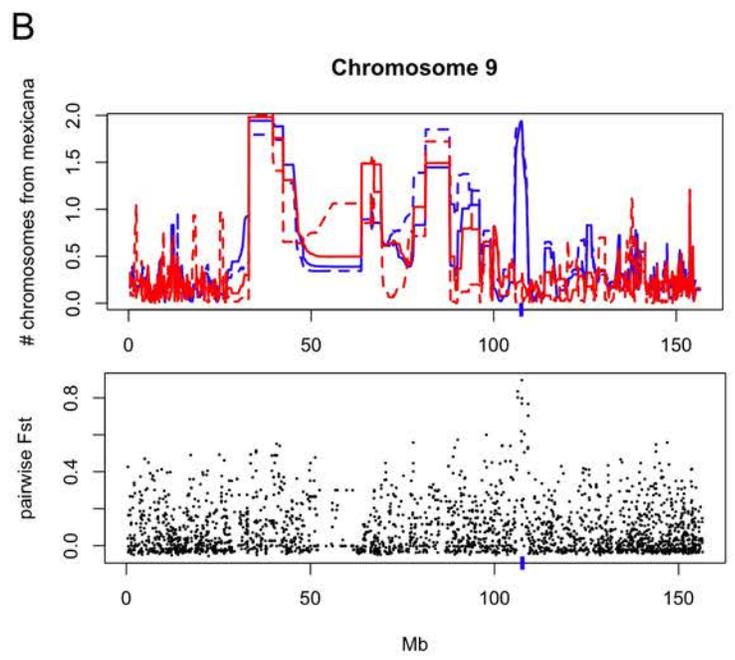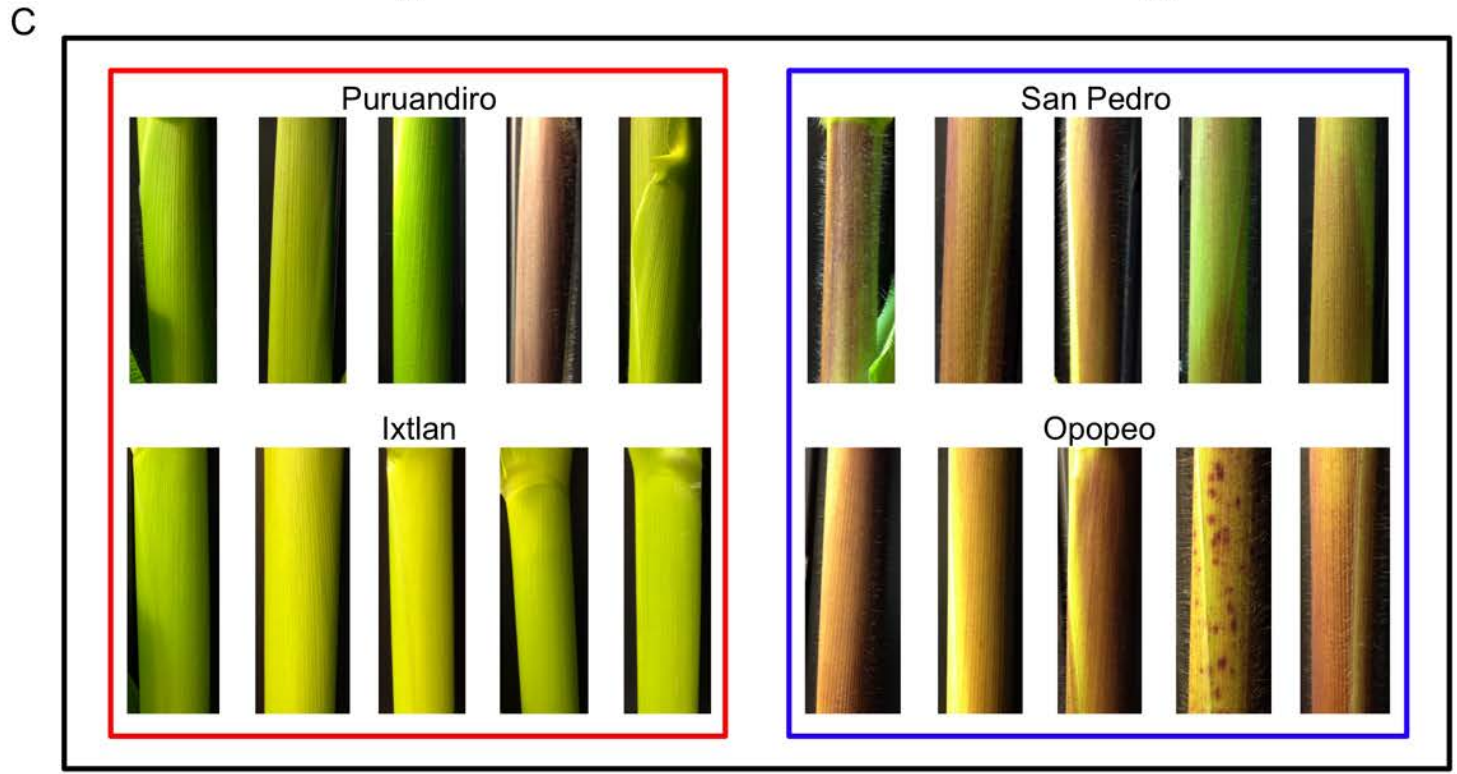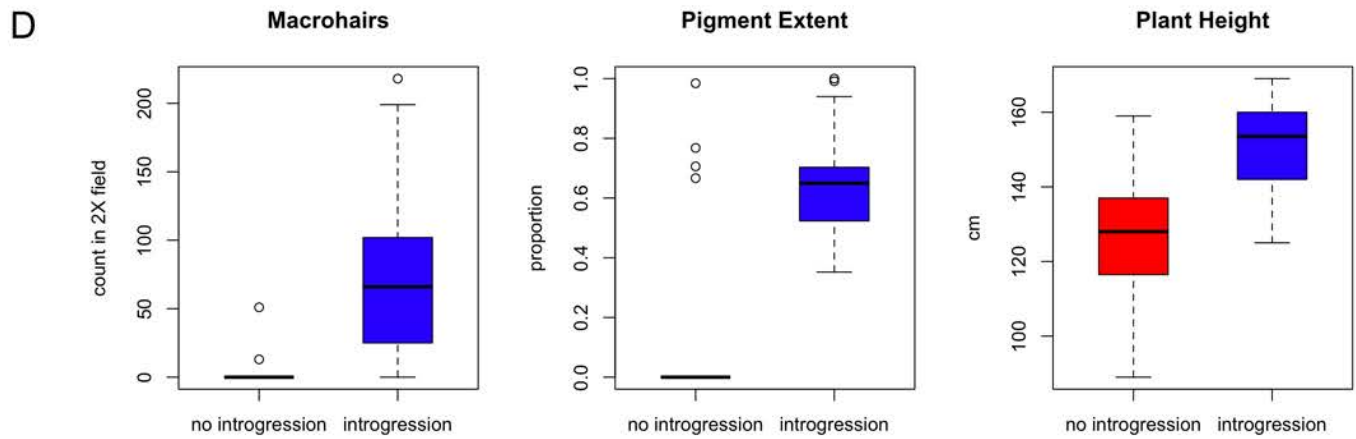

Figure 3

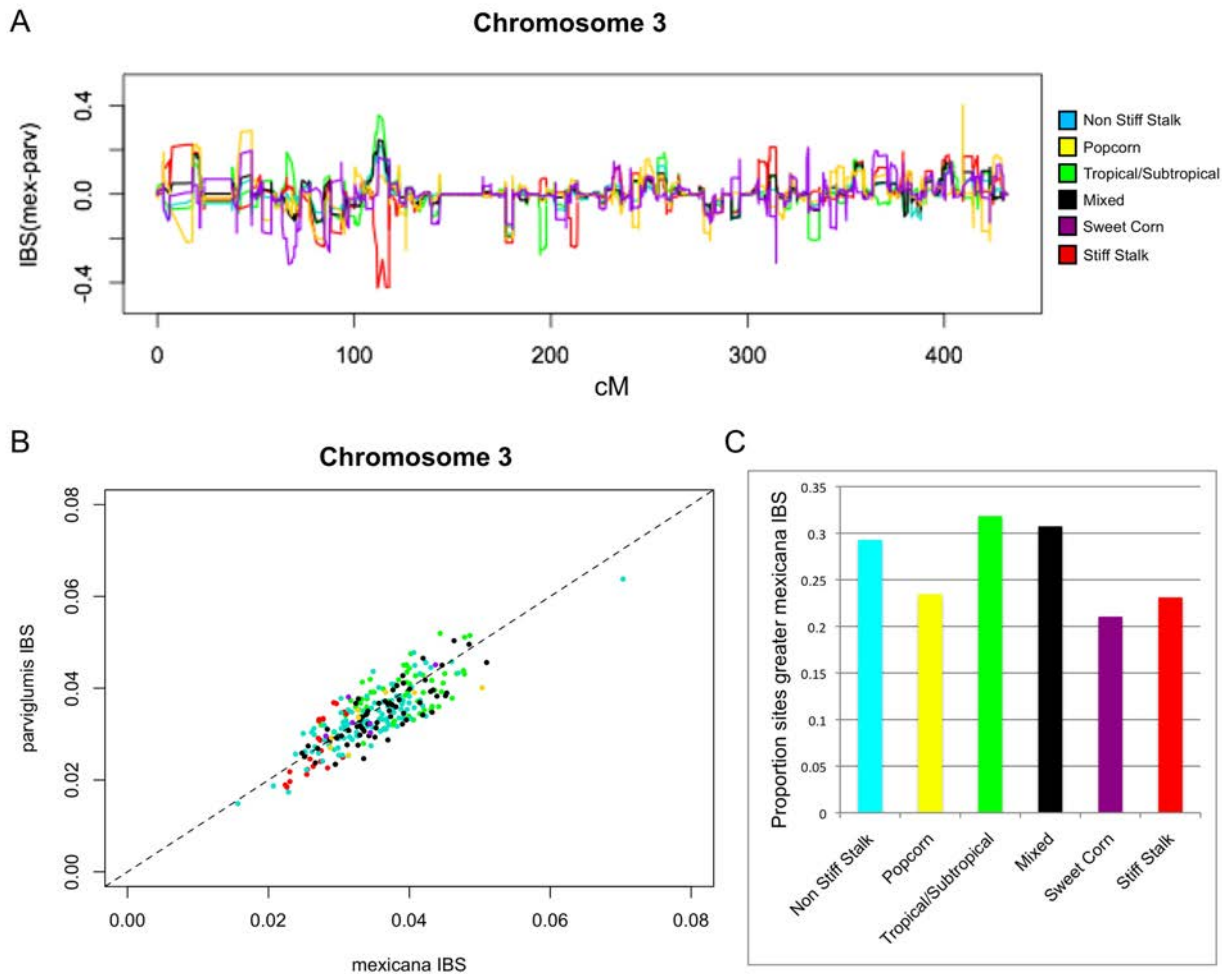

Figure 4

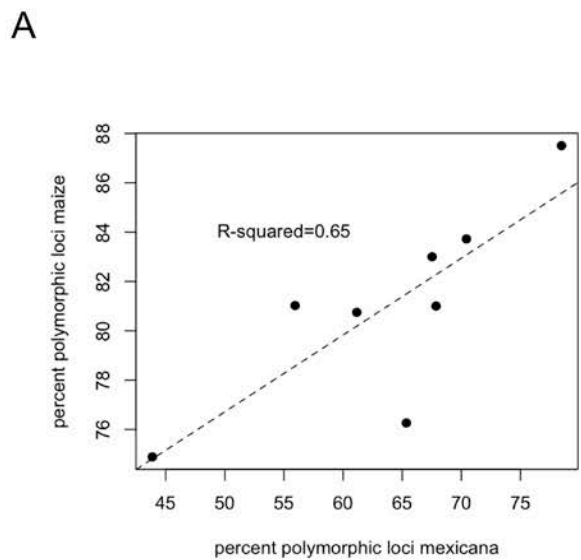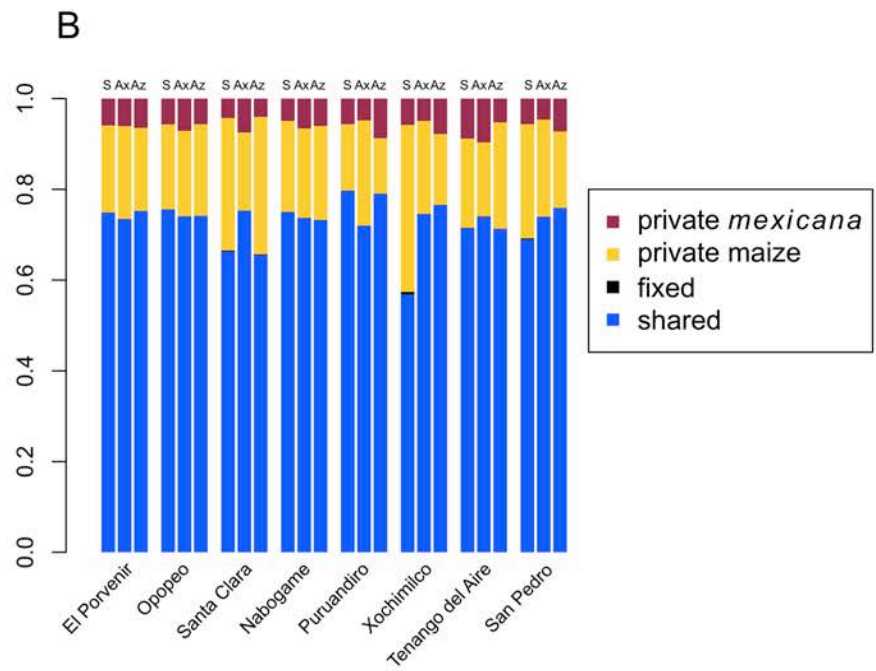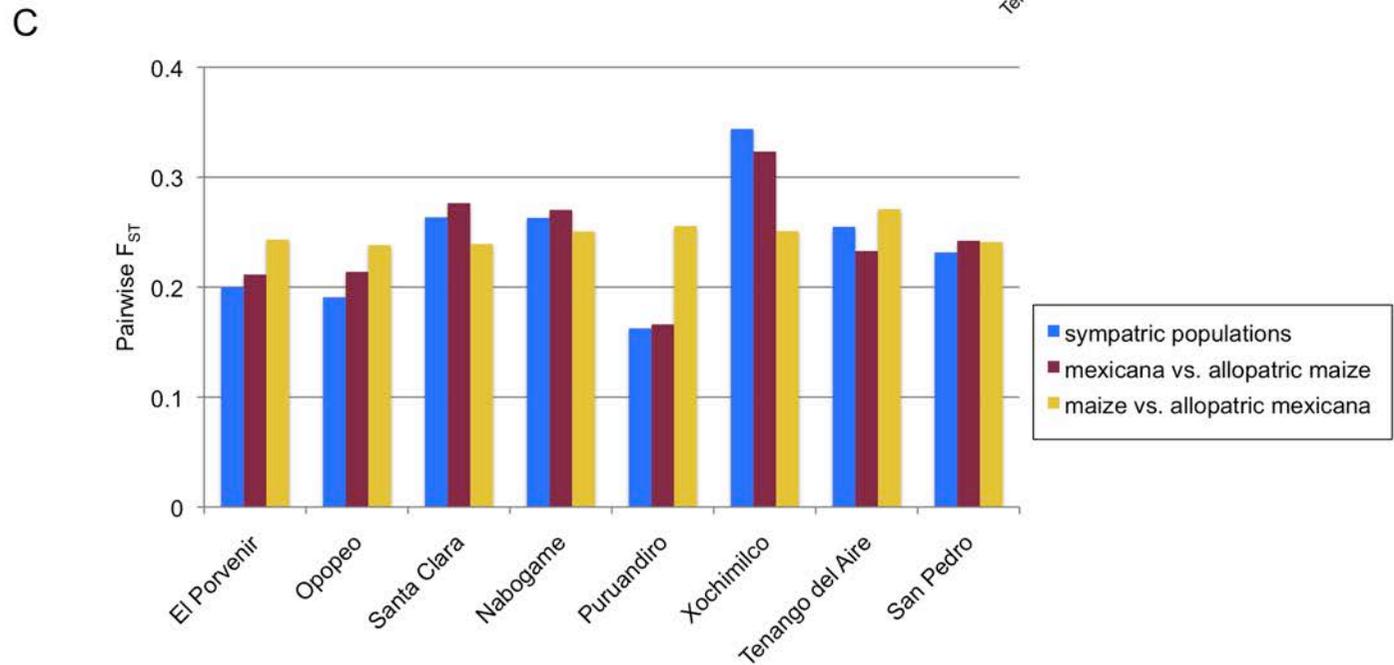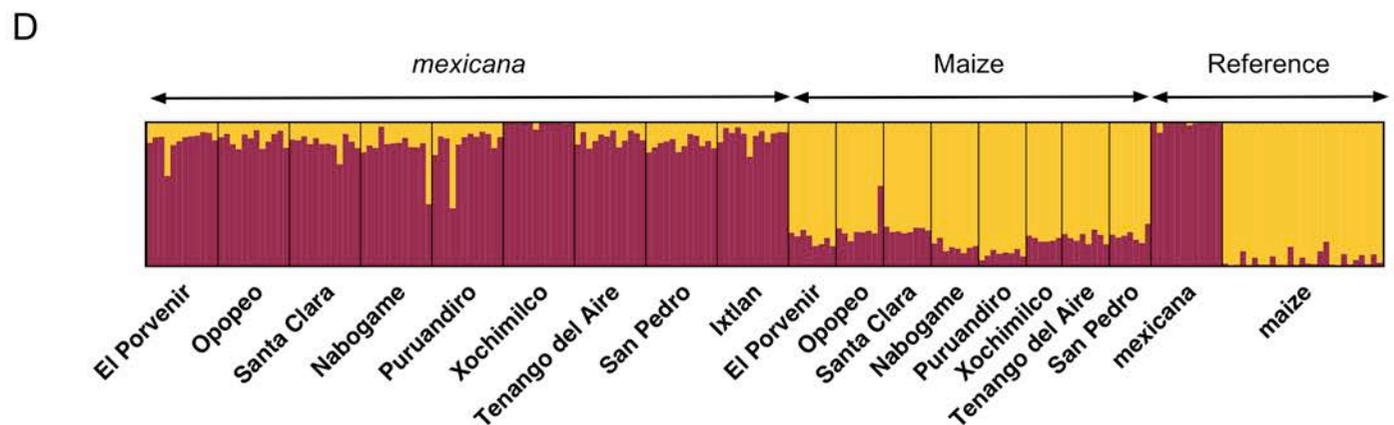

Figure S1

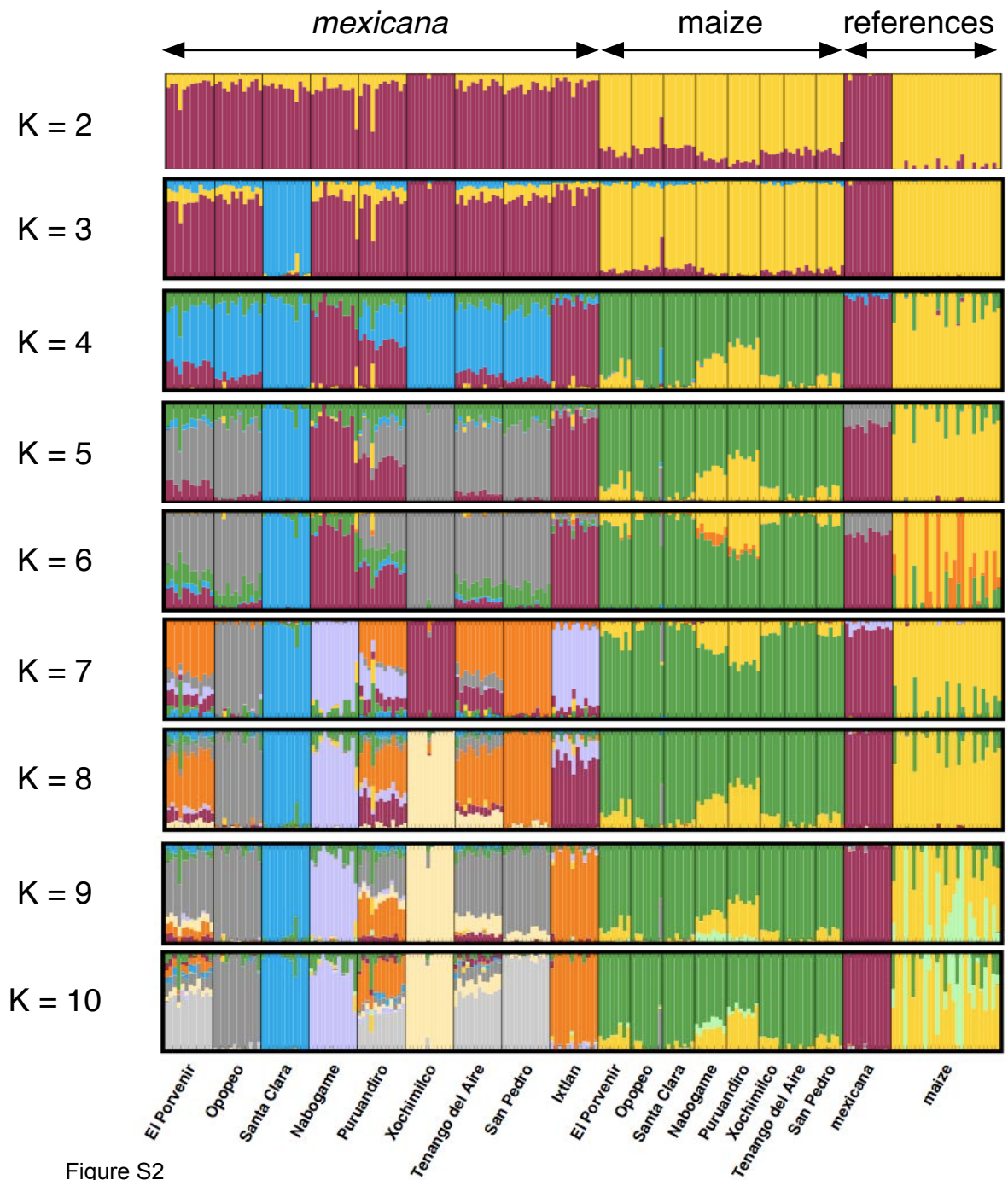

Figure S2

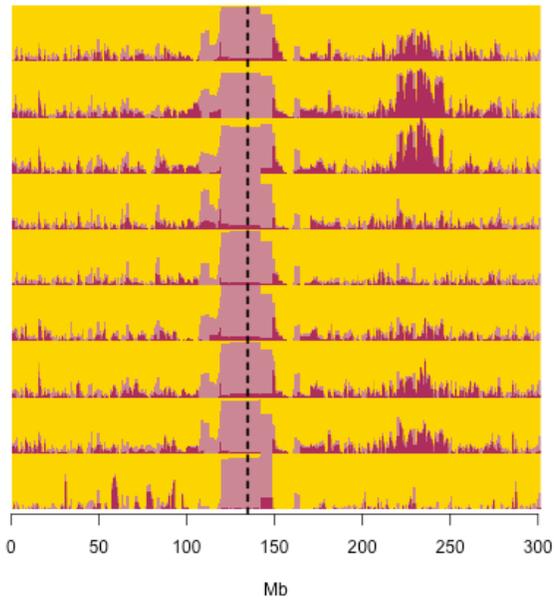
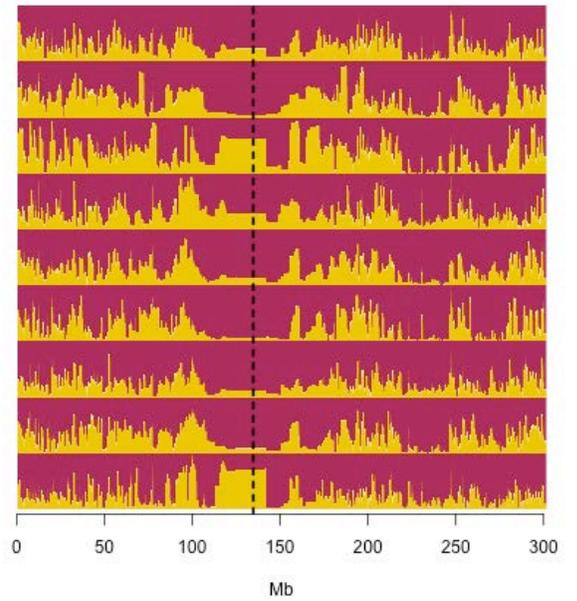
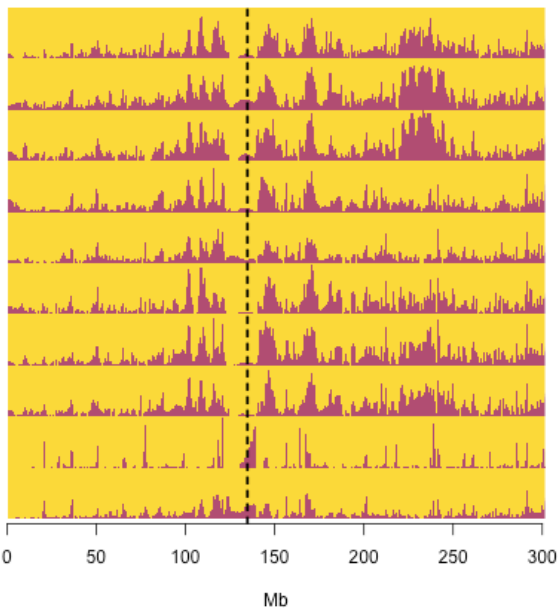
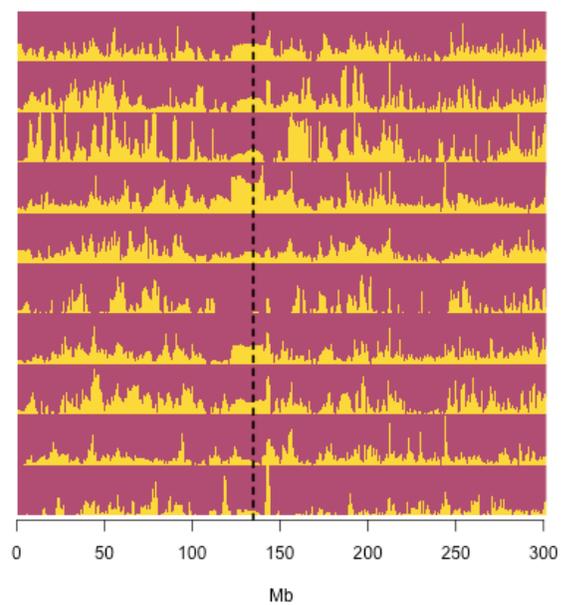
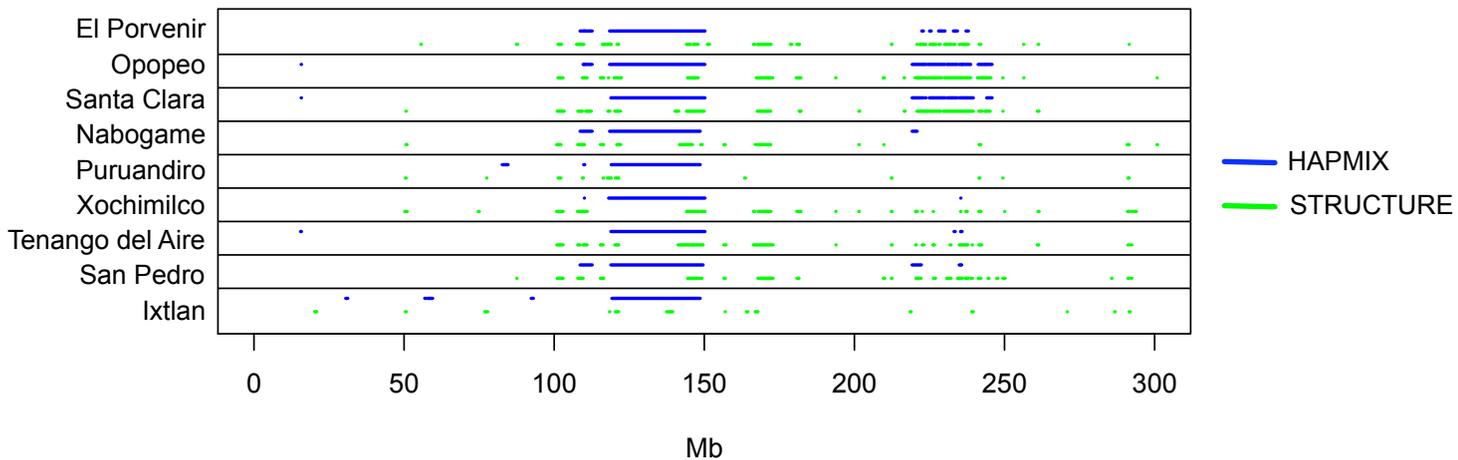

Figure S3

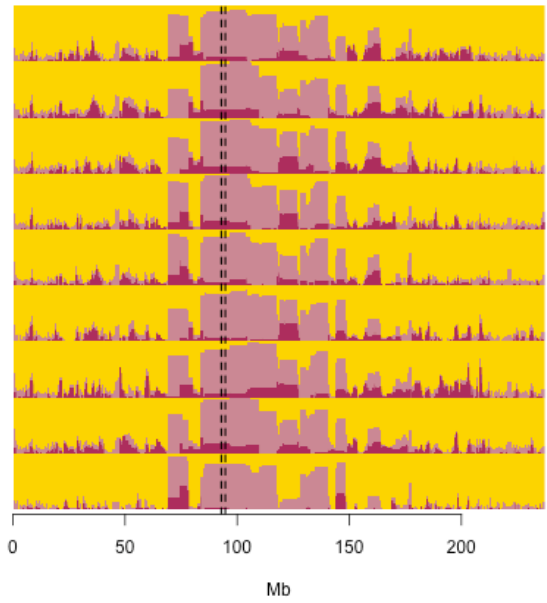
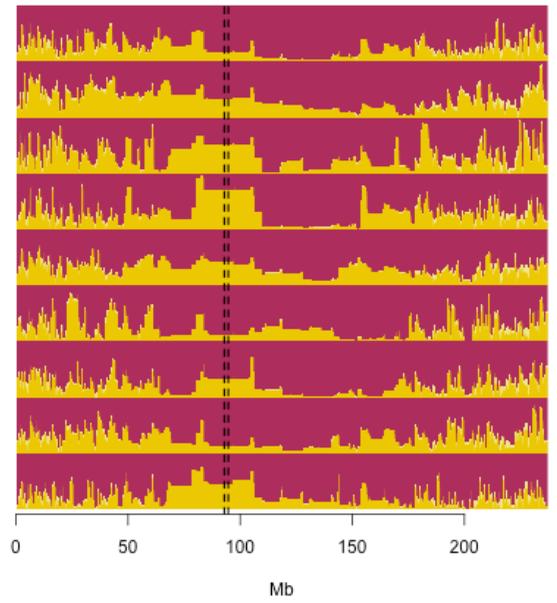
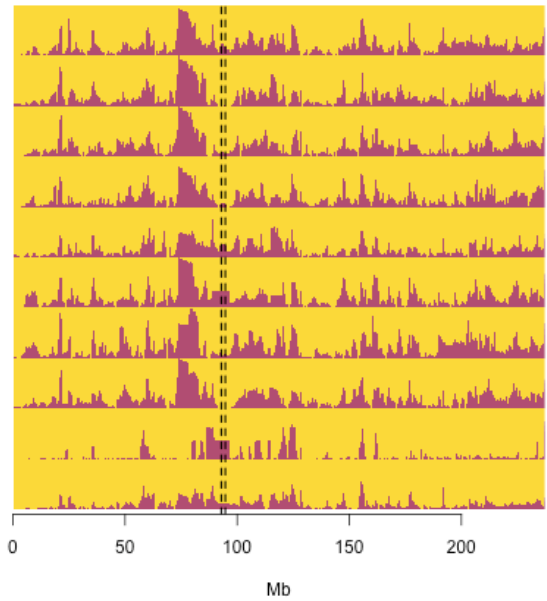
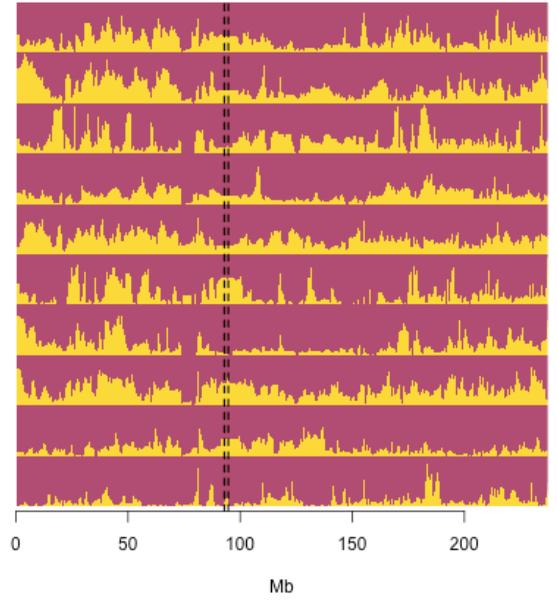
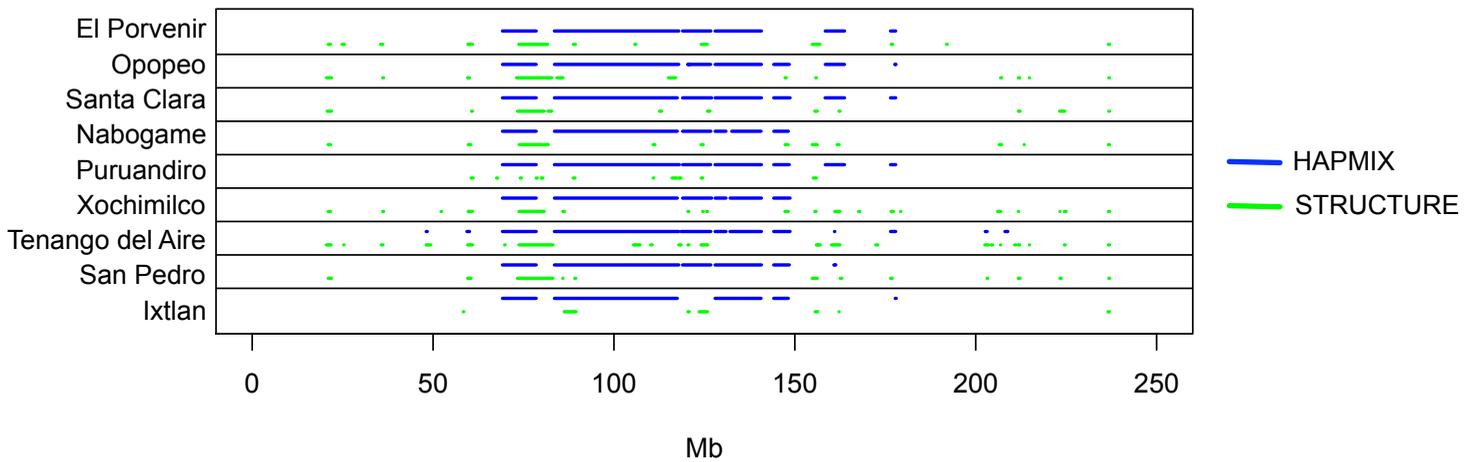

Figure S3

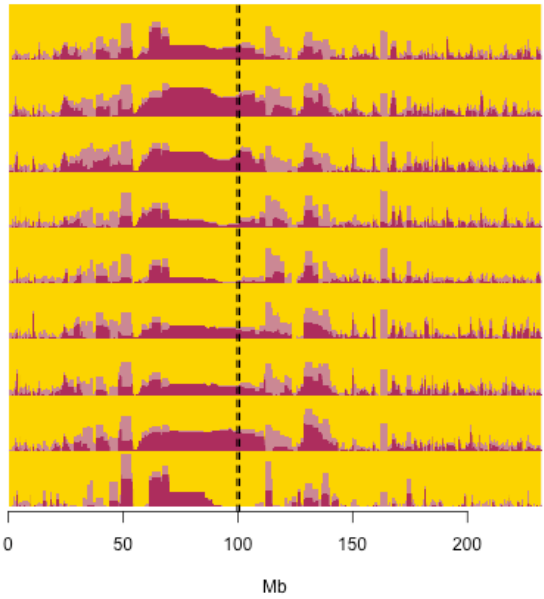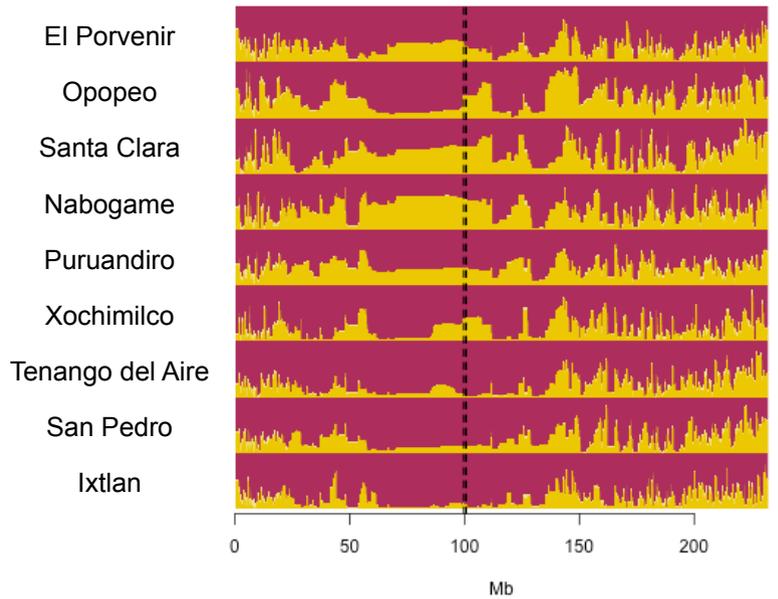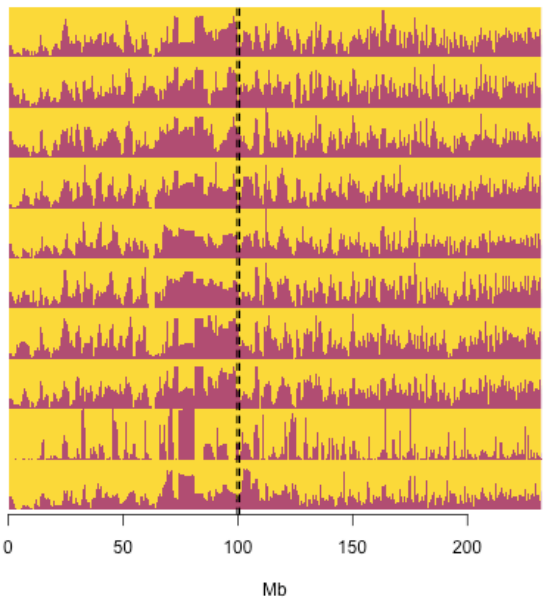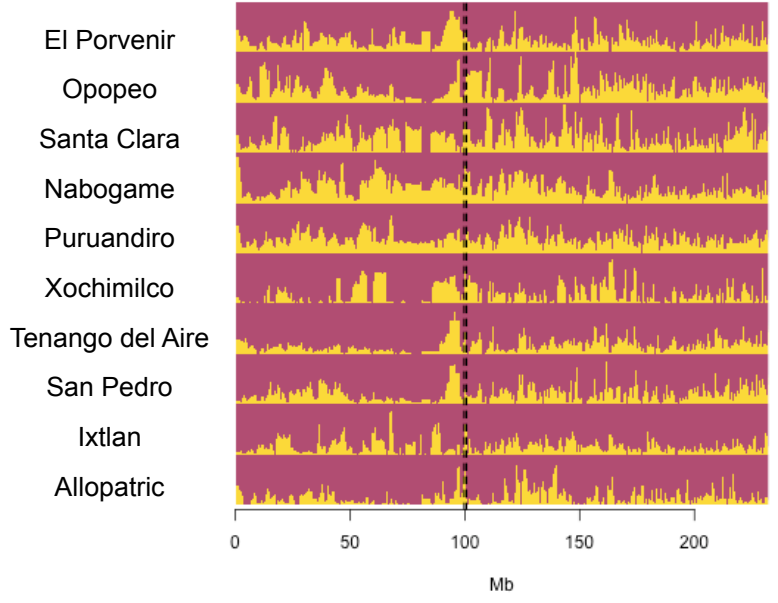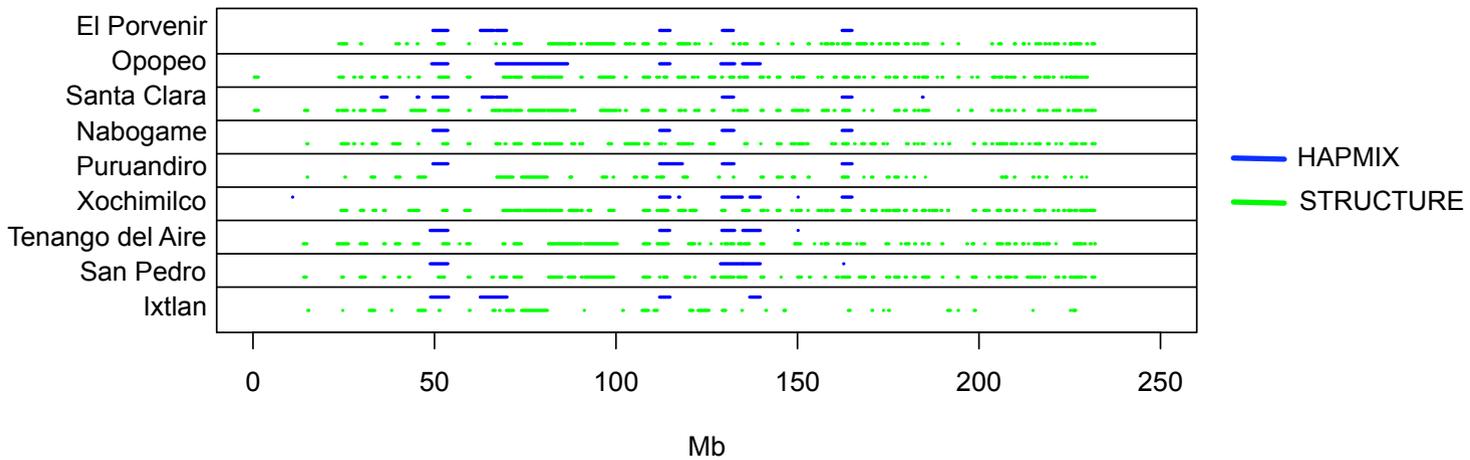

Figure S3

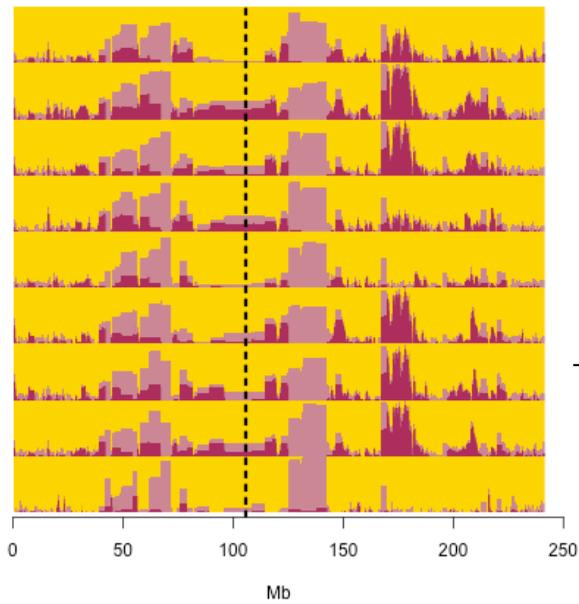
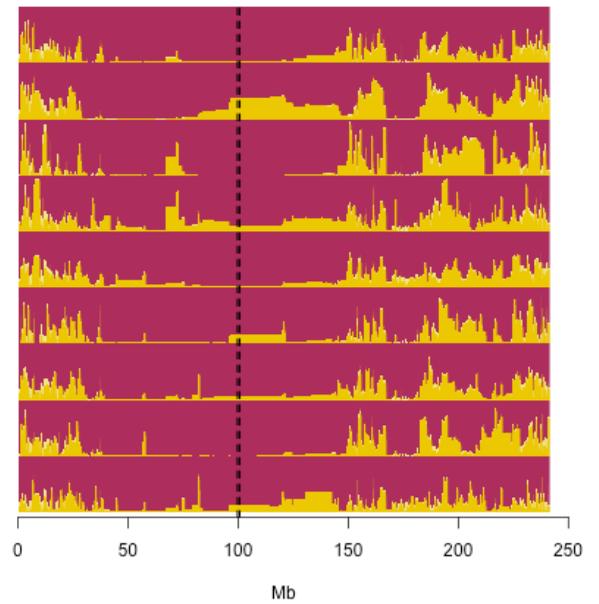
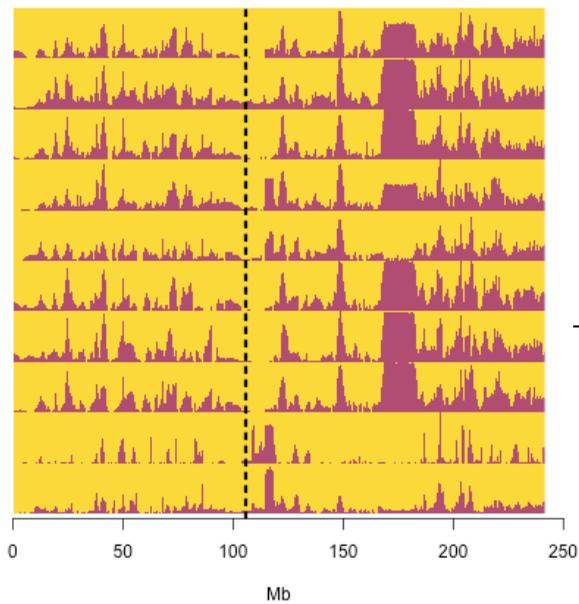
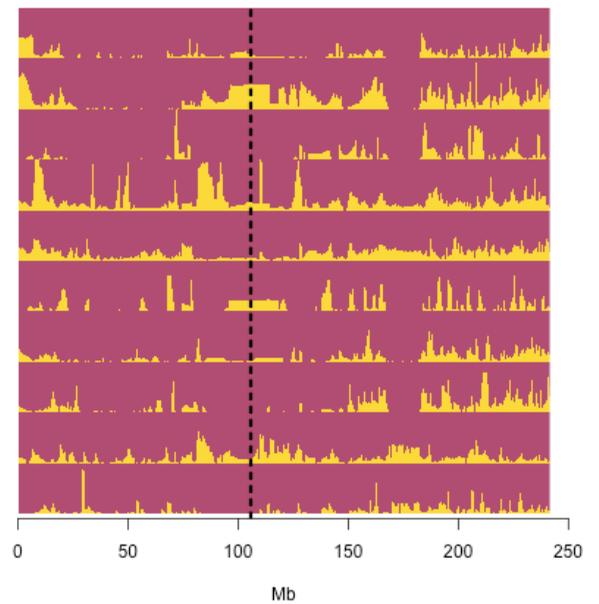

### Chromosome 4: Maize

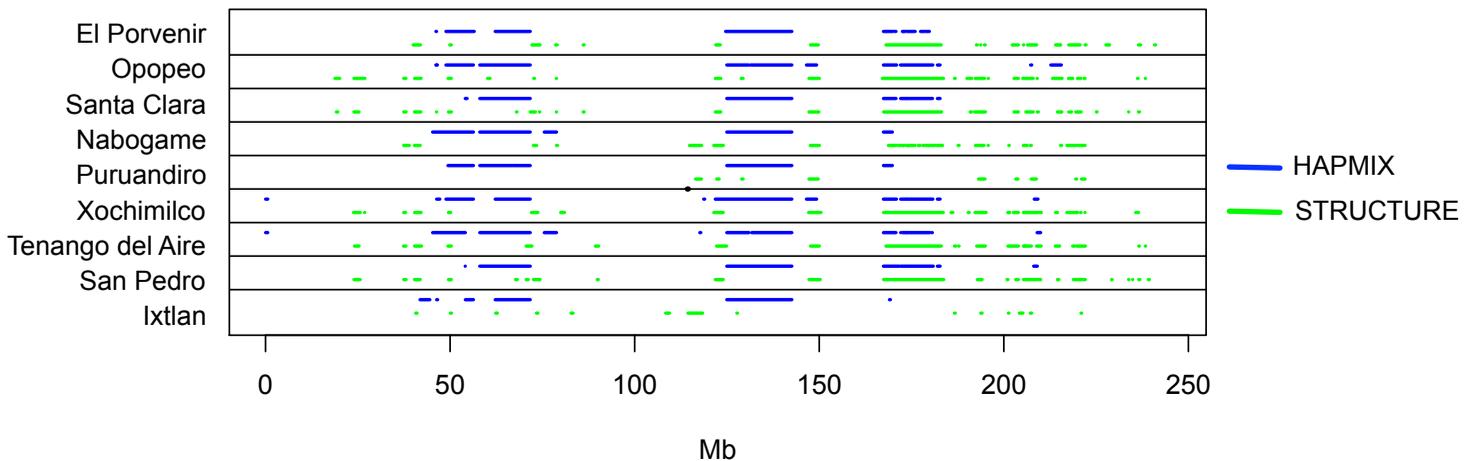

Figure S3

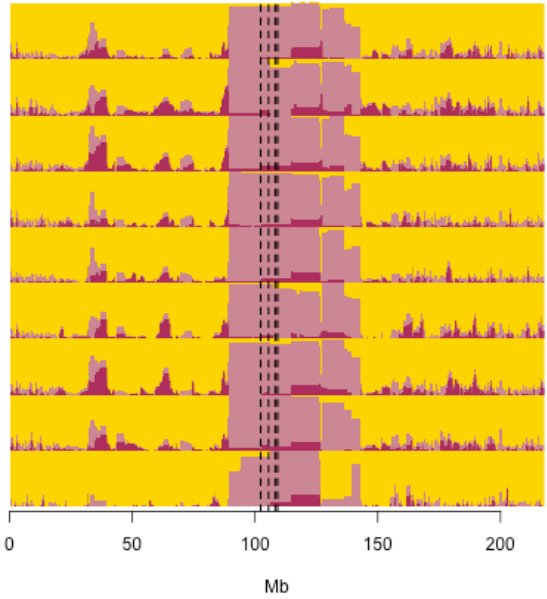
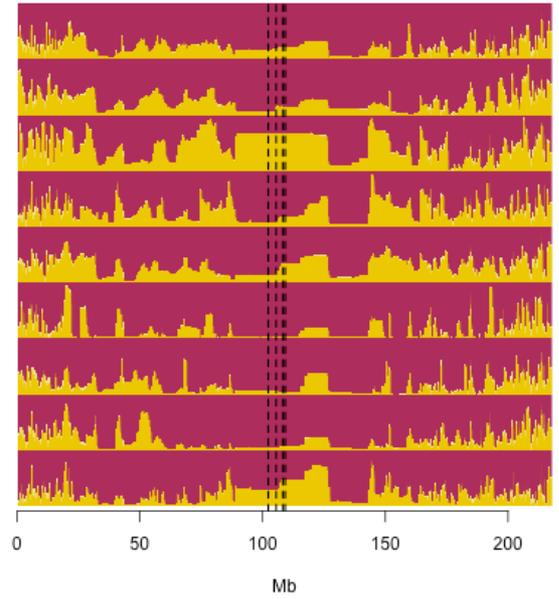
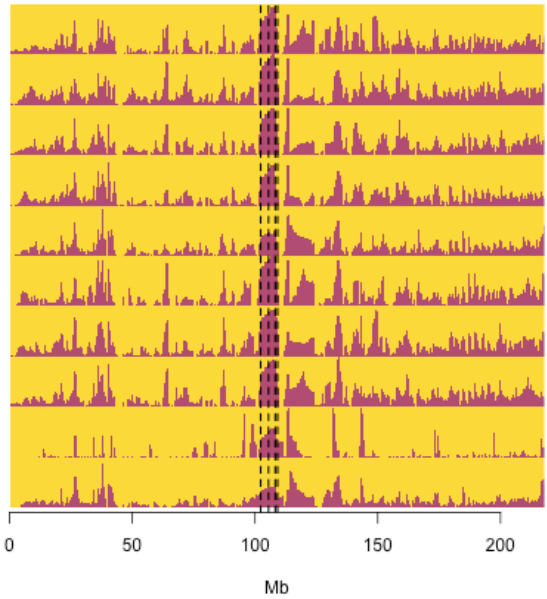
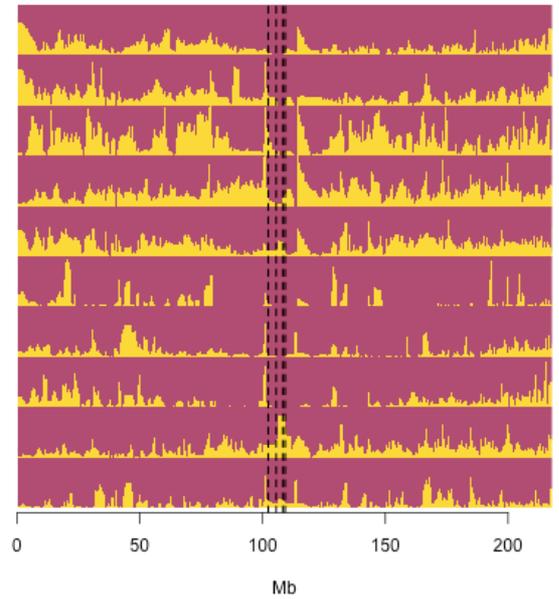
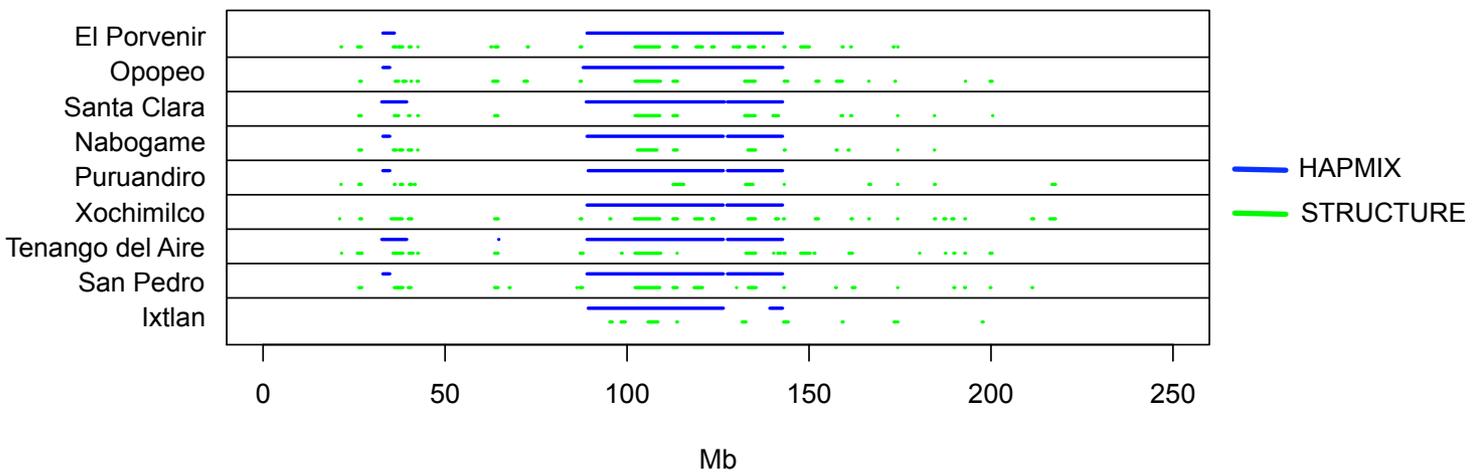

Figure S3

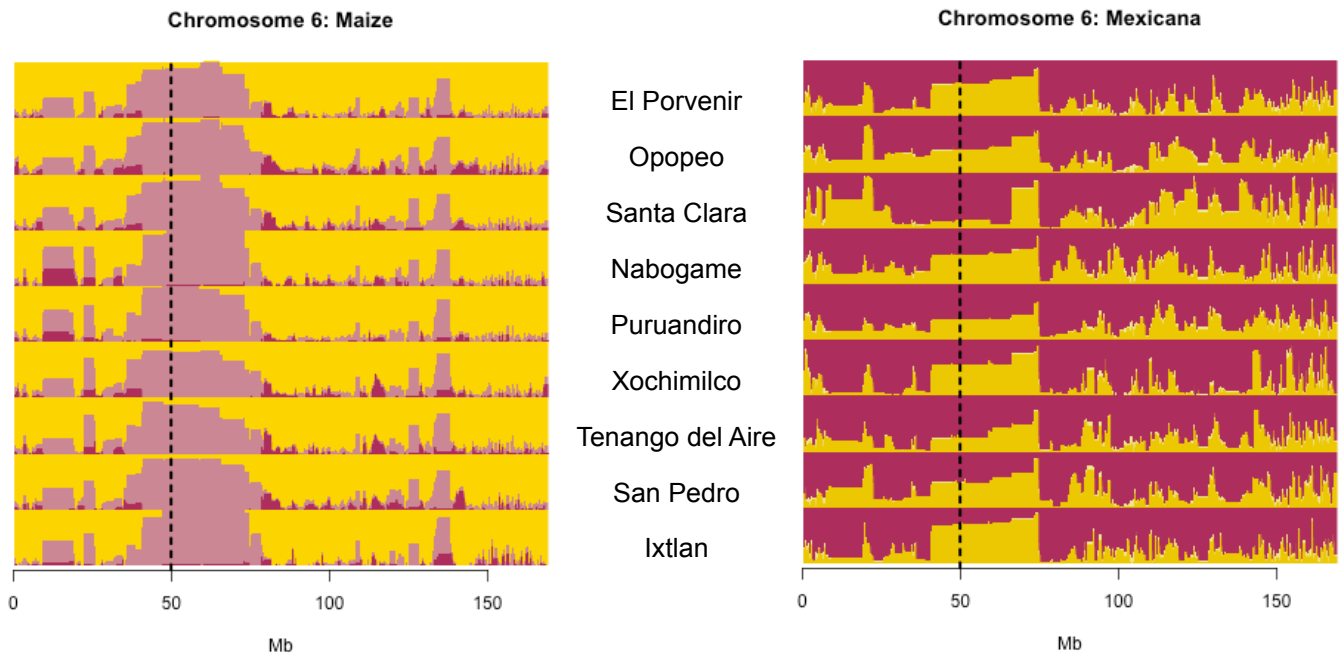
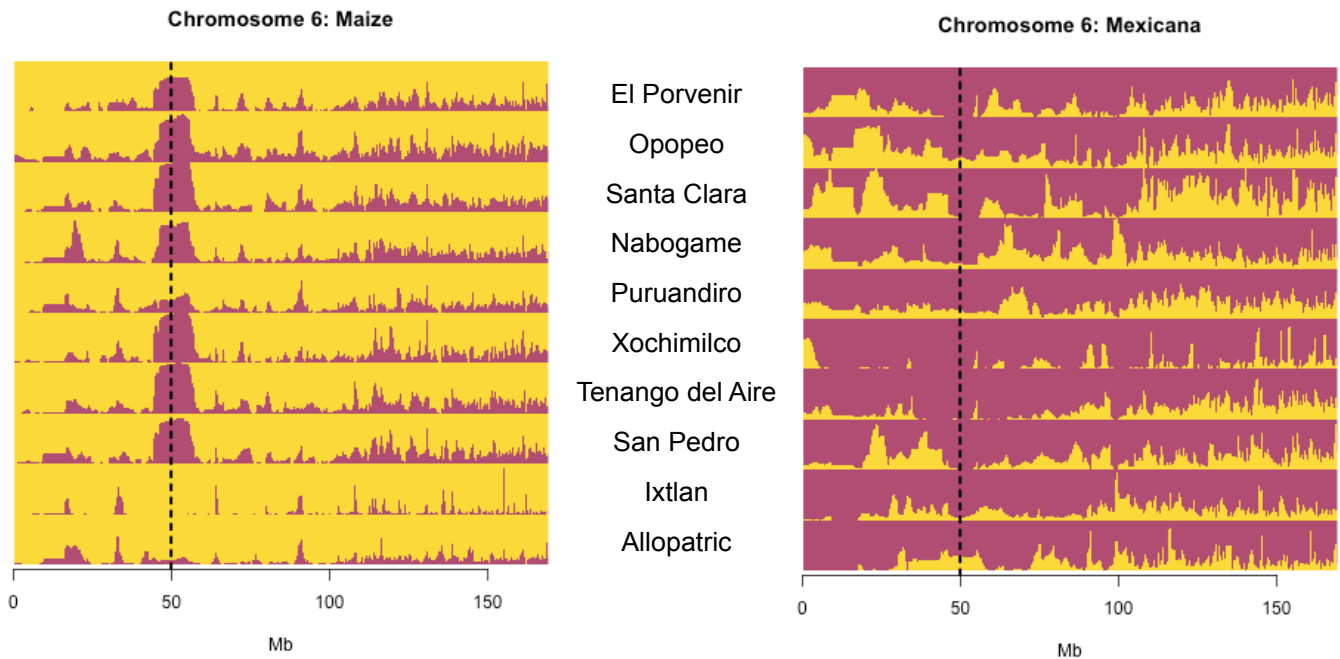
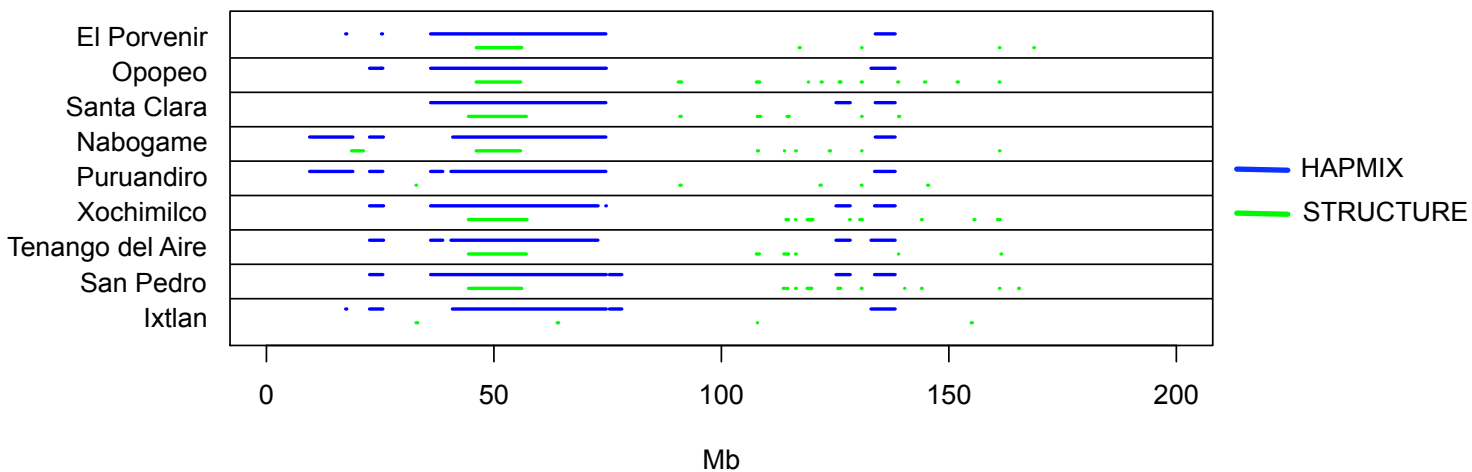

Figure S3

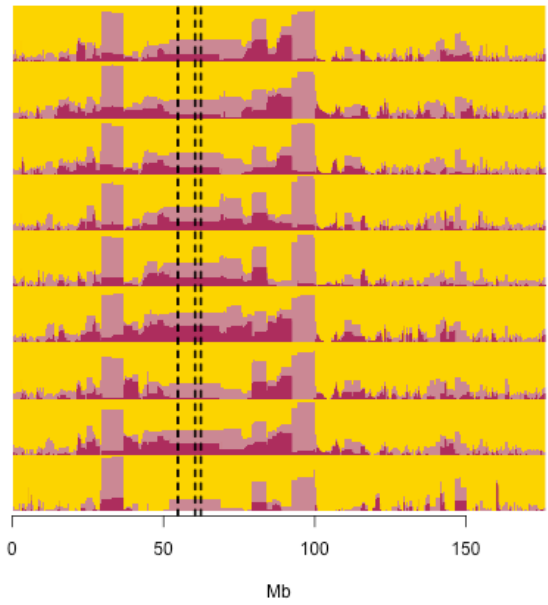
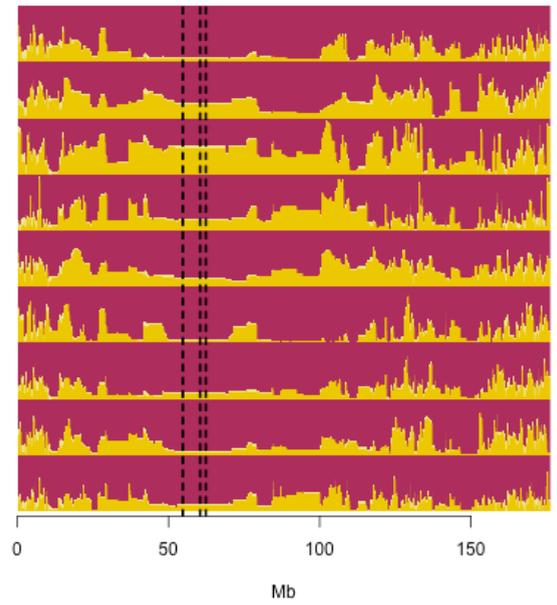
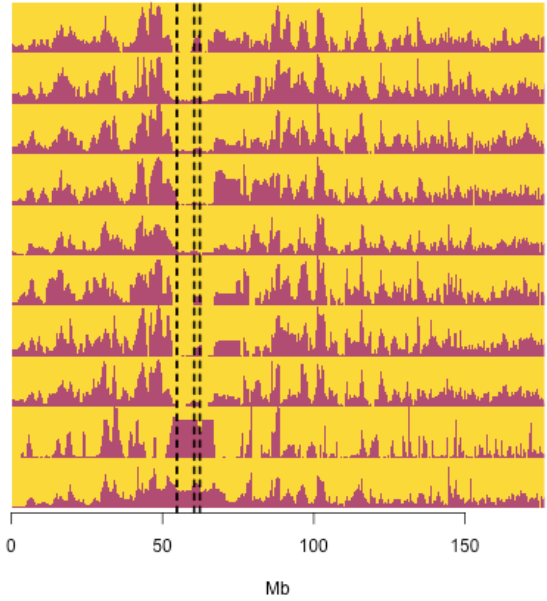
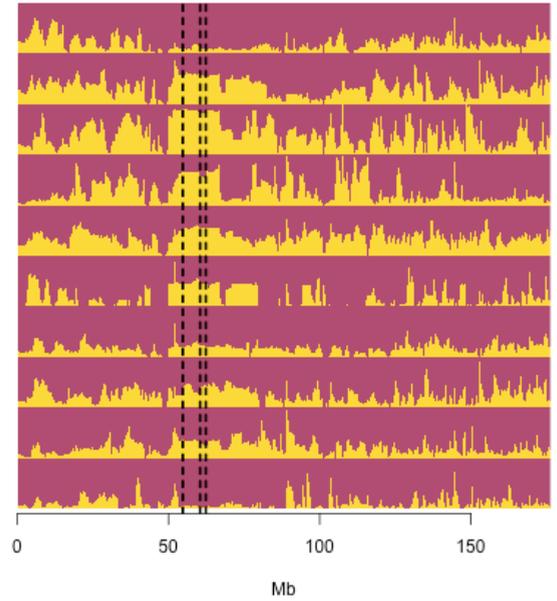
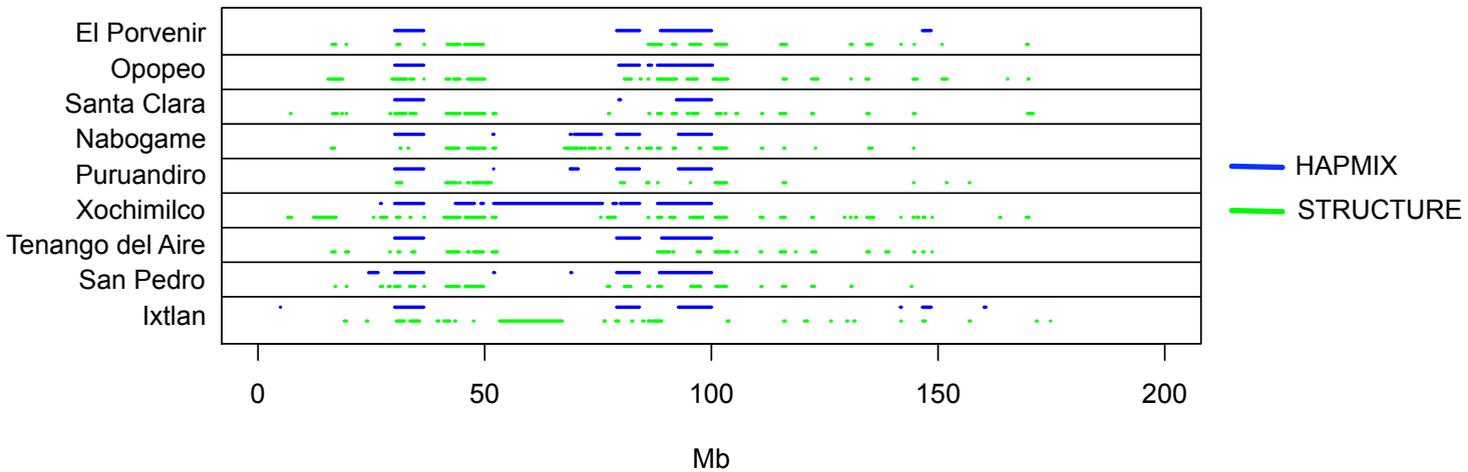

Figure S3

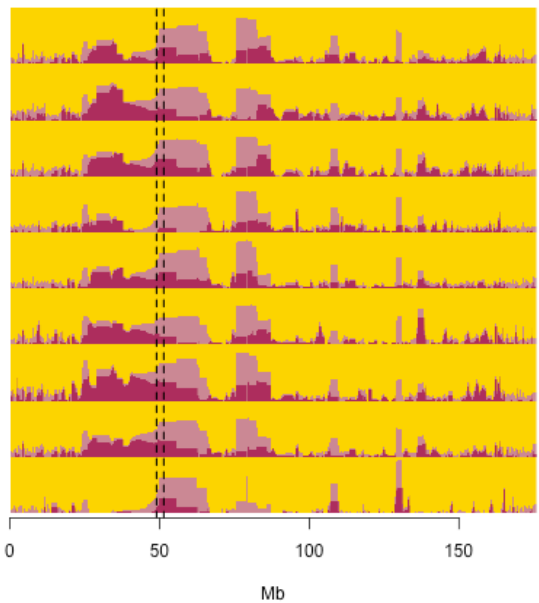
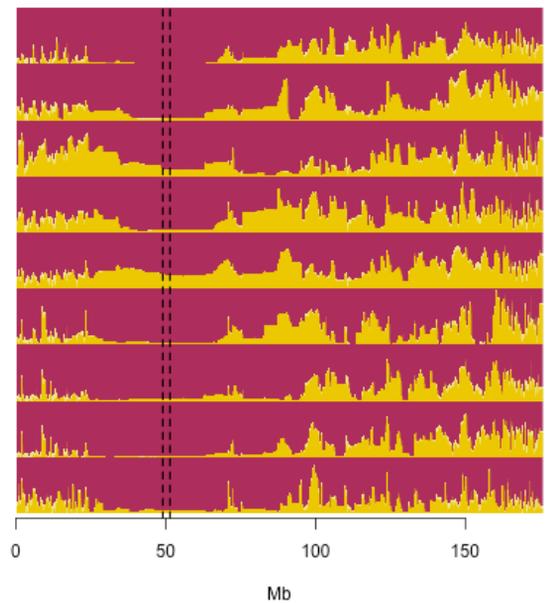
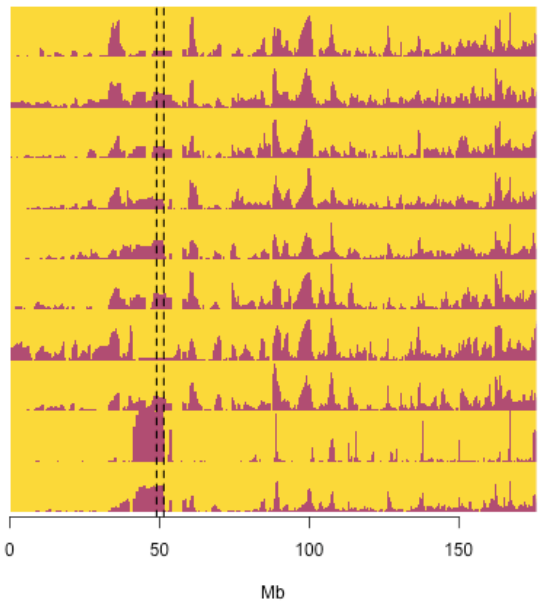
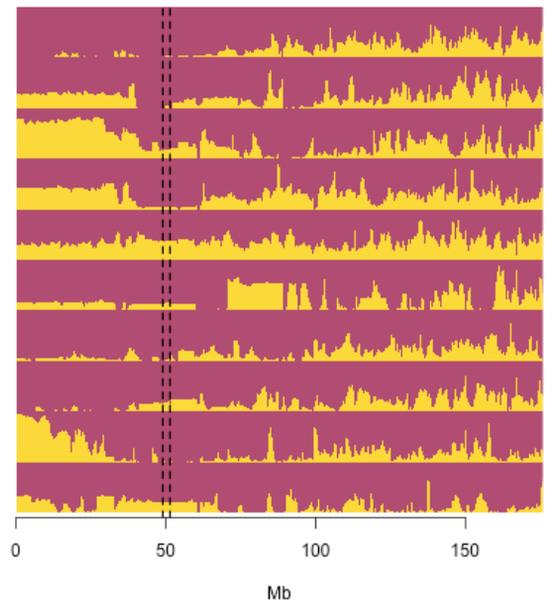
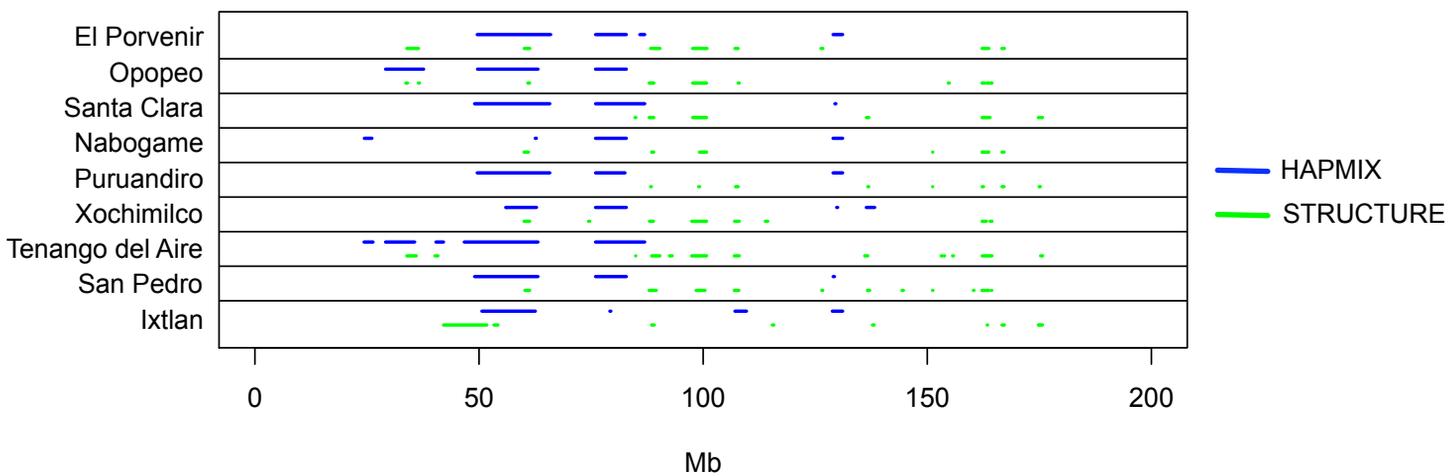

Figure S3

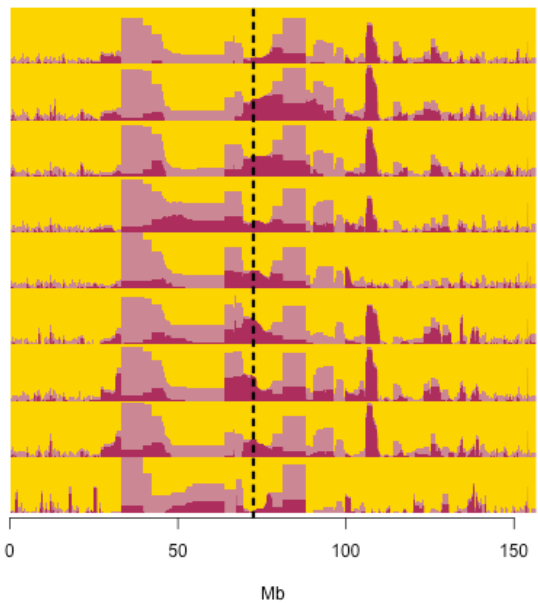
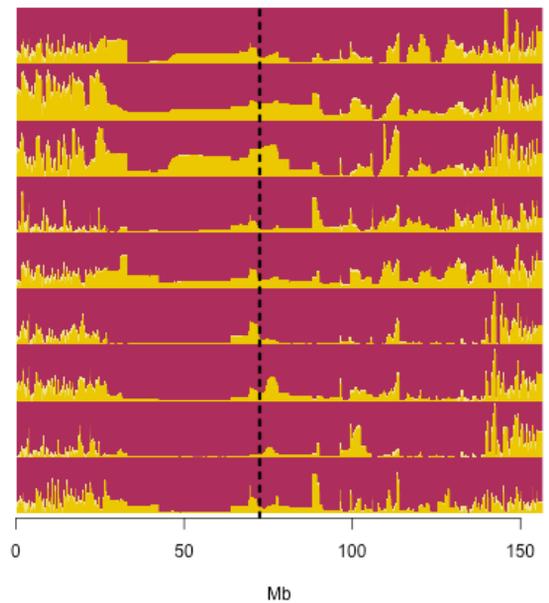
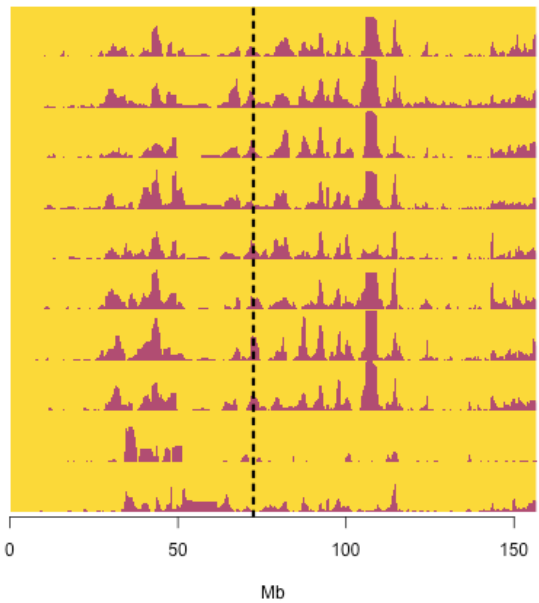
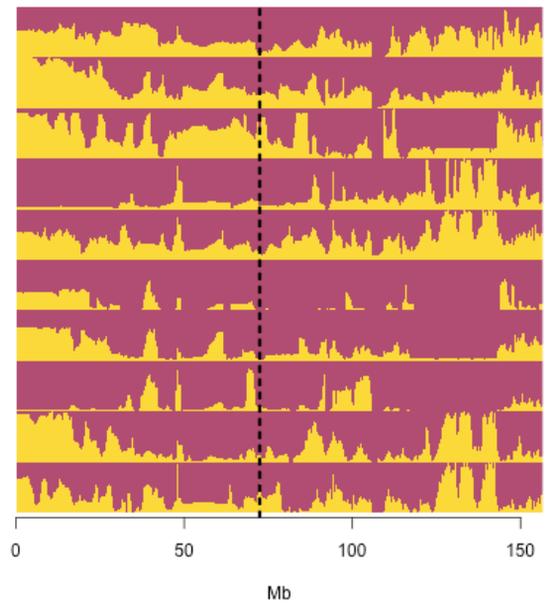
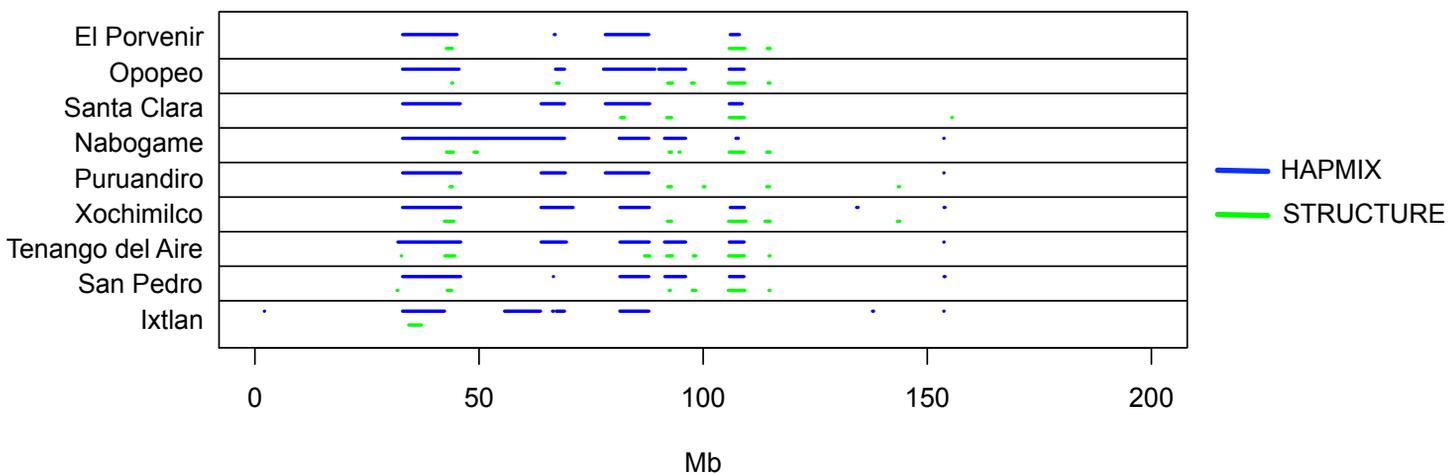

Figure S3

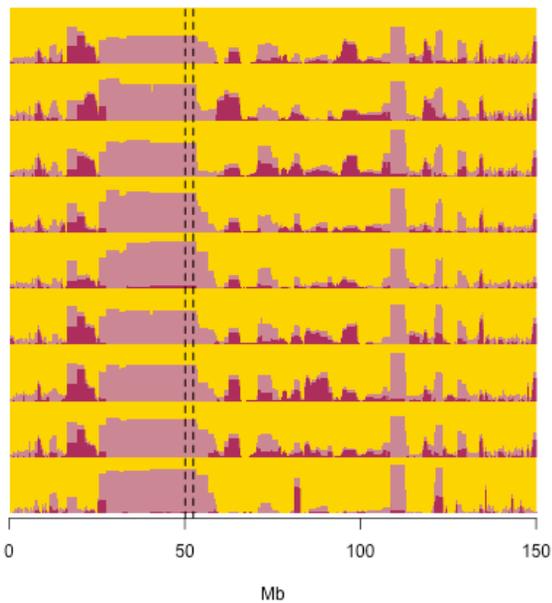
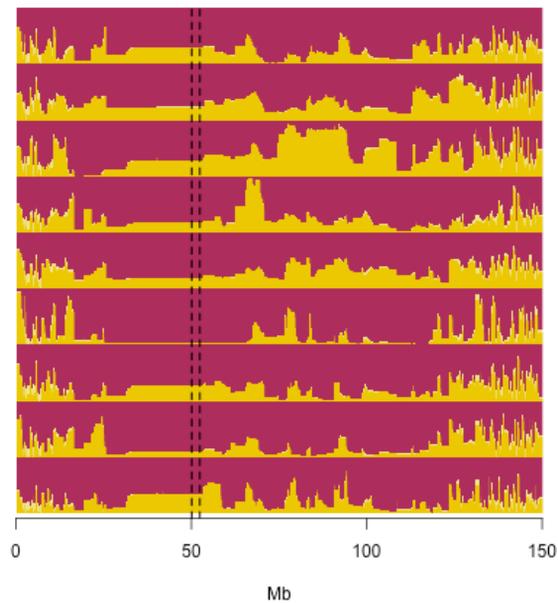
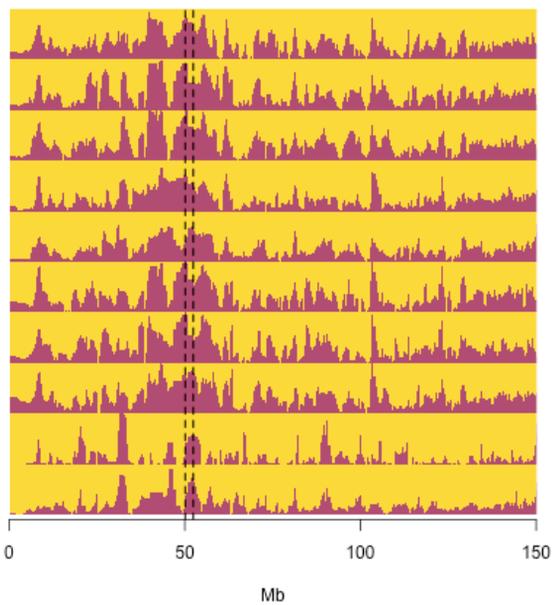
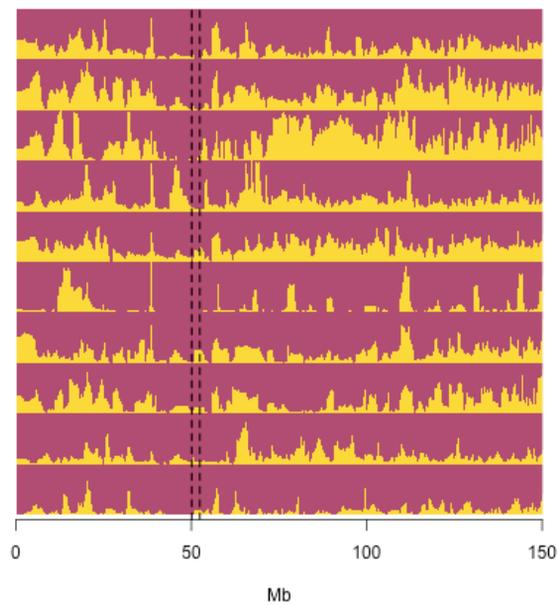
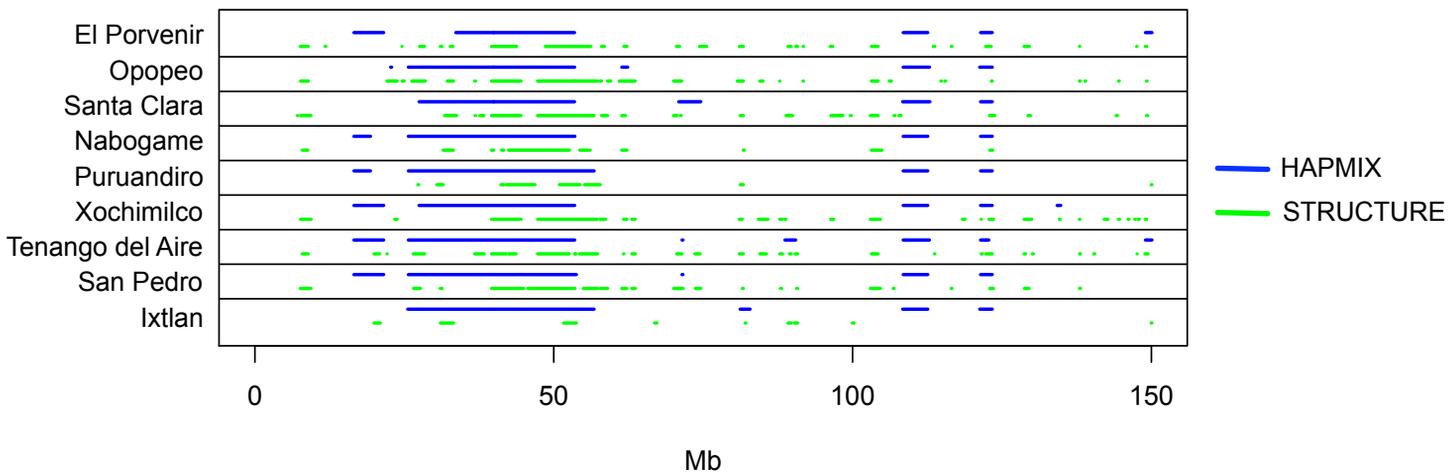

Figure S3

A

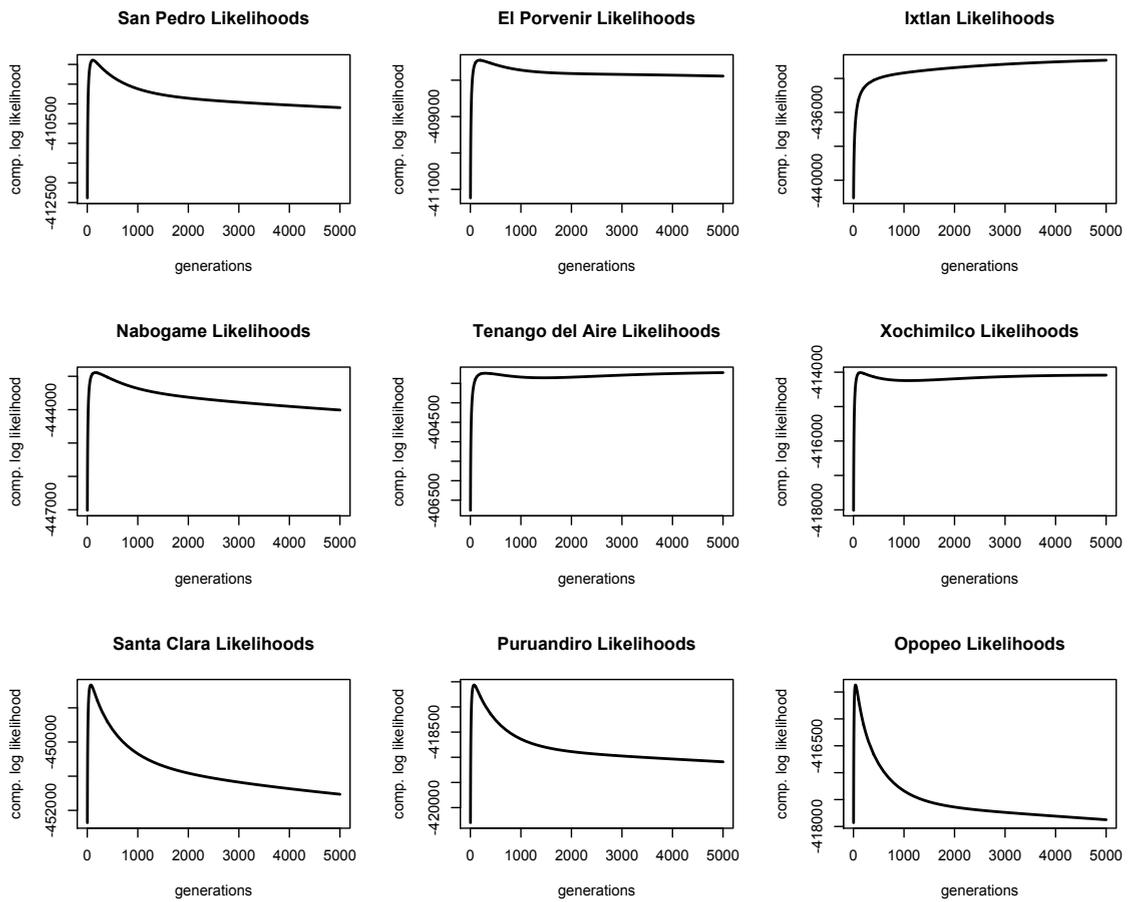

B

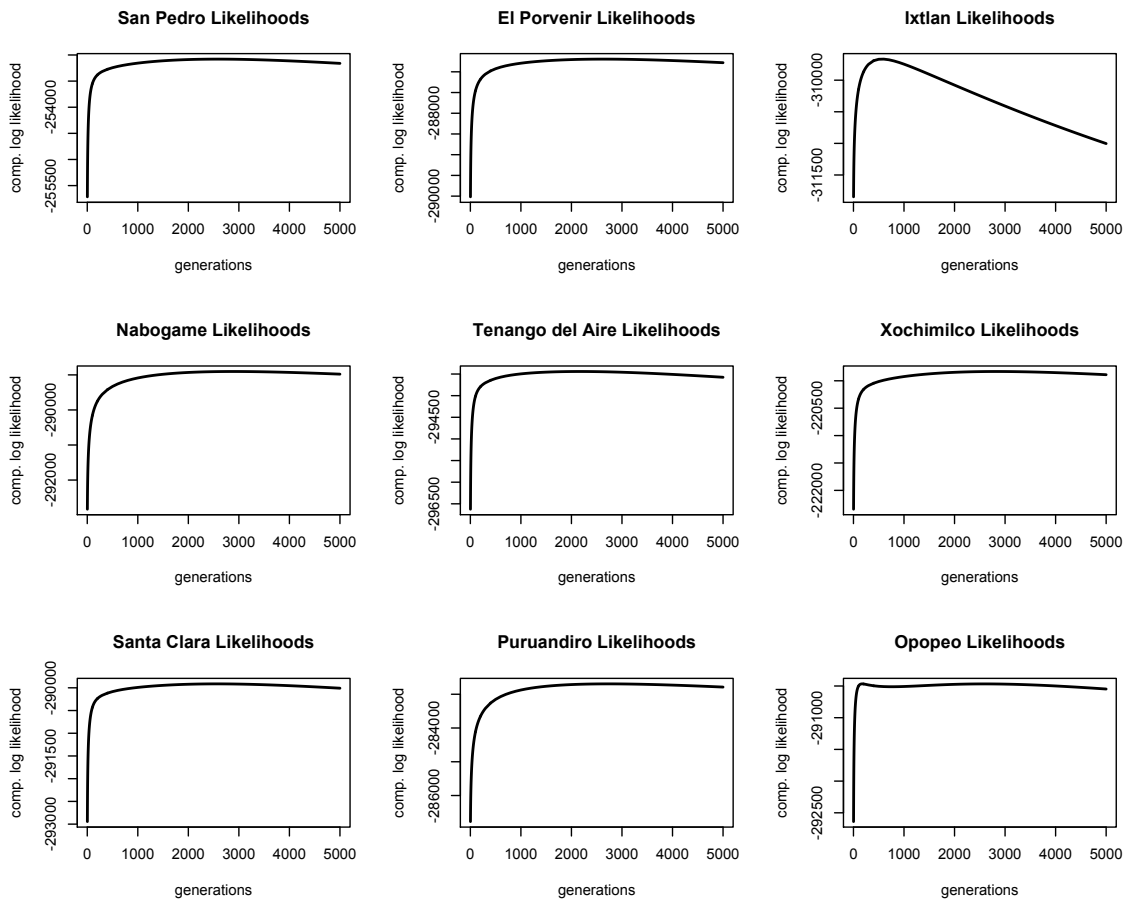

Figure S4

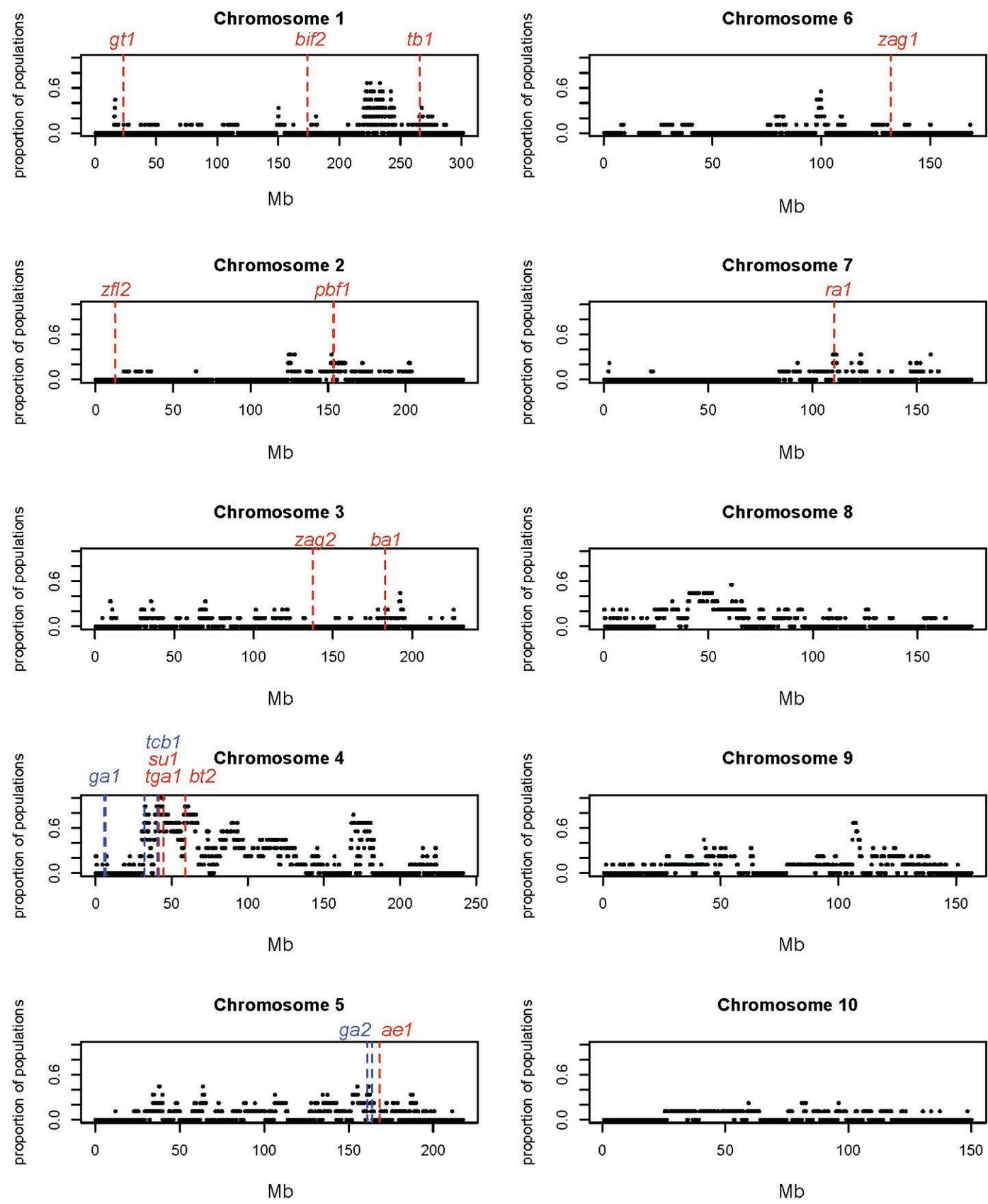

Figure S5

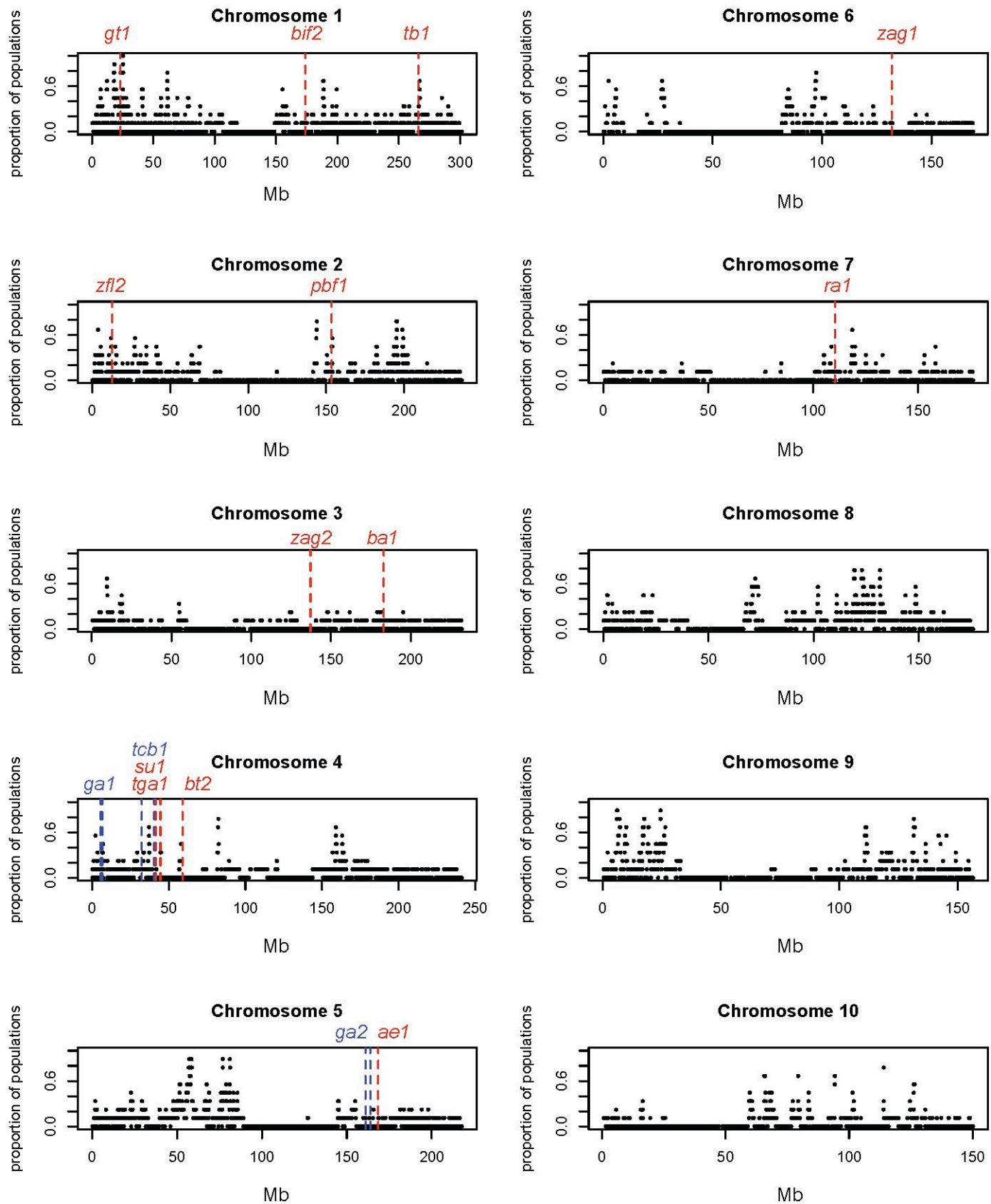

Figure S5

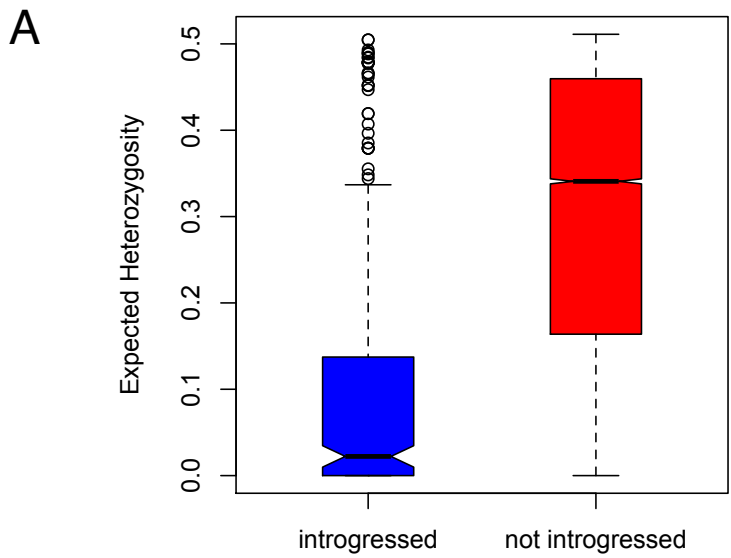
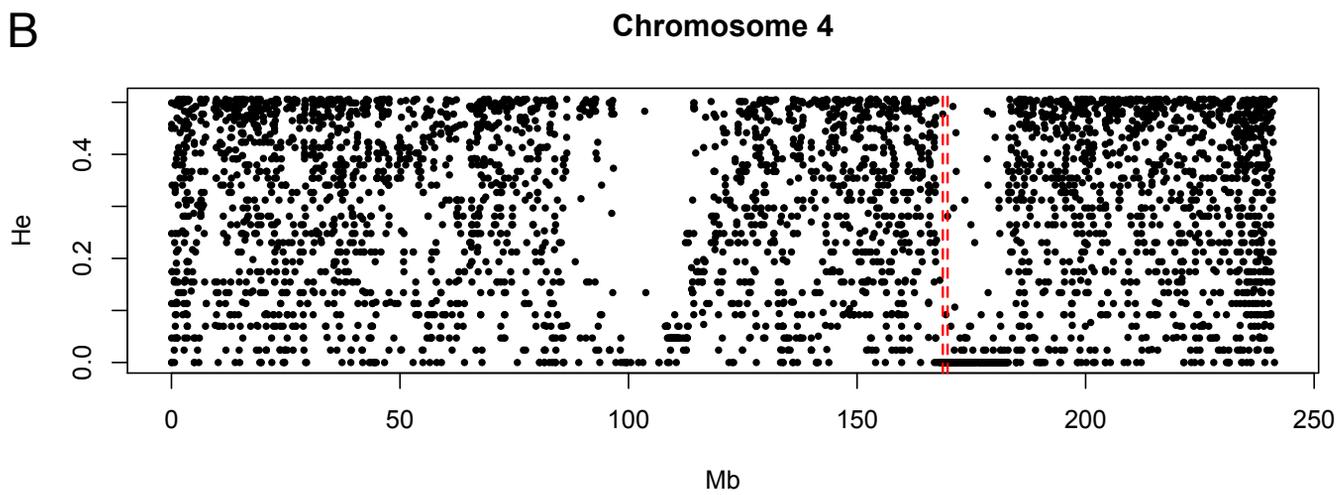
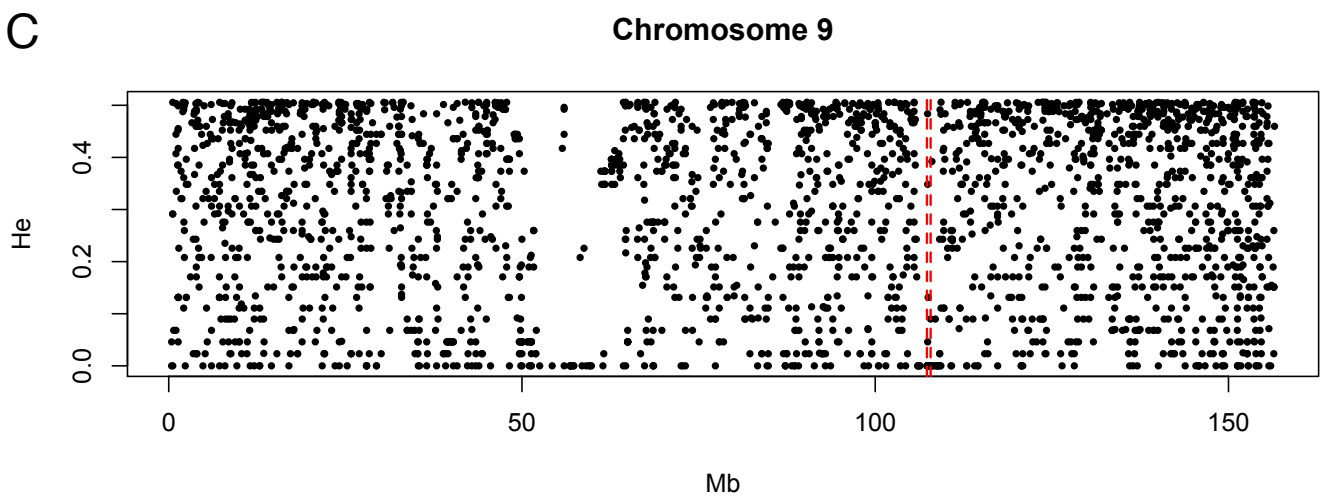

Figure S6

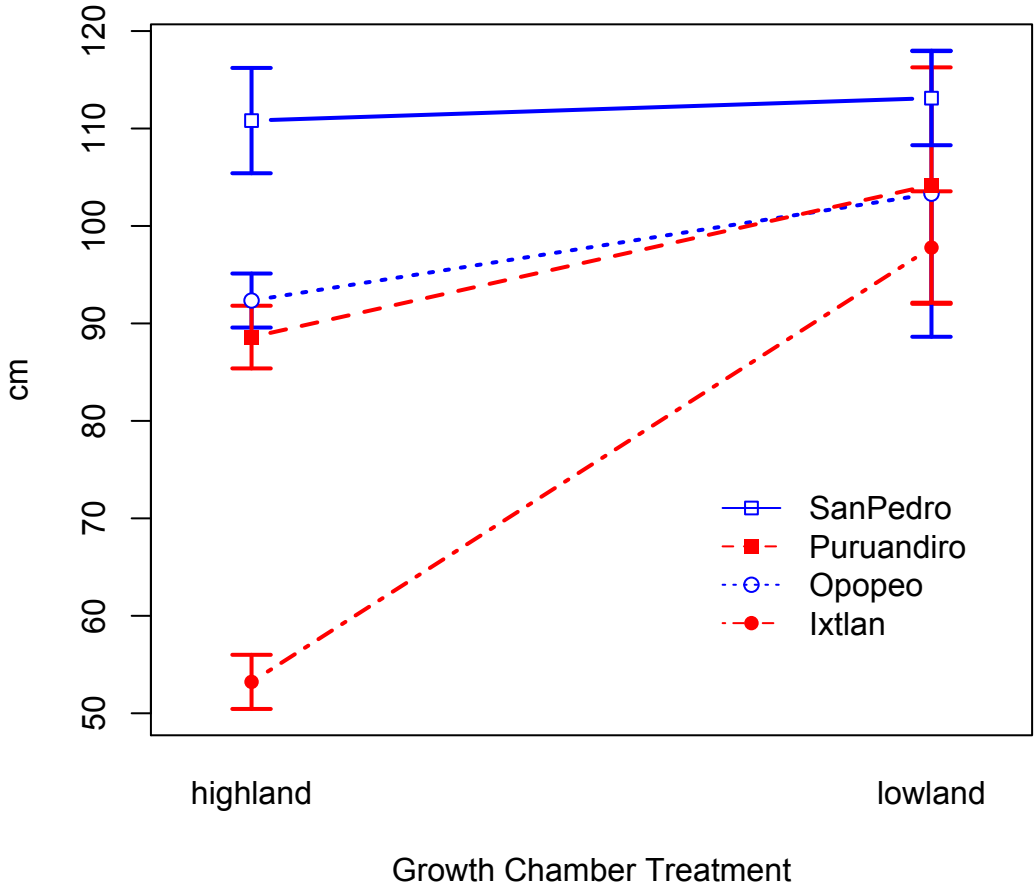
Figure S7

# Chromosome 1

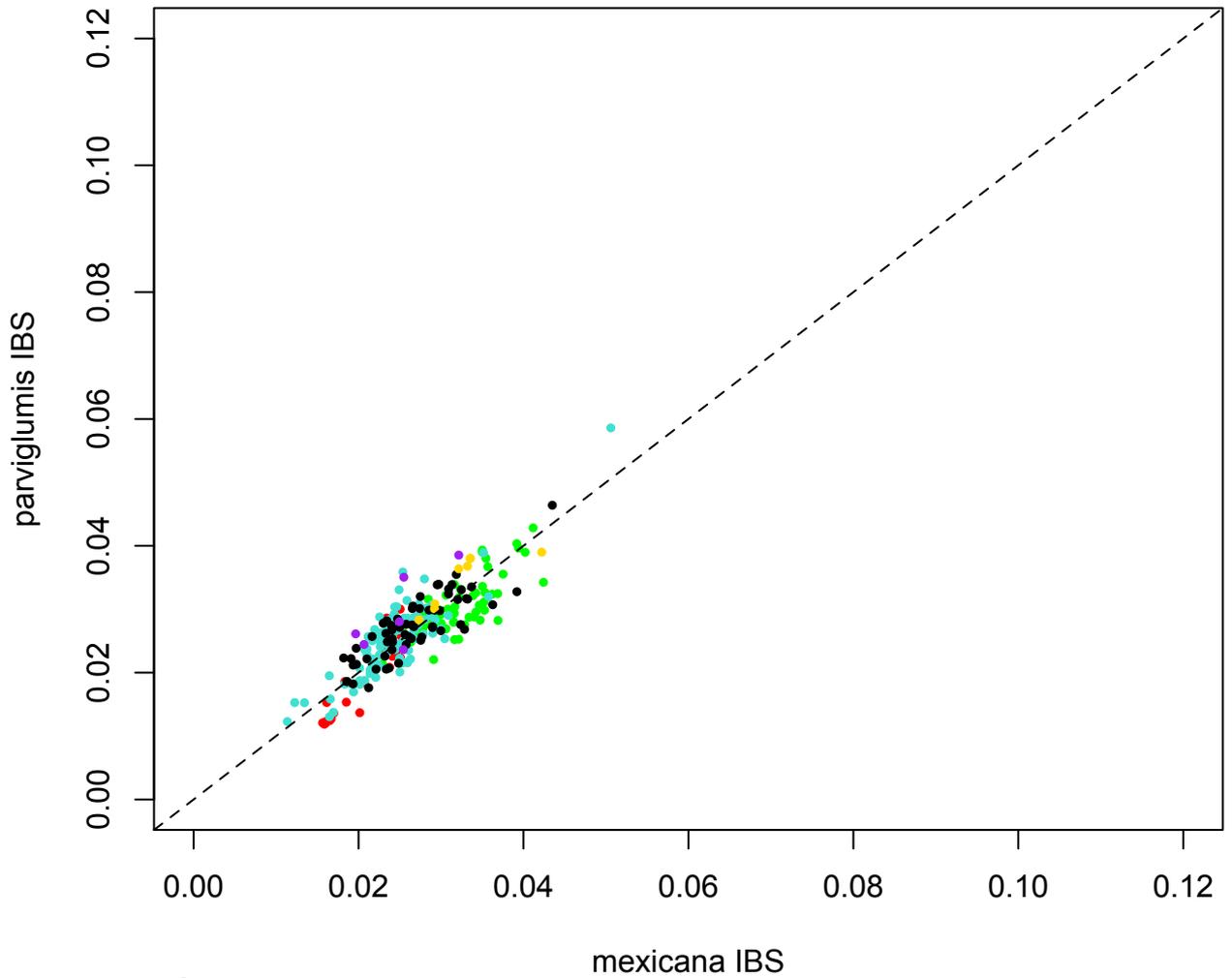

Figure S8

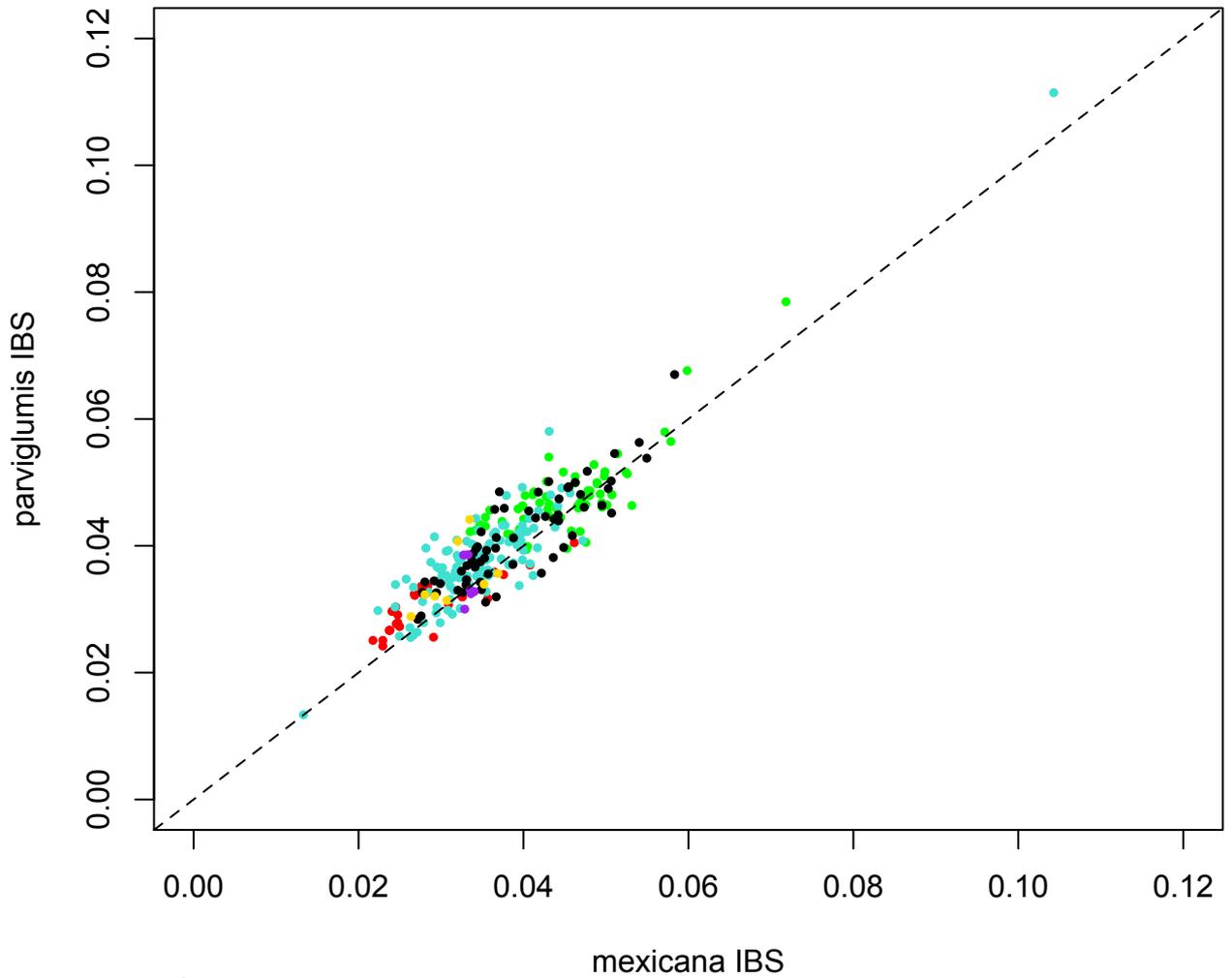

Figure S8

# Chromosome 3

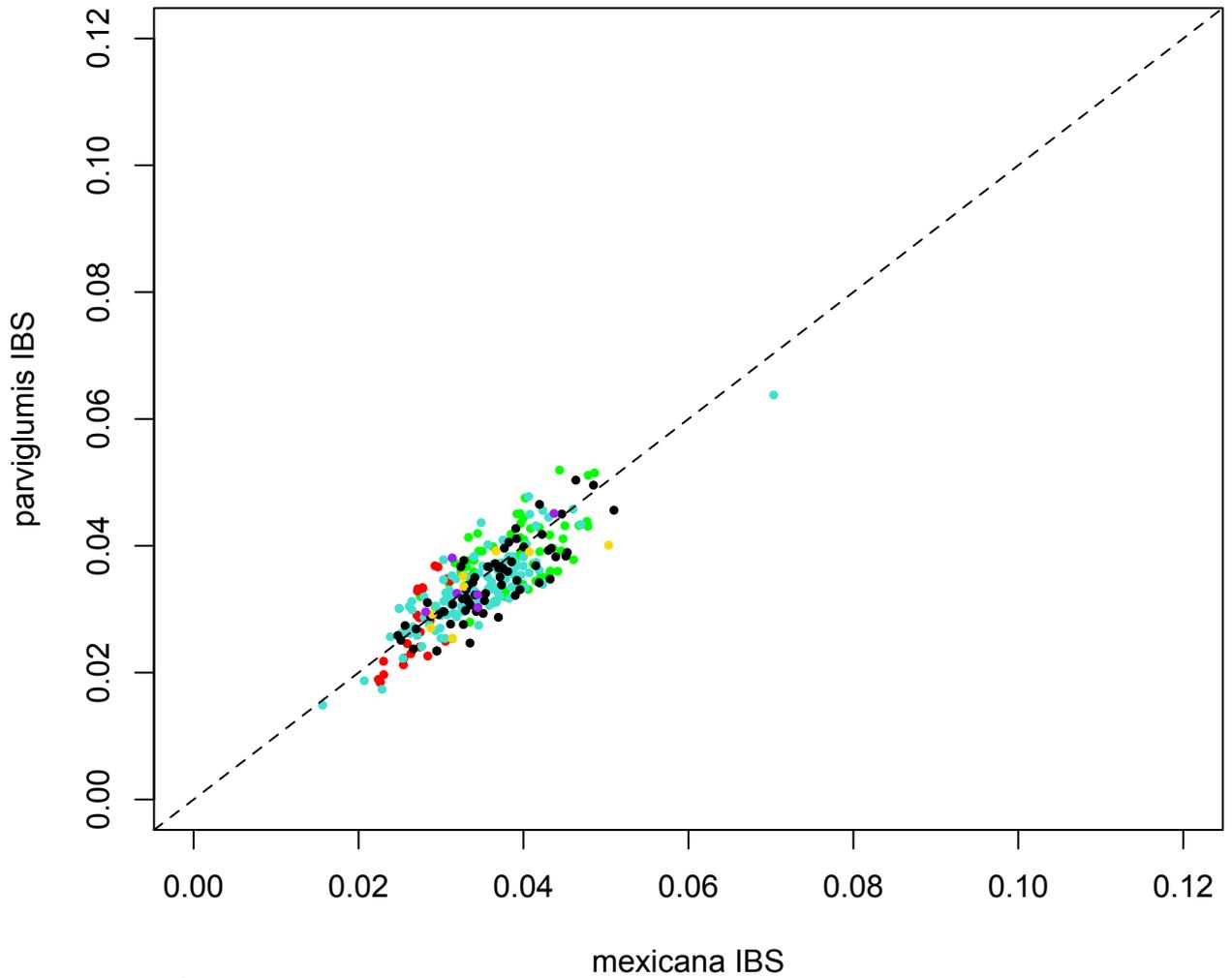

Figure S8

# Chromosome 4

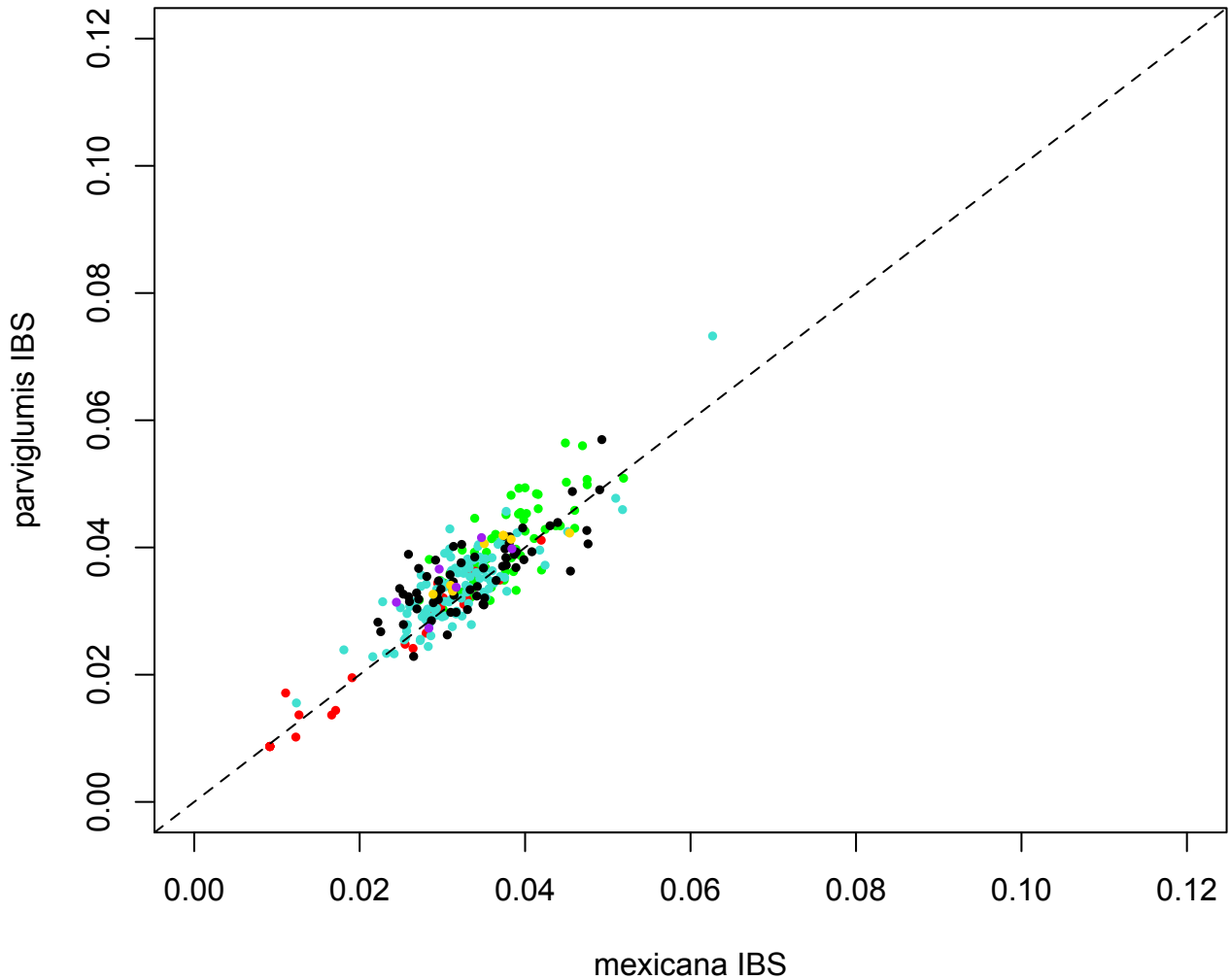

Figure S8

# Chromosome 5

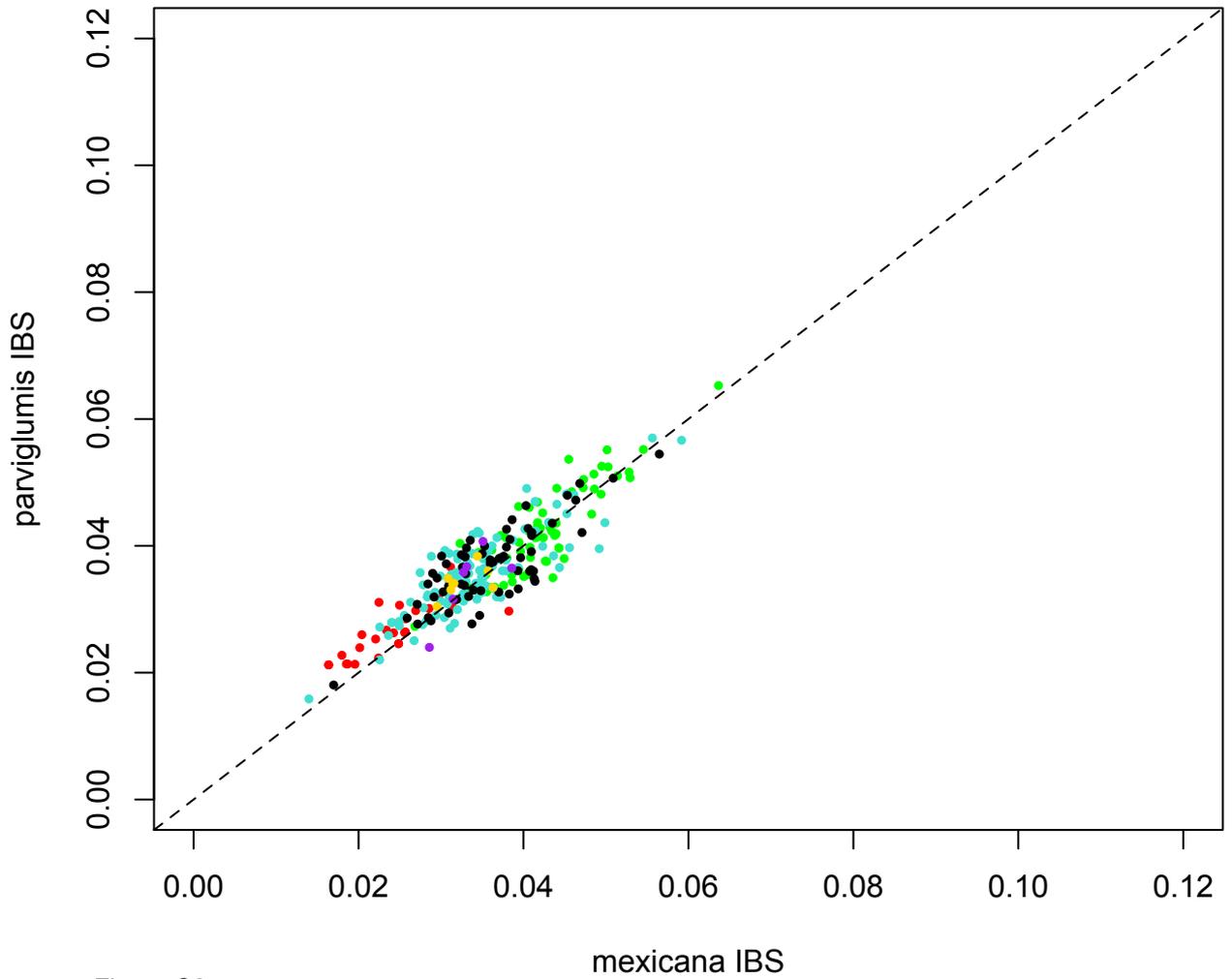

Figure S8

# Chromosome 6

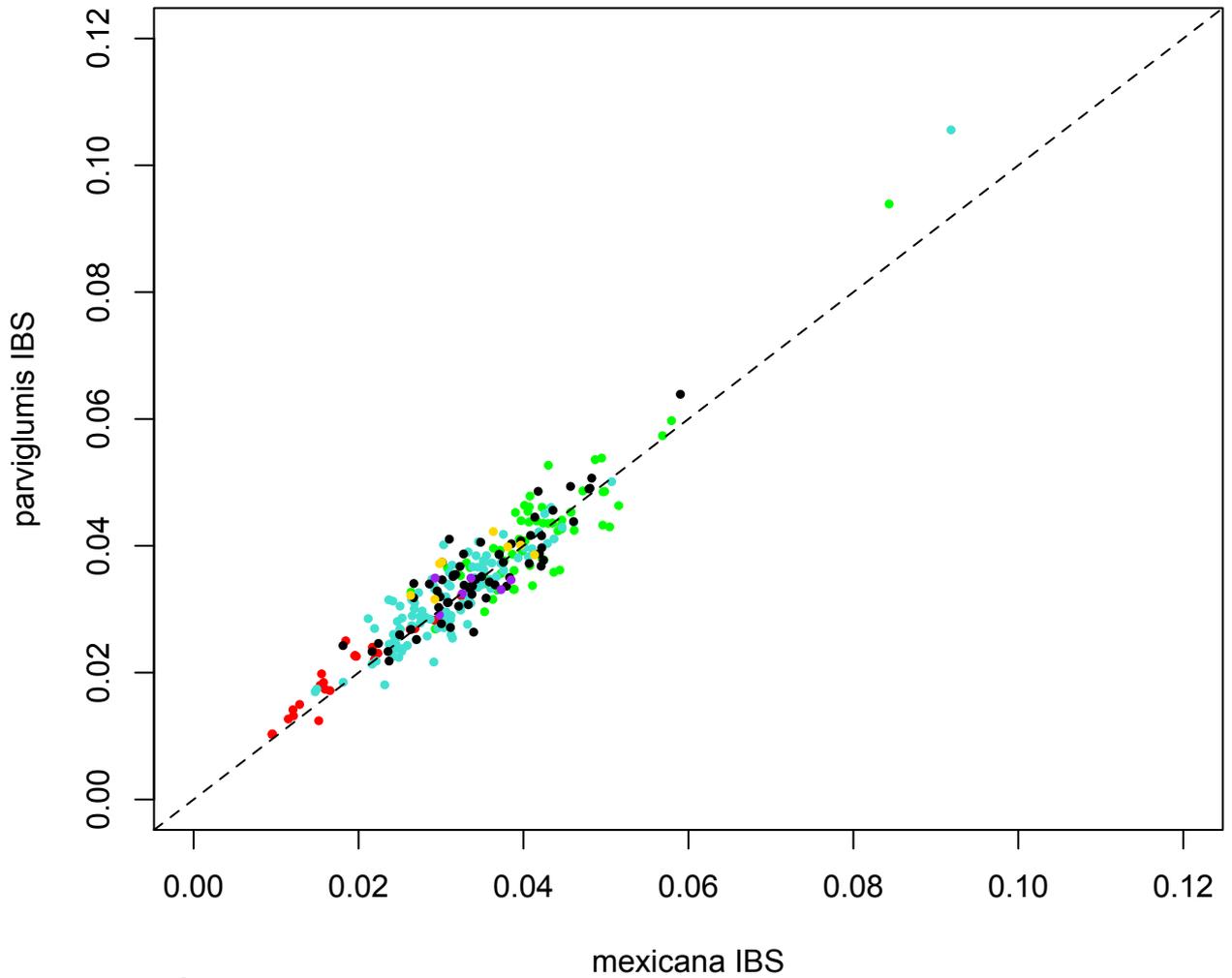

Figure S8

# Chromosome 7

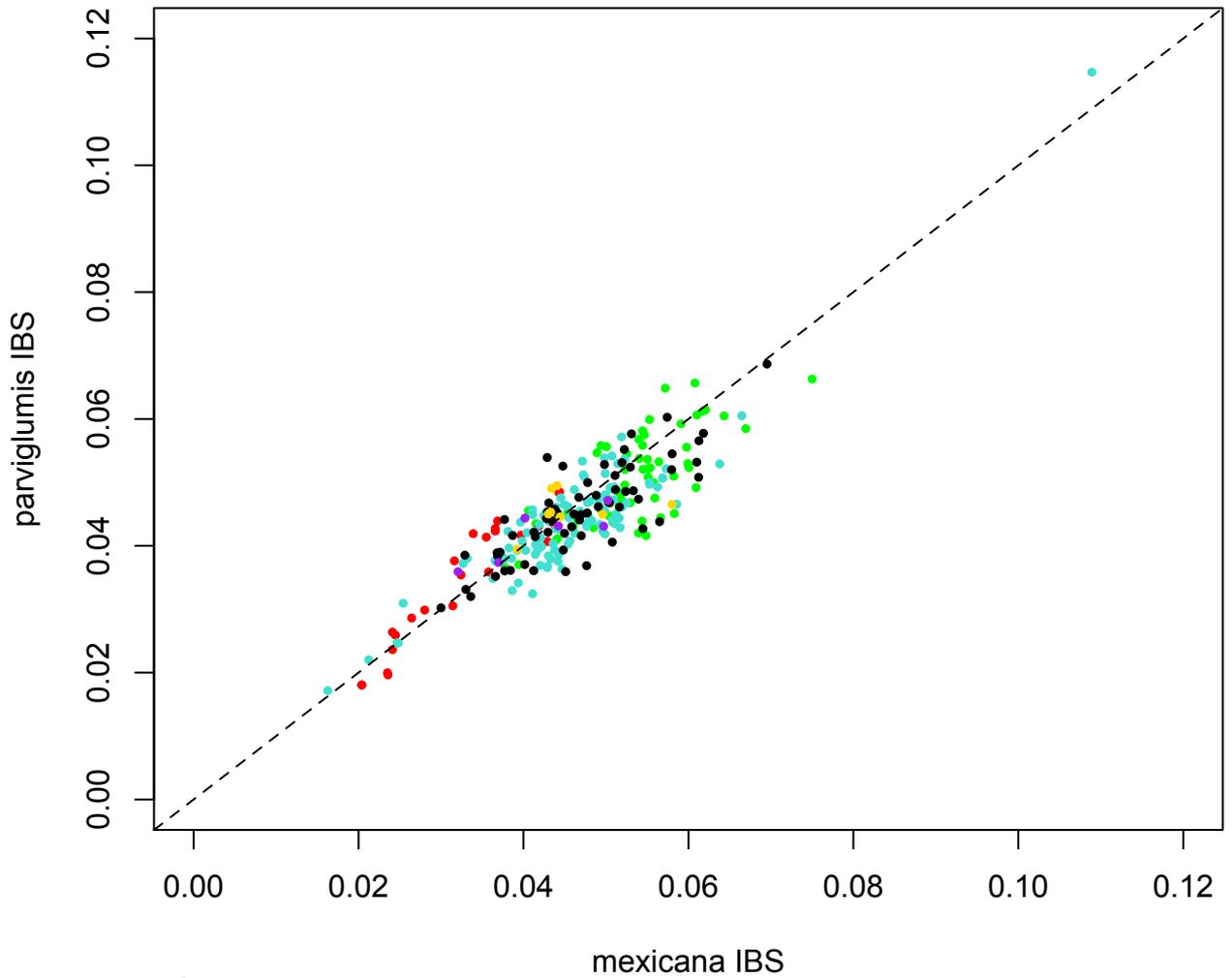

Figure S8

# Chromosome 8

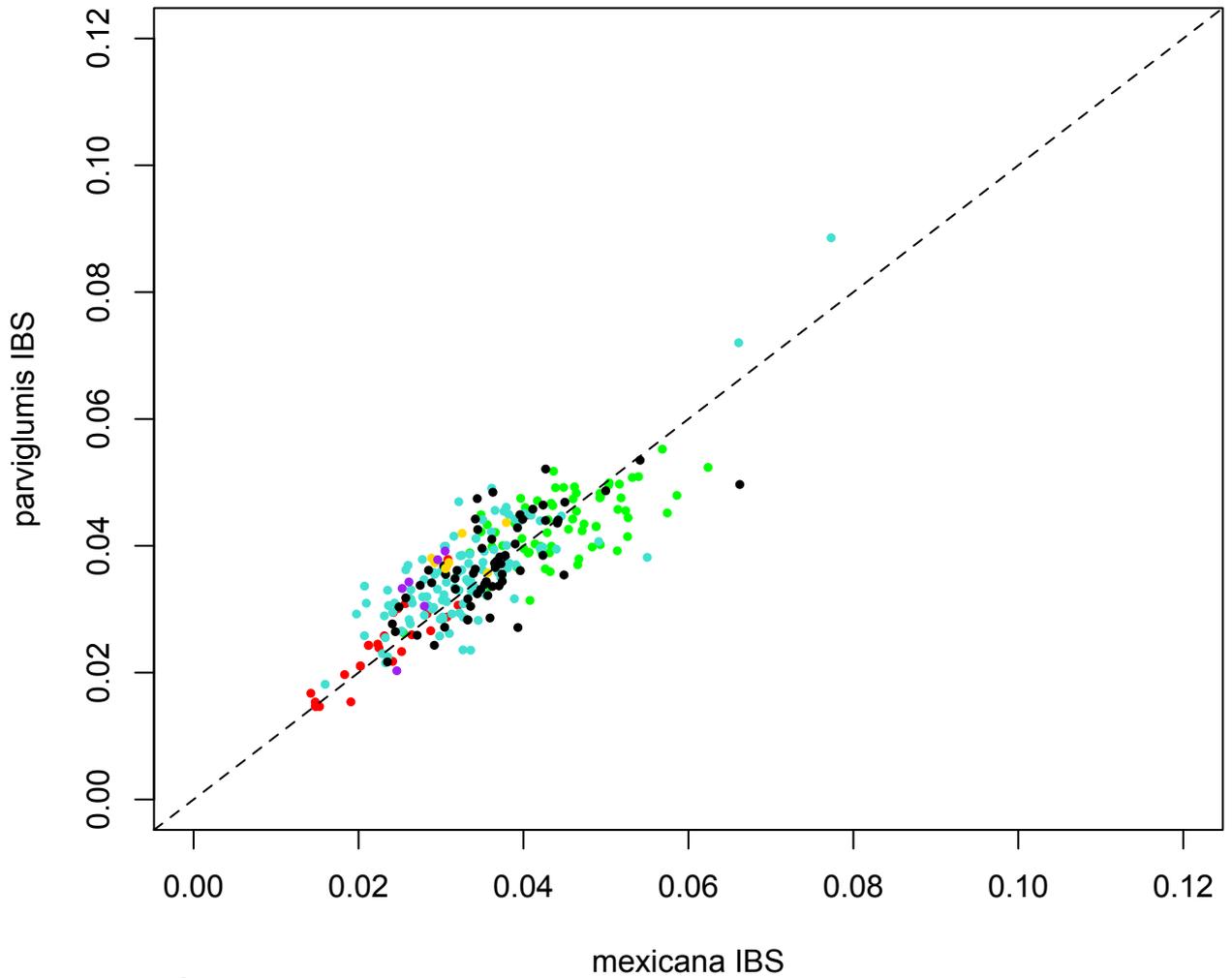

Figure S8

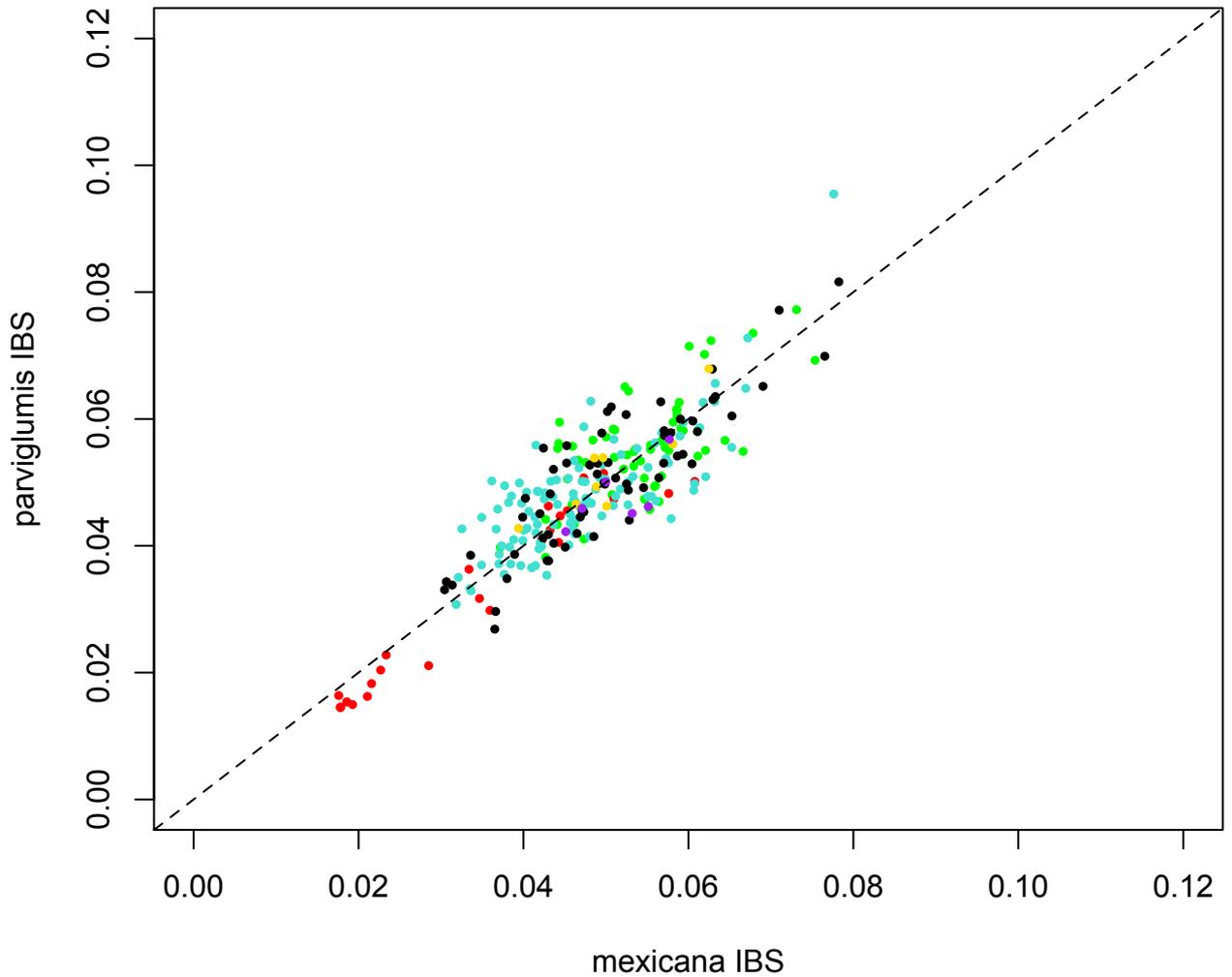

Figure S8

# Chromosome 10

Figure S8

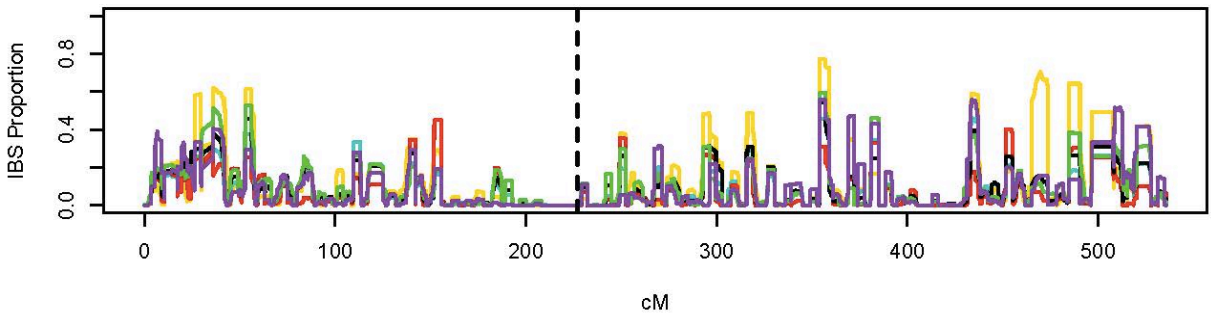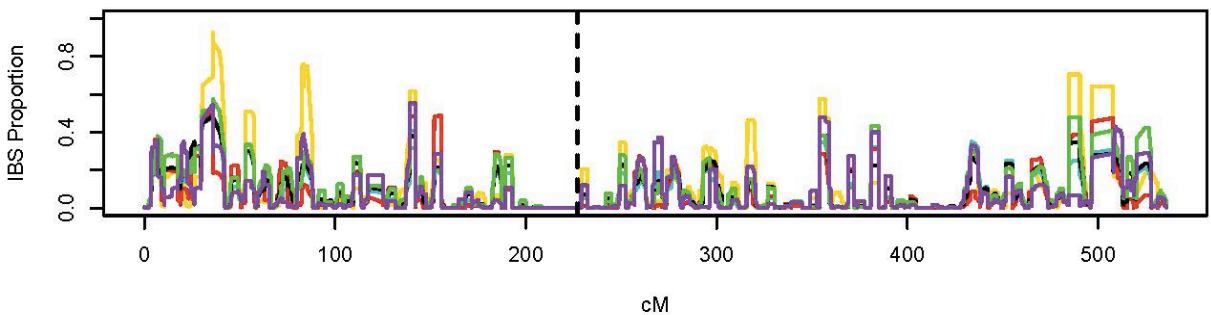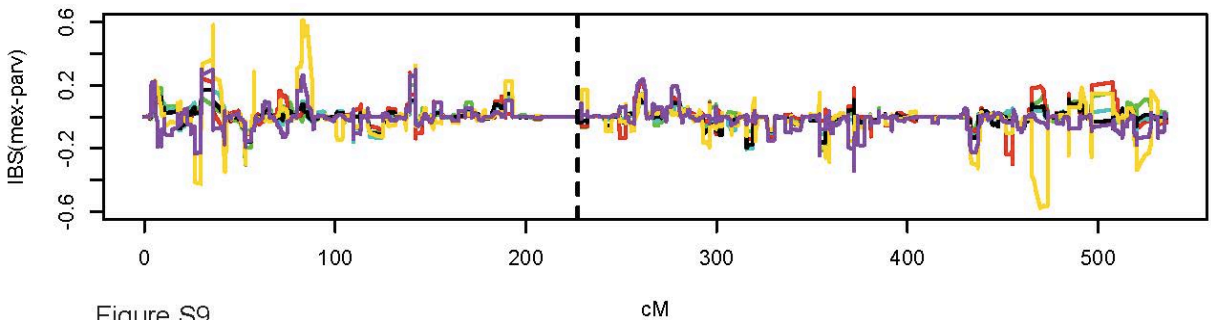

Figure S9

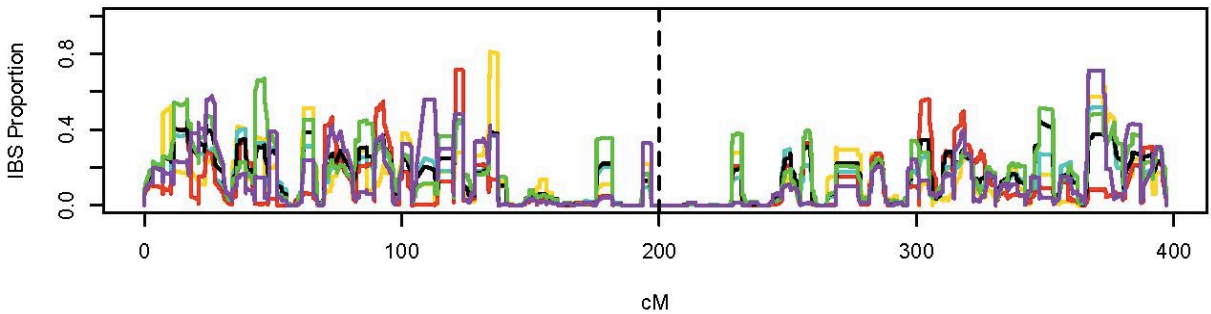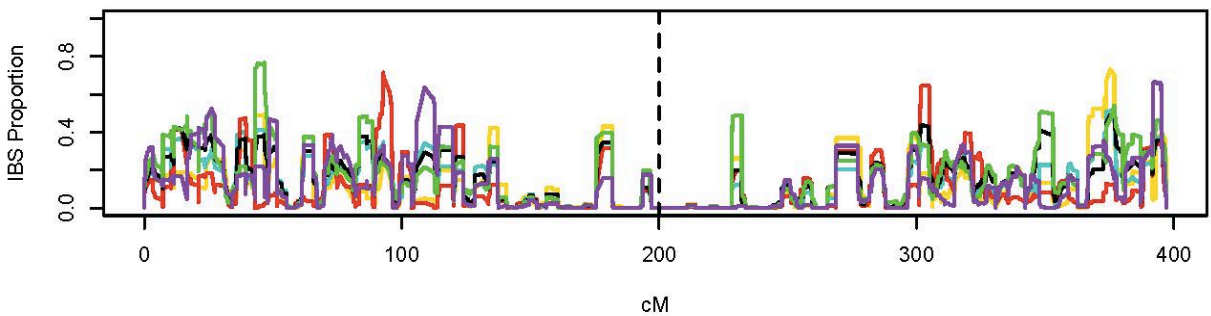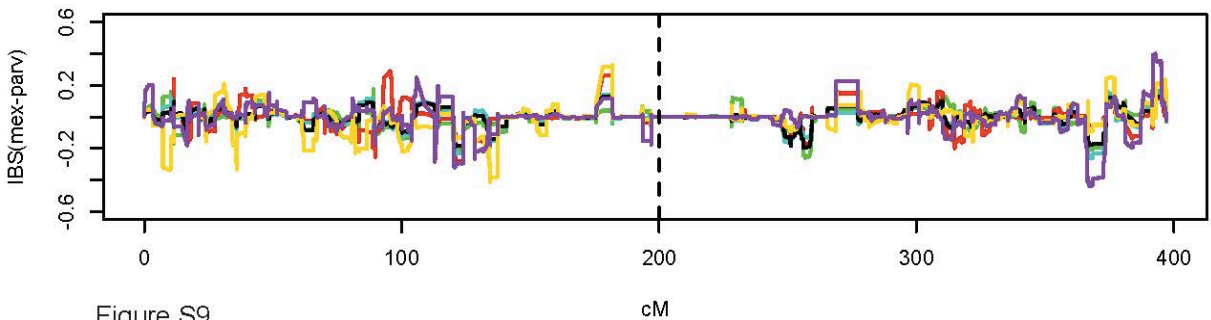
Figure S9

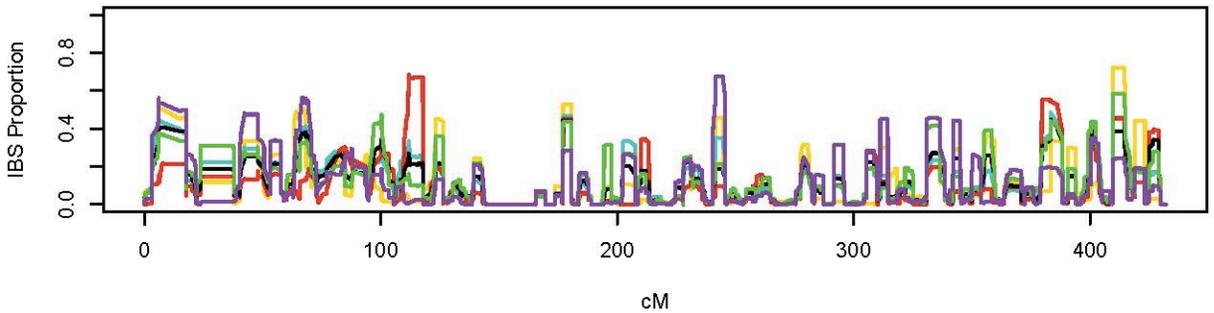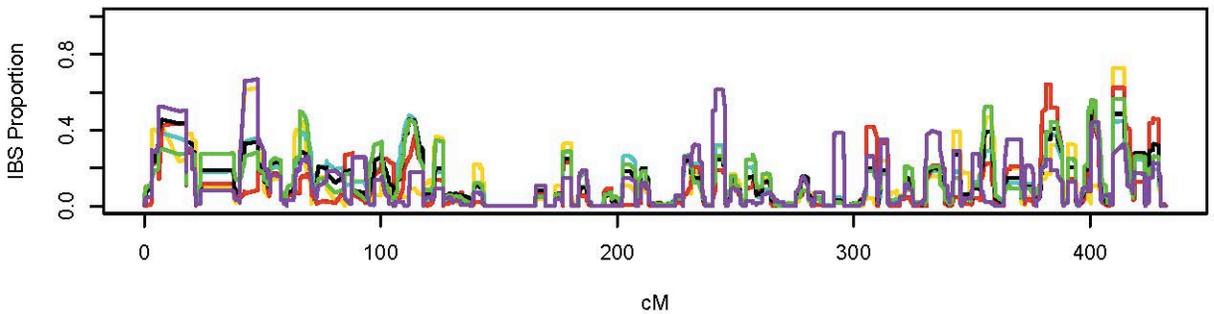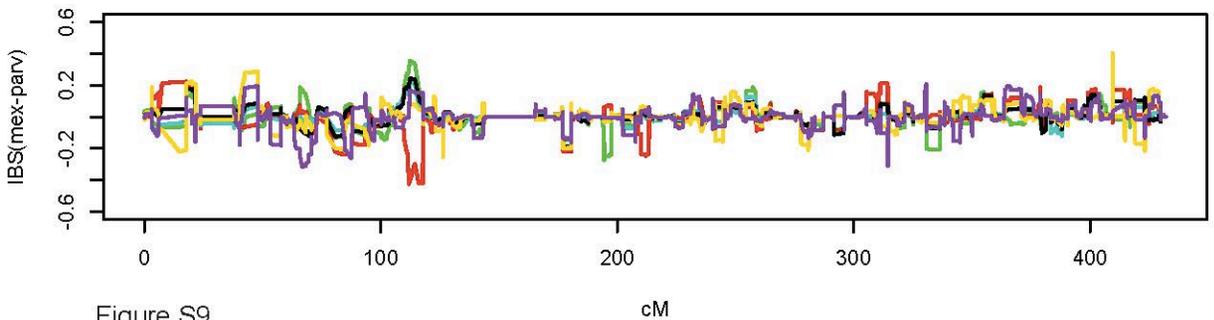

Figure S9

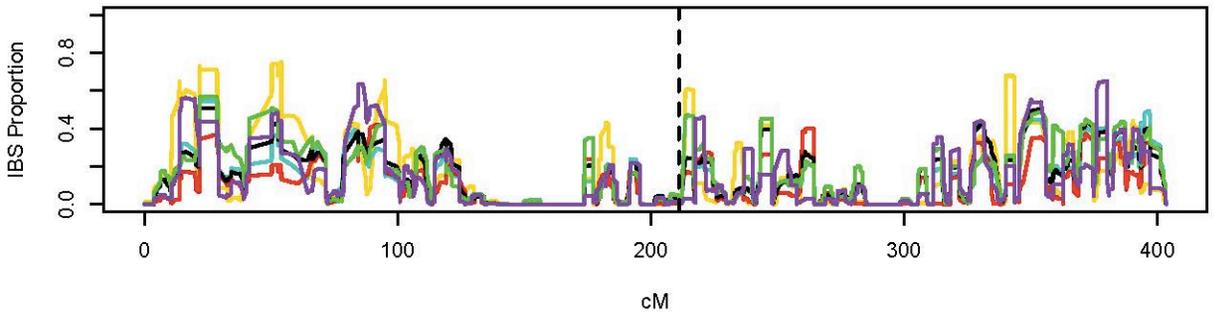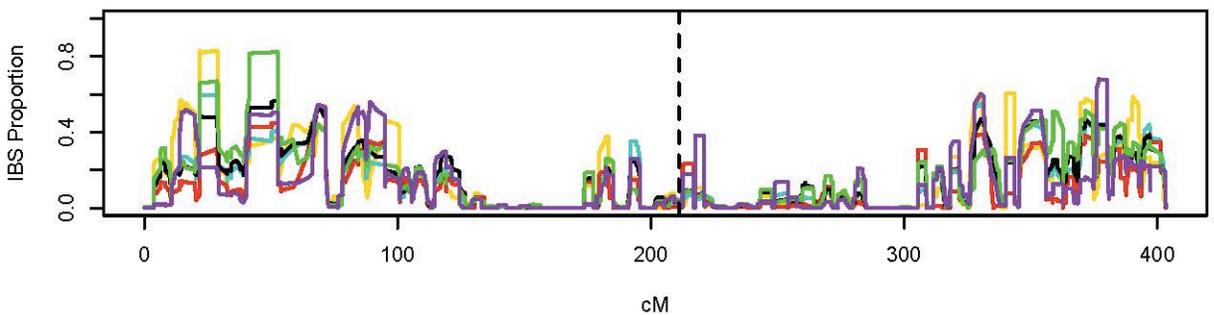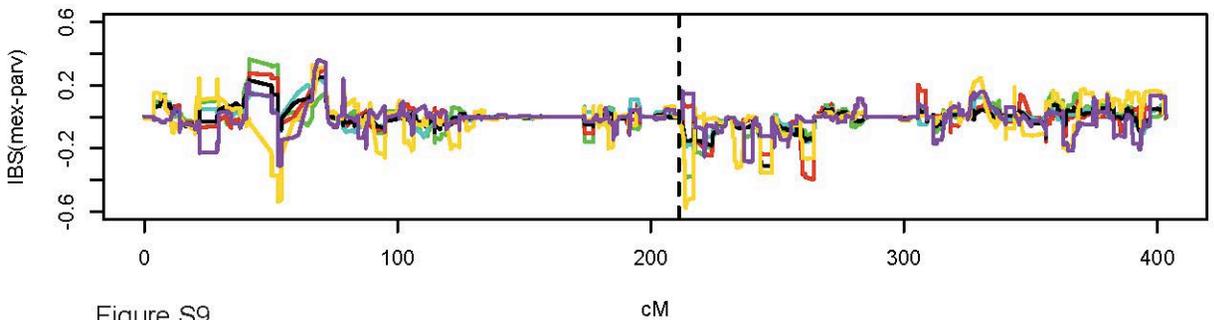

Figure S9

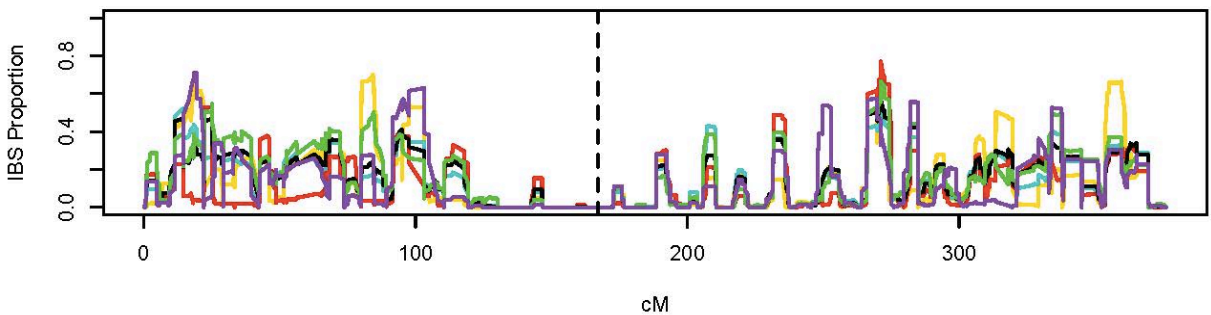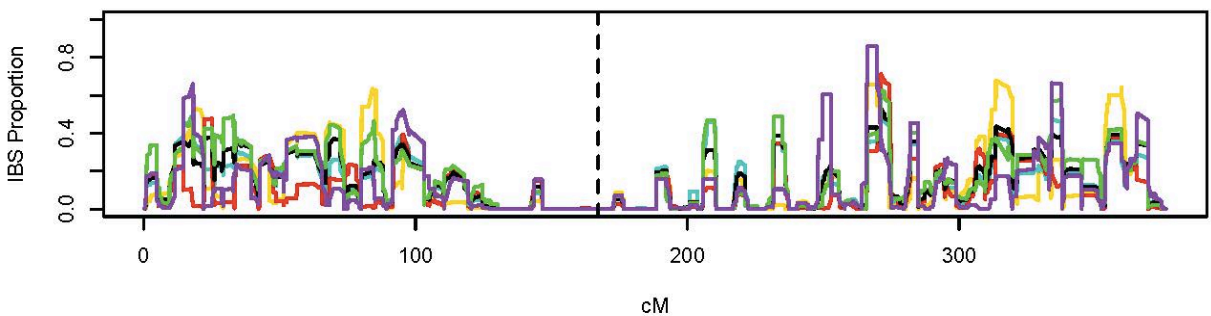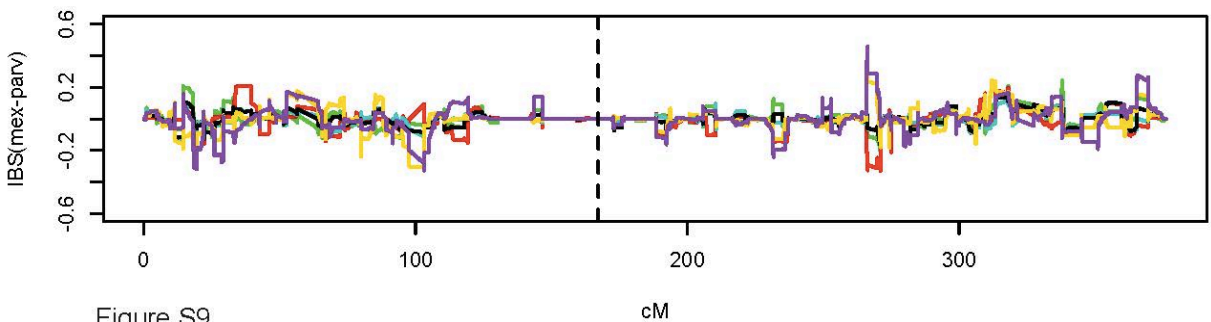

Figure S9

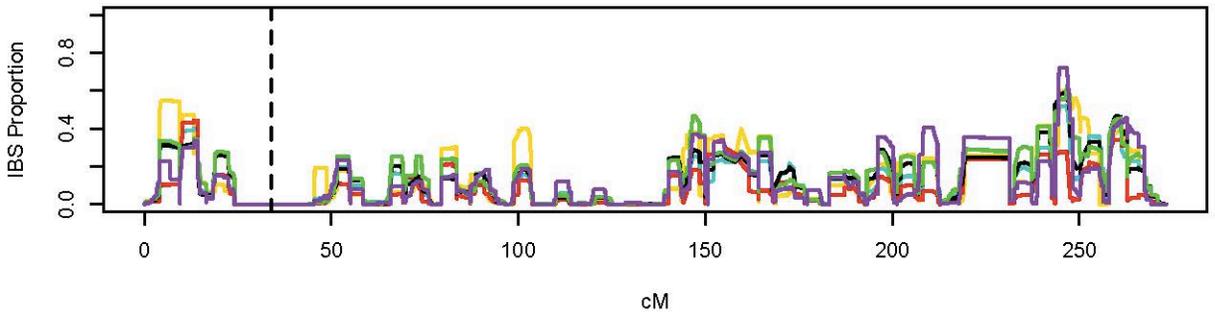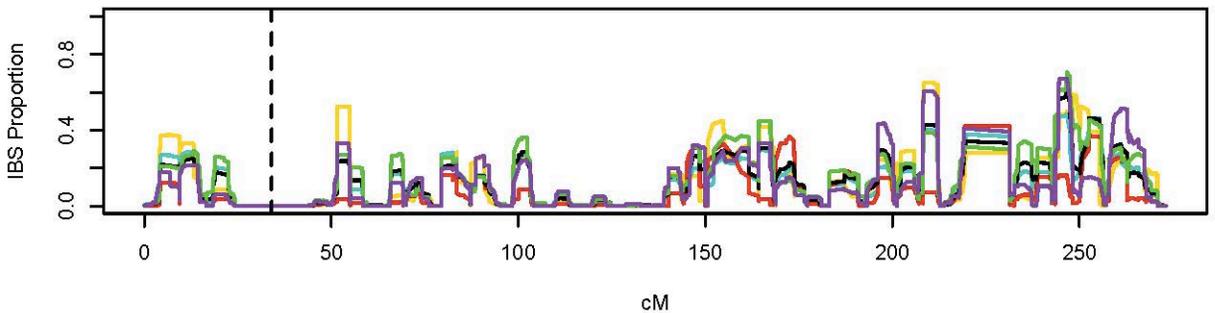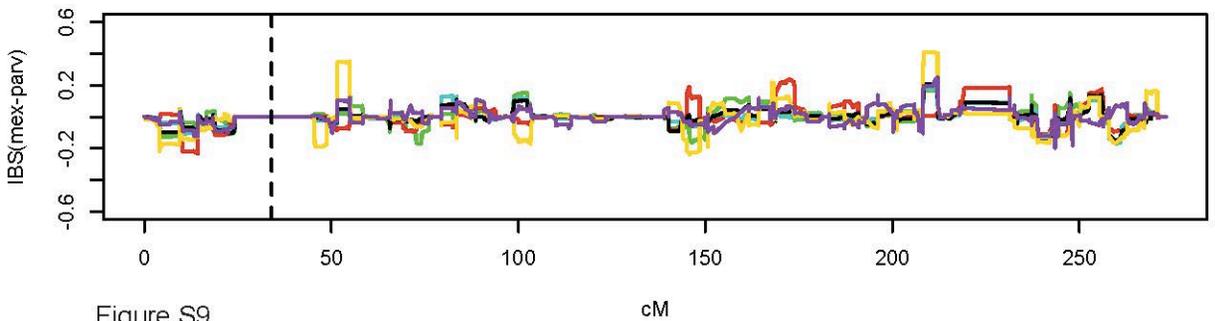

Figure S9

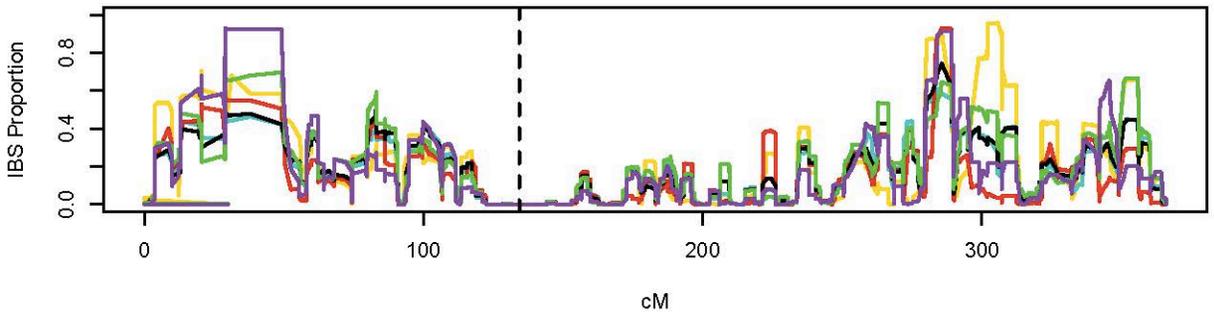
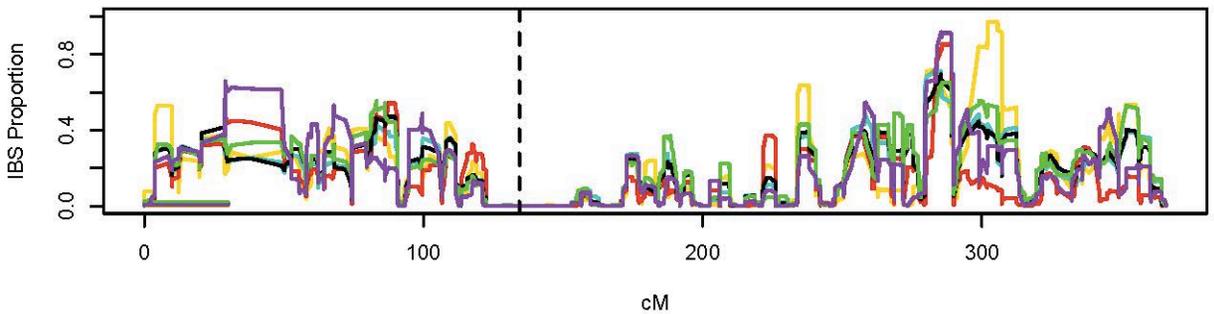
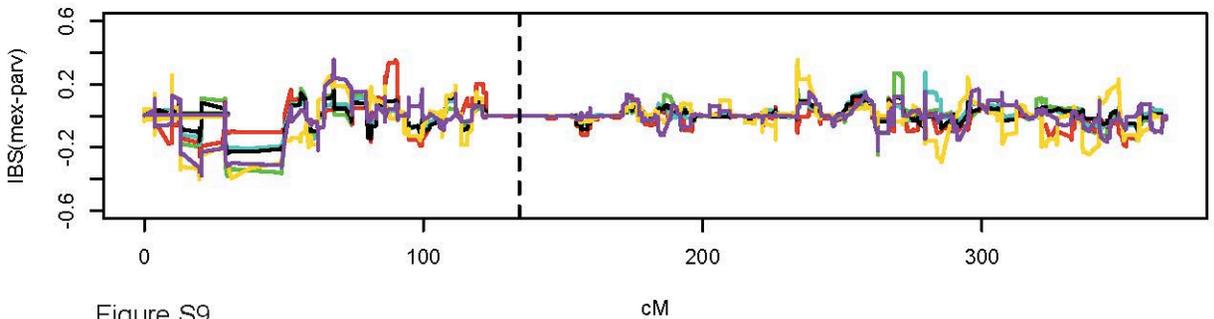

Figure S9

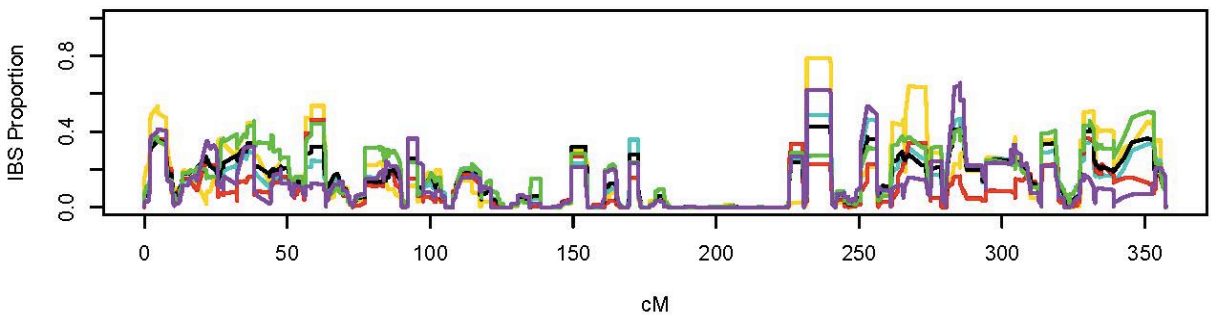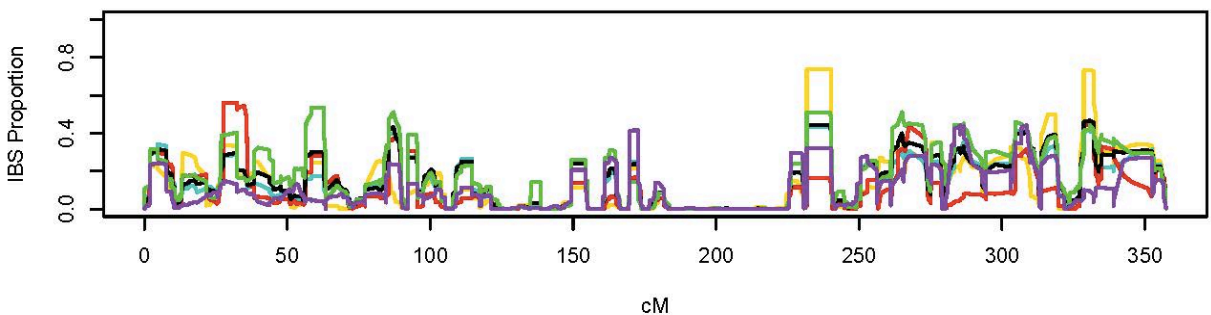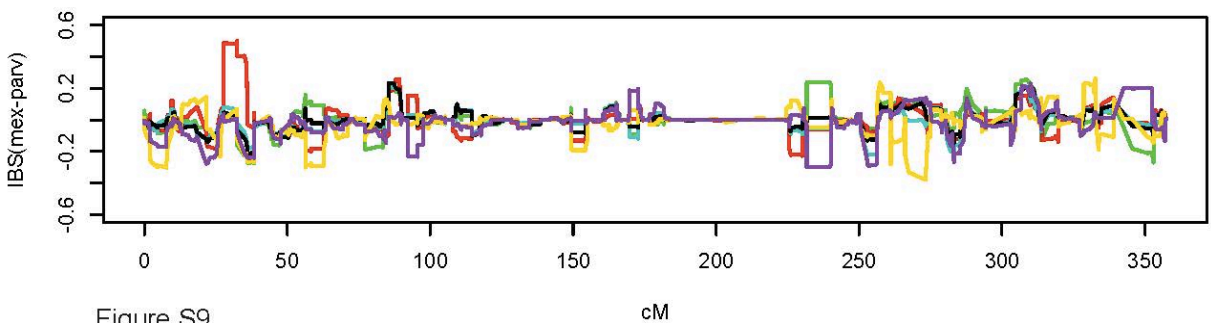

Figure S9

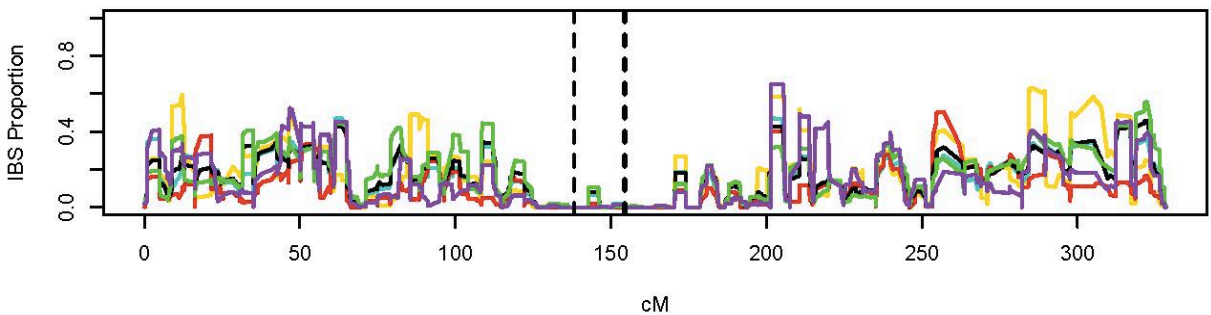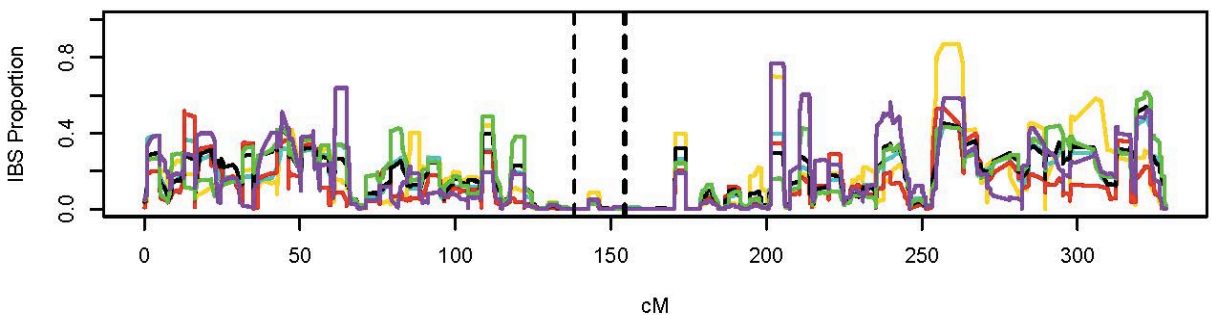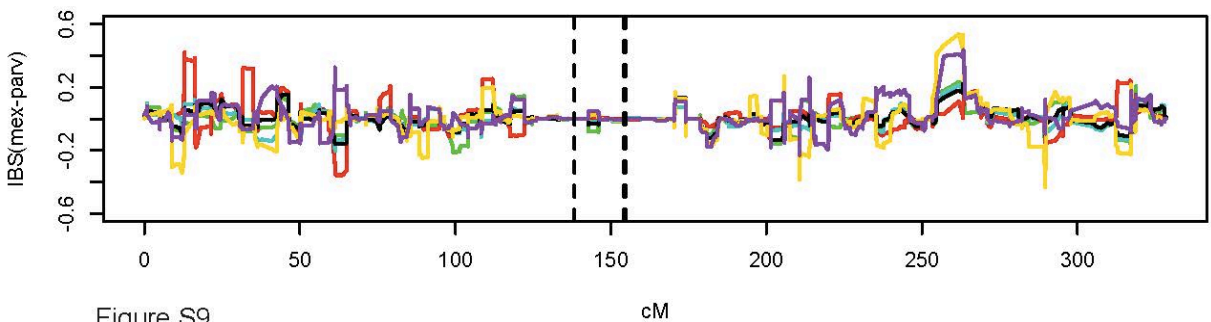

Figure S9

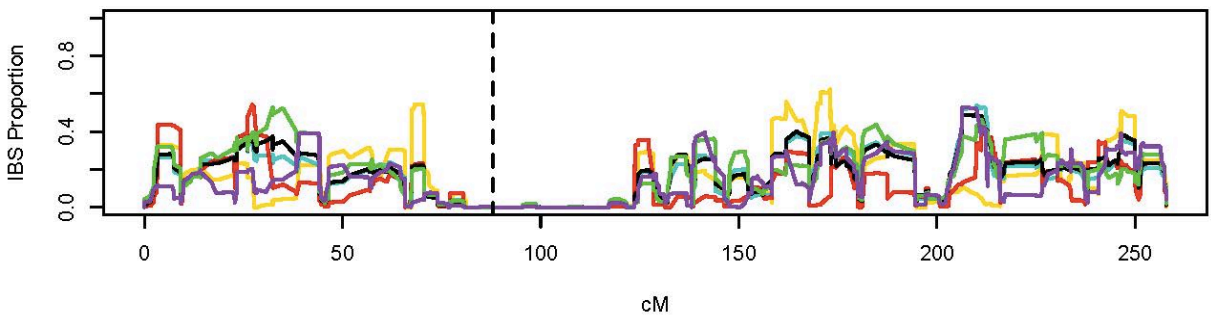
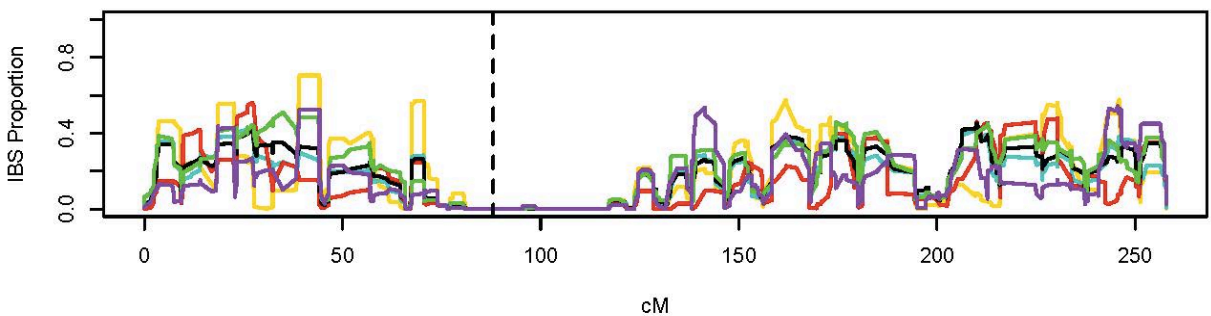
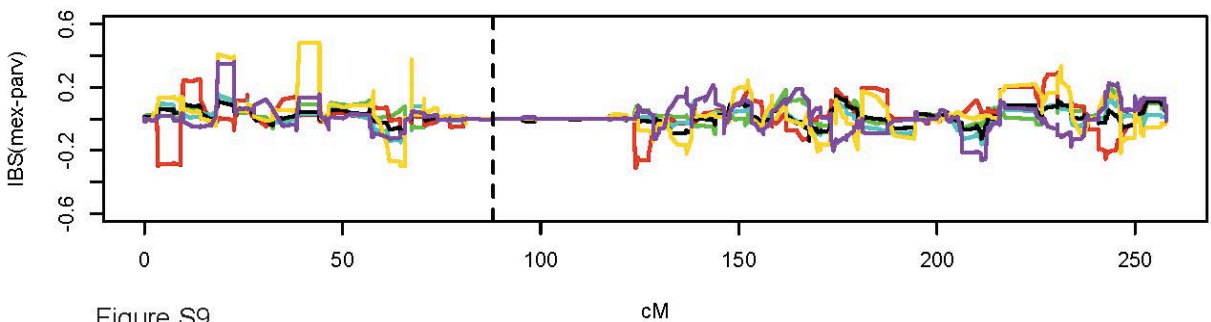

Figure S9

| Sampling Locality | State | Latitude | Longitude | Elevation | Comments |
|---|---|---|---|---|---|
| El Porvenir | Michoacan | 19.68 | -100.64 | 2094 | Sympatric site |
| Ixtlan | Michoacan | 20.17 | -102.37 | 1547 | Sympatric site |
| Nabogame | Chihuahua | 26.25 | -106.92 | 2020 | Sympatric site |
| Opopeo | Michoacan | 19.42 | -101.61 | 2213 | Sympatric site |
| Puruandiro | Michoacan | 20.11 | -101.49 | 1915 | Sympatric site |
| San Pedro | Puebla | 19.09 | -98.49 | 2459 | Sympatric site |
| Santa Clara | Michoacan | 19.42 | -101.64 | 2173 | Sympatric site |
| Tenango del Aire | Mexico | 19.12 | -99.59 | 2609 | Sympatric site |
| Xochimilco | Federal District | 19.29 | -99.08 | 2237 | Sympatric site |
| Puerta Encantada | Morelos | 18.97 | -99.03 | 1658 | Allopatric *mexicana* |

Table S1

| Sampling Locality | $H_E$ Maize | $H_E$ mexicana | %P Maize | %P mexicana | $H_O$ Maize | $H_O$ mexicana | $F_{IS}$ Maize | $F_{IS}$ mexicana |
|---|---|---|---|---|---|---|---|---|
| El Porvenir | 0.308 | 0.215 | 0.837 | 0.704 | 0.307 | 0.199 | 0.004 | 0.073 |
| Ixtlan | 0.224 | 0.202 | 0.515 | 0.668 | 0.210 | 0.172 | 0.063 | 0.148 |
| Nabogame | 0.307 | 0.185 | 0.830 | 0.675 | 0.299 | 0.171 | 0.025 | 0.078 |
| Opopeo | 0.296 | 0.212 | 0.810 | 0.679 | 0.287 | 0.204 | 0.031 | 0.040 |
| Puruandiro | 0.328 | 0.248 | 0.875 | 0.785 | 0.318 | 0.231 | 0.032 | 0.069 |
| San Pedro | 0.303 | 0.198 | 0.808 | 0.612 | 0.297 | 0.190 | 0.021 | 0.042 |
| Santa Clara | 0.298 | 0.175 | 0.810 | 0.559 | 0.294 | 0.163 | 0.014 | 0.070 |
| Tenango del Aire | 0.277 | 0.201 | 0.763 | 0.653 | 0.276 | 0.185 | 0.005 | 0.078 |
| Xochimilco | 0.288 | 0.150 | 0.749 | 0.439 | 0.261 | 0.146 | 0.095 | 0.030 |
| Puerta Encantada | XX | 0.174 | XX | 0.517 | XX | 0.166 | XX | 0.047 |

Table S2

**Summary statistics of introgressed regions**

*mexicana* to maize

| Sampling Locality | $H_E$ Introgressed | $H_E$ Other | $F_{ST}$ Introgressed | $F_{ST}$ Other | Shared Introgressed | Shared Other | Private Maize Introgressed | Private Maize Other | Private *mexicana* Introgressed | Private *mexicana* Other | Fixed Introgressed | Fixed Other |
|---|---|---|---|---|---|---|---|---|---|---|---|---|
| El Porvenir | 0.2753799 | 0.3084655 | 0.1016177 | 0.1392759 | 0.617547807 | 0.75052913 | 0.363329584 | 0.188231035 | 0.01912261 | 0.060710705 | 0 | 0.00052913 |
| Ixtlan | 0.2825794 | 0.221167 | 0.2238203 | 0.3159608 | 0.550239234 | 0.494048764 | 0.169856459 | 0.161054454 | 0.275119617 | 0.318288023 | 0.004784689 | 0.026608758 |
| Nabogame | 0.2788081 | 0.3068175 | 0.1227777 | 0.1826417 | 0.749134948 | 0.749026754 | 0.1816609 | 0.201247964 | 0.069204152 | 0.049559623 | 0 | 0.000165659 |
| Opopeo | 0.2047337 | 0.2993648 | 0.1067488 | 0.1347984 | 0.741559239 | 0.75489081 | 0.195211786 | 0.188239308 | 0.063228975 | 0.056841447 | 0 | 2.84349E-05 |
| Puruandiro | 0.3442468 | 0.3268419 | 0.08557987 | 0.1158877 | 0.807926829 | 0.795771751 | 0.161585366 | 0.146672516 | 0.030487805 | 0.057500891 | 0 | 5.48411E-05 |
| San Pedro | 0.1838618 | 0.3054311 | 0.1234848 | 0.1669365 | 0.749023438 | 0.686886758 | 0.200195313 | 0.254158164 | 0.05078125 | 0.056802616 | 0 | 0.002152461 |
| Santa Clara | 0.1782895 | 0.3018164 | 0.1522303 | 0.188364 | 0.711607787 | 0.659502965 | 0.227108868 | 0.296102796 | 0.061283345 | 0.042276193 | 0 | 0.002118046 |
| Tenango del Aire | 0.1979898 | 0.2788172 | 0.1171819 | 0.1712267 | 0.755832037 | 0.711723128 | 0.169517885 | 0.198530033 | 0.074650078 | 0.089239954 | 0 | 0.000506885 |
| Xochimilco | 0.1925975 | 0.2906366 | 0.2122725 | 0.2481682 | 0.636909871 | 0.565194207 | 0.317596567 | 0.370621913 | 0.042918455 | 0.058514818 | 0.002575107 | 0.005669062 |
| **Mean Value** | **0.23760961** | 0.293262 | **0.13841265** | 0.184806656 | **0.70219791** | 0.68528603 | 0.220673636 | 0.22276202 | **0.0763107** | 0.087748252 | **0.00081776** | 0.004203697 |

maize to *mexicana*

| Sampling Locality | $H_E$ Introgressed | $H_E$ Other | $F_{ST}$ Introgressed | $F_{ST}$ Other | Shared Introgressed | Shared Other | Private Maize Introgressed | Private Maize Other | Private *mexicana* Introgressed | Private *mexicana* Other | Fixed Introgressed | Fixed Other |
|---|---|---|---|---|---|---|---|---|---|---|---|---|
| El Porvenir | 0.2945587 | 0.2130144 | 0.06931251 | 0.1399546 | 0.847309136 | 0.74509695 | 0.088861076 | 0.19476082 | 0.060075094 | 0.080120796 | 0.003754693 | 0.000444469 |
| Ixtlan | 0.2491171 | 0.2011588 | 0.2213745 | 0.3155903 | 0.587591241 | 0.493990088 | 0.102189781 | 0.161596802 | 0.302919708 | 0.64355393 | 0.00729927 | 0.026503847 |
| Nabogame | 0.2481527 | 0.1820343 | 0.1163573 | 0.1836003 | 0.761710794 | 0.74868072 | 0.156822811 | 0.202149937 | 0.081466395 | 0.065450884 | 0 | 0.000167528 |
| Opopeo | 0.2756859 | 0.2054324 | 0.07556263 | 0.1391488 | 0.82228206 | 0.748329687 | 0.112756648 | 0.195204872 | 0.064961292 | 0.075415794 | 0 | 2.95631E-05 |
| Puruandiro | 0.3113247 | 0.2453547 | 0.05417671 | 0.1178099 | 0.830535572 | 0.79466044 | 0.091926459 | 0.148736848 | 0.07753797 | 0.071157999 | 0 | 5.62651E-05 |
| San Pedro | 0.2787583 | 0.1941888 | 0.08787657 | 0.1690437 | 0.790951638 | 0.684921864 | 0.147425897 | 0.256455019 | 0.061622465 | 0.082425488 | 0 | 0.002168098 |
| Santa Clara | 0.2229133 | 0.1659465 | 0.1355008 | 0.1954262 | 0.743426459 | 0.648971155 | 0.20336894 | 0.307244198 | 0.053204601 | 0.06384827 | 0 | 0.002348962 |
| Tenango del Aire | 0.3045455 | 0.1982717 | 0.06728964 | 0.1712072 | 0.830227743 | 0.711708983 | 0.097308489 | 0.198848929 | 0.072463768 | 0.124975817 | 0 | 0.000495677 |
| Xochimilco | 0.2340694 | 0.146462 | 0.1590767 | 0.2507646 | 0.723773585 | 0.561626071 | 0.220377358 | 0.374492558 | 0.05509434 | 0.103503664 | 0.000754717 | 0.005751015 |
| **Mean Value** | **0.26879173** | 0.194651511 | **0.10961415** | 0.186949511 | **0.77086758** | 0.68199844 | **0.13567083** | 0.226609998 | **0.09214951** | 0.145605849 | **0.00131208** | 0.00421838 |

**Summary statistics of regions resistant to introgression**

*mexicana* to maize

| Sampling Locality | $H_E$ Resistant | $H_E$ Other | $F_{ST}$ Resistant | $F_{ST}$ Other | Shared Resistant | Shared Other | Private Maize Resistant | Private Maize Other | Private *mexicana* Resistant | Private *mexicana* Other | Fixed Resistant | Fixed Other |
|---|---|---|---|---|---|---|---|---|---|---|---|---|
| El Porvenir | 0.2749012 | 0.3087061 | 0.2502804 | 0.1348532 | 0.696113074 | 0.748941539 | 0.148409894 | 0.193859526 | 0.151943463 | 0.056778354 | 0.003533569 | 0.00042058 |
| Ixtlan | 0.2054375 | 0.2311528 | 0.3705682 | 0.2827458 | 0.448216167 | 0.520947642 | 0.159716995 | 0.161966739 | 0.349014 | 0.300157373 | 0.043052838 | 0.016928246 |
| Nabogame | 0.2891554 | 0.3070446 | 0.2606789 | 0.1786472 | 0.733965015 | 0.749611856 | 0.180029155 | 0.201750176 | 0.086005831 | 0.048468596 | 0 | 0.000169372 |
| Opopeo | 0.292168 | 0.2952088 | 0.1726507 | 0.1333318 | 0.750577367 | 0.754344957 | 0.140877598 | 0.189115609 | 0.108545035 | 0.056511935 | 0 | 2.74997E-05 |
| Puruandiro | 0.2976903 | 0.3279484 | 0.2126225 | 0.1124372 | 0.740708729 | 0.797671156 | 0.136560069 | 0.147138047 | 0.121866897 | 0.055162738 | 0.000864304 | 2.80584E-05 |
| San Pedro | 0.2663732 | 0.3028015 | 0.271226 | 0.1639294 | 0.612352168 | 0.69022644 | 0.269382392 | 0.252303252 | 0.110381078 | 0.055500056 | 0.007884363 | 0.001970252 |
| Santa Clara | 0.2769316 | 0.2975818 | 0.2246167 | 0.186638 | 0.684420772 | 0.660988737 | 0.225033289 | 0.294928702 | 0.089214381 | 0.042029629 | 0.001331558 | 0.002052932 |
| Tenango del Aire | 0.2461619 | 0.2768872 | 0.2913862 | 0.1659853 | 0.627798507 | 0.715829251 | 0.186567164 | 0.197844647 | 0.184701493 | 0.085850245 | 0.000932836 | 0.000475857 |
| Xochimilco | 0.2558858 | 0.2886745 | 0.3565807 | 0.2433566 | 0.505882353 | 0.569675131 | 0.346666667 | 0.369742695 | 0.127058824 | 0.055543044 | 0.020392157 | 0.005039131 |
| **Mean Value** | **0.26718943** | 0.292889522 | **0.26784559** | 0.177991611 | **0.64444824** | 0.689804079 | **0.19924925** | 0.223183266 | **0.14763678** | 0.084000219 | 0.008665736 | 0.003012437 |

maize to *mexicana*

| Sampling Locality | $H_E$ Resistant | $H_E$ Other | $F_{ST}$ Resistant | $F_{ST}$ Other | Shared Resistant | Shared Other | Private Maize Resistant | Private Maize Other | Private *mexicana* Resistant | Private *mexicana* Other | Fixed Resistant | Fixed Other |
|---|---|---|---|---|---|---|---|---|---|---|---|---|
| El Porvenir | 0.08641262 | 0.2220906 | 0.3091455 | 0.1290763 | 0.45683998 | 0.76384421 | 0.486622918 | 0.175723805 | 0.04997476 | 0.060259651 | 0.006562342 | 0.000172335 |
| Ixtlan | 0.114875 | 0.2053405 | 0.4593768 | 0.3086477 | 0.363636364 | 0.500496665 | 0.300256082 | 0.154987938 | 0.222151088 | 0.322037747 | 0.113956466 | 0.02247765 |
| Nabogame | 0.05491979 | 0.1864328 | 0.3956578 | 0.1773904 | 0.509536785 | 0.753902892 | 0.435967302 | 0.196156726 | 0.047683924 | 0.049912653 | 0.006811989 | 2.77293E-05 |
| Opopeo | 0.08097689 | 0.216191 | 0.267618 | 0.1293434 | 0.501445087 | 0.764182645 | 0.479768786 | 0.177166577 | 0.018786127 | 0.05862254 | 0 | 2.82382E-05 |
| Puruandiro | 0.1197766 | 0.2487197 | 0.3609585 | 0.1135291 | 0.578125 | 0.797790388 | 0.35625 | 0.144968062 | 0.0625 | 0.057214135 | 0.003125 | 2.74145E-05 |
| San Pedro | 0.08837457 | 0.2024909 | 0.3106023 | 0.1601032 | 0.511291257 | 0.697348161 | 0.428488709 | 0.057542059 | 0.038216561 | 0.243997719 | 0.022003474 | 0.001112062 |
| Santa Clara | 0.03887639 | 0.1797279 | 0.3720051 | 0.1802606 | 0.406001225 | 0.673330679 | 0.545009186 | 0.281822318 | 0.016533987 | 0.044221363 | 0.032455603 | 0.00062564 |
| Tenango del Aire | 0.09347815 | 0.2080118 | 0.3133192 | 0.1593214 | 0.533208955 | 0.727408623 | 0.405970149 | 0.181141366 | 0.056716418 | 0.091244834 | 0.004104478 | 0.000205176 |
| Xochimilco | 0.07597781 | 0.1673766 | 0.3584865 | 0.2214519 | 0.421604248 | 0.602678722 | 0.527110117 | 0.330758072 | 0.035774176 | 0.063391923 | 0.015511459 | 0.003171283 |
| **Mean Value** | **0.08374087** | 0.204042422 | **0.34968552** | 0.175458222 | **0.47574321** | 0.697886998 | **0.44060481** | 0.188918547 | **0.06092634** | 0.110100285 | **0.02272565** | 0.00309417 |

Table S3

| chr | region of introgression | in Lauter QTL? | window 1 start | window 1 end | window 2 start | window 2 end |
|---|---|---|---|---|---|---|
| 1 | 120-145Mb | no | 120904094 | 121469572 | 144425995 | 145312340 |
| 2 | 73-78Mb | yes | 73807029 | 78500359 | NA | NA |
| 4 | 169-180Mb | yes | 168753601 | 169803287 | NA | NA |
| 5 | 102-135Mb | yes | 102877443 | 113778281 | 133333397 | 135180623 |
| 6 | 46-56Mb | no | 46111110 | 55854813 | NA | NA |
| 7 | 30-31Mb | no | 30664165 | 31314057 | NA | NA |
| 9 | 107-125Mb | yes | 107315114 | 107840288 | NA | NA |
| 9 | 43Mb | yes | 43903996 | 43903996 | 43287718 | 43287718 |
| 10 | 39-54Mb | yes | 39590245 | 53484238 | NA | NA |

Table S4